\documentclass[11pt]{article}
\pdfoutput=1
\usepackage{jheppub}
\usepackage{amsmath}
\usepackage{bm}
\usepackage{slashed}
\usepackage{graphicx}

\newcommand{\tr}{\mathrm{tr}\,}

\newlength{\dummysp}
\newcommand{\T}{\mathbb{T}}

\newcommand{\beq}{\begin{eqnarray}}
\newcommand{\eeq}{\end{eqnarray}}

\newcommand{\gappeq}{\mathrel{\rlap {\raise.5ex\hbox{$>$}}
{\lower.5ex\hbox{$\sim$}}}}
\newcommand{\lappeq}{\mathrel{\rlap{\raise.5ex\hbox{$<$}}
{\lower.5ex\hbox{$\sim$}}}}

\newcommand{\ben}{\begin{enumerate}}
\newcommand{\een}{\end{enumerate}}

\newcommand{\bit}{\begin{itemize}}
\newcommand{\eit}{\end{itemize}}

\def\[{\left [}
\def\]{\right ]}
\def\({\left (}
\def\){\right )}
\def\R{{\mathbb R}}
\def\S{{\mathbb S}}
\def\Z{{\mathbb Z}}

\title{Higher-order gaugino condensates  on a  twisted $\mathbb T^4$\\ 
{ \hspace{60 mm}\small\it In the beginning was semi-classics....}}

 \author[a]{Mohamed M. Anber,}\author[b]{Erich Poppitz} 
  
\affiliation[a]{Centre for Particle Theory, Department of Mathematical Sciences, Durham University, South Road, Durham DH1 3LE, UK}
\affiliation[b]{Department of Physics,   University of Toronto, 60 St George St., 
Toronto, ON M5S 1A7, Canada}
\emailAdd{mohamed.anber@durham.ac.uk}\emailAdd{poppitz@physics.utoronto.ca}   
   
\abstract{

{\flushleft{W}}e compute the gaugino condensates, $\left\langle \prod_{i=1}^k \text{tr}(\lambda\lambda)(x_i) \right\rangle $ for $1$ $\leq$ $k$ $\le$ $N-1$, in $SU(N)$ super Yang-Mills theory on a small four-dimensional torus $\mathbb{T}^4$, subject to 't Hooft twisted boundary conditions.  Two recent advances are crucial to performing the calculations and interpreting the result: the understanding of generalized anomalies involving $1$-form center symmetry and the construction of multi-fractional instantons on the twisted $\T^4$. These self-dual classical configurations have topological charge $k/N$ and can be described as a sum over $k$ closely packed lumps in an instanton liquid.  Using the path integral formalism, we perform the condensate calculations in the semi-classical limit and find, assuming gcd$(k,N)=1$, $\left\langle \prod_{i=1}^k \text{tr}(\lambda\lambda)(x_i) \right\rangle = {\cal N}^{-1} \; N^2\left(16\pi^2 \Lambda^3\right)^k$, where $\Lambda$ is the strong-coupling scale and ${\cal N}$ is a normalization constant. We determine the normalization constant, using path integral, as ${\cal N} = N^2$, which is $N$ times larger than the normalization used in our earlier publication \cite{Anber:2022qsz}. This finding resolves the extra-factor-of-$N$ discrepancy encountered there, aligning our results with those obtained through direct supersymmetric methods on $\R^4$.  The normalization constant ${\cal N}$ can be understood within the Euclidean path-integral framework as the Witten index $I_W$. From the Hamiltonian approach, it is well-established that $I_W = N$. While the value ${\cal N} = N^2$ correctly reproduces the condensate result, this discrepancy between the Hamiltonian and path-integral formulations calls for reconciliation. We attempt to provide a potential solution we outline in our discussion.}

\begin{document}

\maketitle

\flushbottom

%%%%%%%%%%%%%%%%%%%%%%%%%%%%
\section{Introduction}
%%%%%%%%%%%%%%%%%%%%%%%%%%%%
 
Dynamical mass generation in $4$-dimensional strongly coupled gauge theories is a notoriously difficult problem. Supersymmetry can provide an exceptional means of overcoming this hurdle. The simplest  such gauge theory is ${\cal N}=1$ super Yang-Mills (SYM) theory, which exhibits  a $\mathbb Z_N^{(1)}$ 1-form (center) symmetry and a $\mathbb Z_{2N}^\chi$ chiral (discrete $R$) symmetry. If $\mathbb Z_{2N}^\chi$ fully breaks, a bilinear fermion condensate, the gaugino condensate, must form. Based on dimensional analysis,  the condensate must scale as $\langle\mbox{tr}(\lambda\lambda) \rangle = c \Lambda^3$, where $\Lambda$ represents the strong-coupling scale and $c$ is a dimensionless number. Efforts to determine the exact value of $c$ (given a definition of $\Lambda$) via instanton calculus date back to the 1980s. Two primary methods have been employed in this pursuit: strongly-coupled and weakly-coupled instanton techniques. For comprehensive reviews, see \cite{Shifman:1999mv,Shifman:1999kf,Amati:1988ft,Dorey:2002ik,Vandoren:2008xg,Terning:2006bq}. 

The weak coupling instanton method successfully determines the precise value of the constant.\footnote{We will not discuss the strongly-coupled instanton calculation, whose validity has been questioned many times, see \cite{Dorey:2002ik} for discussion and references. The weakly-coupled result has recently received independent confirmation via a large-$N$ lattice determination \cite{Bonanno:2024bqg}.} One finds $|c| = 16\pi^2$ for $SU(N)$ gauge group (the numerical coefficient was obtained in \cite{Cordes:1985um} and corrected in \cite{Finnell:1995dr}). These calculations are performed on $\mathbb{R}^4$ by starting with super QCD: this is SYM endowed with additional $N-1$ massive fundamental chiral supermultiplets. In the small-mass limit, the vacuum expectation values of the scalars are much larger than the strong scale, leading the gauge group to fully break and pushing the theory into the weak coupling regime. The superpotential of this theory is constructed and minimized. Subsequently, the masses are increased beyond the strong scale, causing the fundamental flavours to decouple and giving  SYM as the limiting theory. Utilizing the power of holomorphy then yields the value of $c$ in SYM. This method can also be used to calculate higher-order gaugino condensates: one finds $\langle \prod_{i=1}^k\mbox{tr}(\lambda\lambda)(x_i)\rangle=\left(\langle\mbox{tr}(\lambda\lambda) \rangle\right)^k=\left(16\pi^2\Lambda^3\right)^k$. This result is remarkable as it features two important aspects. The first is clustering, a generic property of any local and Lorentz-invariant quantum field theory: the expectation value of the connected correlator of two operators  $\langle {\cal O}_1(x_1){\cal O}_2(0)\rangle$ must decompose as $\langle {\cal O}_1\rangle\langle {\cal O}_2\rangle$ in the limit $|x_1|\rightarrow\infty$, with obvious generalization for more than two operators. The second feature is specific to supersymmetric theories:  the correlation functions are independent of the insertion points $x_i$. 

Although the method based on the superpotential yields the correct result, it falls short of providing an understanding of the microscopic origin of the mechanisms driving dynamical mass generation. To address this, an approach was considered in \cite{Davies:1999uw,Davies:2000nw}, where one of the spatial directions is compactified on a small circle $\mathbb{S}^1$ with a circumference smaller than $\Lambda^{-1}$. This setup places the theory on $\mathbb{R}^3 \times \mathbb{S}^1$. The compactification pushes the theory into the weak-coupling regime, revealing monopole-instantons as the semi-classical microscopic objects responsible for catalyzing symmetry breaking.\footnote{Confinement and chiral symmetry breaking in this theory are due to the magnetic-bion mechanism  \cite{Unsal:2007jx}.} In this context, the bilinear gaugino condensate calculations yield a value of $|c|=16\pi^2$, consistent with the results obtained via supersymmetry and holomorphy.

Exploring geometries beyond $\mathbb{R}^3 \times \mathbb{S}^1$ to understand the origin of dynamical symmetry breaking and calculate the exact value of $c$ was also pursued as early as 1984. In ref.~\cite{Cohen:1983fd}, the gaugino condensate in the background of a $4$-dimensional torus, $\T^4$, with 't Hooft twisted boundary conditions was considered. However, the coefficient $c$ was not calculated\footnote{The calculation of $c$  was not possible then, for reasons reviewed in  \cite{Anber:2022qsz} and further below.} until our recent work  \cite{Anber:2022qsz}, where we studied the $SU(2)$ case using the path integral formalism.

The renewed interest in the $\T^4$ geometry was motivated by the growing interest in generalized global symmetries \cite{Gaiotto:2014kfa}, operations that extend beyond those acting on local fields to those acting on higher-dimensional objects. A particular example is the $\mathbb Z_N^{(1)}$ $1$-form symmetry that acts on Wilson's lines in $SU(N)$ SYM. One important application of this extended symmetry is that one can examine the behavior of the partition function as we perform a discrete chiral transformation in the background gauge field of $\mathbb Z_N^{(1)}$. This yields a $\mathbb Z_N$ phase, which is interpreted as a mixed 't Hooft anomaly between $\mathbb Z_N^{(1)}$ and $\mathbb Z_{2N}^{\chi}$. Assuming SYM confines in the IR, the anomaly implies that the chiral symmetry must break. Turning on a background gauge field of a discrete $1$-form symmetry can only be performed on manifolds with non-trivial $2$-cycles, and $\mathbb T^4$ is the most natural and simplest example of such manifold. In this context, the non-trivial twists on $\mathbb{T}^4$ induce discrete $2$-form fluxes on its $2$-cycles, leading to instantons with fractional topological charges $Q = 1/N$. According to the index theorem, an adjoint fermion must have two zero modes in such a background, resembling a bilinear condensate. This suggests that the fractional instantons responsible for the mixed 't Hooft anomaly could also provide the microscopic origin of dynamical mass generation---a situation when you have the cake and eat it too.

Our calculations in \cite{Anber:2022qsz} (which focused on the $SU(2)$ case) yielded $\langle\mbox{tr}(\lambda\lambda) \rangle = 2\times(16\pi^2 \Lambda^3)$,  twice the value computed using supersymmetry technology\footnote{For brevity, in the rest of the paper, we use the phrase ``$\R^4$ result'' to refer to the result of the weakly-coupled instanton calculation of the gaugino condensate on $\R^4$ or $\R^3 \times \S^1$.}   on $\mathbb R^4$. This created a puzzle that warranted further examination of the situation. This paper extends our calculations on the twisted $\mathbb T^4$ to $SU(N)$ aiming to:
\begin{enumerate}
 \item Understand the origin of the mismatch between the $\mathbb R^4$ and the $\mathbb T^4$ results  for the bilinear condensate. 
  \item Examine higher-order condensates and check the clustering in the infinite volume limit. 
 \end{enumerate}
 One critical requirement for Yang-Mills instantons is that these solutions must be self-dual. Without self-duality, fluctuations in such a background could have negative modes, leading to instabilities. As shown by 't Hooft \cite{tHooft:1981nnx}, there exist simple self-dual abelian solutions to the full non-abelian Yang-Mills equations of motion on $\mathbb{T}^4$ that carry fractional topological charges $Q = k/N$ for $1 \leq k \leq N-1$. They can be obtained by turning on discrete 't Hooft fluxes (or, in other words, by applying twisted boundary conditions) along two of the $2$-cycles of $\mathbb T^4$, say along the 12 and 34 planes. Self-duality is then ensured if the periods of $\mathbb{T}^4$, denoted by $L_\mu$ ($\mu = 1, 2, 3, 4$), satisfy the condition $L_1L_2 = (N-k)L_3L_4$. However, as noted in \cite{Anber:2022qsz}, these solutions admit more fermion zero modes than necessary to saturate the condensates. Additionally, in this case, the adjoint matter contributes a source term to the Yang-Mills equations of motion, rendering these solutions invalid as legitimate backgrounds. To address these issues and lift the extra fermion zero modes, we detune the $\mathbb{T}^4$ periods by introducing a small detuning parameter $\Delta\equiv((N-k)kL_3L_4 - kL_1L_2)/\sqrt{L_1L_2L_3L_4}$. This adjustment allows for identifying an approximate self-dual solution to the Yang-Mills equations of motion as a series expansion in $\Delta$. The price one pays, however, is that such solutions are fully nonabelian. This method, which originated in \cite{GarciaPerez:2000aiw, Gonzalez-Arroyo:2019wpu} for instantons with topological charge $Q=1/N$, was further developed by the authors in \cite{Anber:2023sjn} for  $Q = k/N$, $1 \leq k \leq N-1$.

\begin{figure}[t] %  figure placement: here, top, bottom, or page
   \centering
   \includegraphics[width=3in]{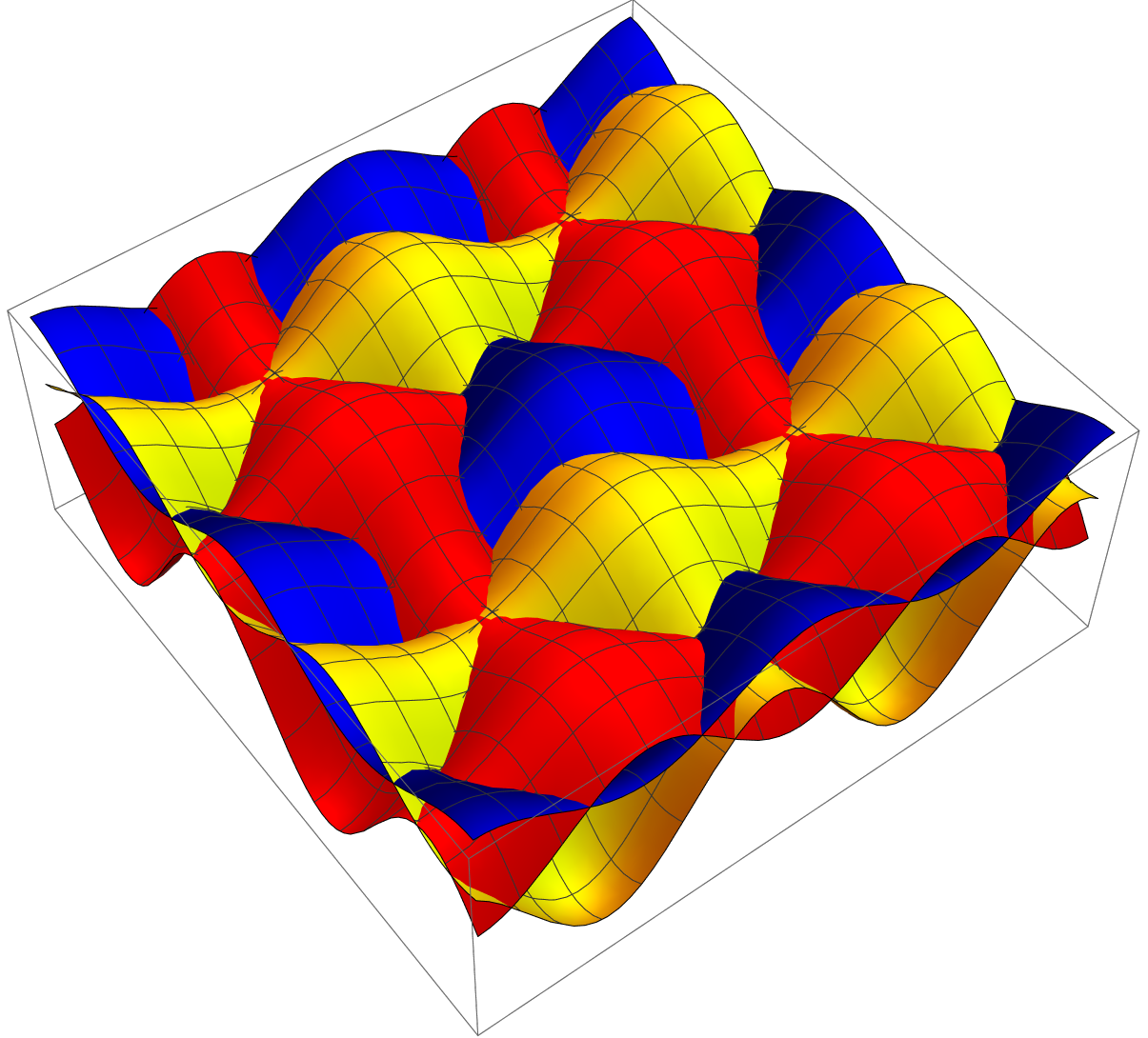} 
   \caption{The multi-fractional instanton solution of charge $Q=k/N$.
Displayed is a 3D plot of the profile described by eqn.~(\ref{multilump1}) with $k = 3$, plotted as a function of $x_1, x_2$ while keeping  $x_3, x_4$ fixed. To enhance visualization, the plot extends to double the periods in $x_1$ and $x_2$. The graph reveals three lumps, each one described by the function $F$ of (\ref{multilump1}) (itself defined in (\ref{fractional1})) but with a different center. These are represented by red, yellow, and blue, clustered (lumped) around the three distinct centers. These lumps, however, are closely packed, more akin to a liquid than a dilute gas. Previously, similar configurations were generated numerically to investigate confinement, as detailed in \cite{Gonzalez-Arroyo:1996eos} and further explored in \cite{Gonzalez-Arroyo:2023kqv}. }
   \label{visual of liquid}
\end{figure}

 The nonabelian solution of topological charge $Q=k/N$ can be represented as a sum over $k$ closely packed lumps, resembling instanton-liquid on $\T^4$, see Figure \ref{visual of liquid} for a visualization. It admits $k$ distinct holonomies in each spacetime direction (the holonomies are along the Cartan generators of the group $U(k)$) for a total of $4k$ holonomies. These constitute a compact bosonic moduli space of dimension $4k$, as per the index theorem. Identifying the symmetries and determining the shape and volume of this space is crucial for computing the condensates. Additionally, each lump supports two adjoint fermion zero modes, for a total of $2k$  zero modes needed to saturate the higher-order gaugino condensates $\langle \prod_{i=1}^k\mbox{tr}(\lambda\lambda)(x_i)\rangle$. 
 
 %%%%%%%%%%%
 %%%%%%%%%%%%%%%

 In calculating these condensates, we limit our analysis to order-$\Delta^0$, as the explicit form of the full solution to order-$\Delta$, while in principle obtainable in a systematic manner,  is complicated and has not yet been found. Yet, using supersymmetric Ward identities, one can show that the condensates  can depend neither on $\Delta$ nor on the insertions $x_i$. Thus, even though the calculations are performed to ${\cal O}(\Delta^0)$, they must be exact. Our path integral computations of the condensates give:
 \begin{eqnarray}\label{our main result of the cond}
\left\langle \prod_{i=1}^k\mbox{tr}(\lambda\lambda)(x_i) \right\rangle
={\cal N}^{-1} \; N^2\left(16\pi^2 \Lambda^3\right)^k\,,
\end{eqnarray}
 where ${\cal N}$ is a normalization constant, a path integral without operator insertion.
 
  To obtain a meaningful non-zero normalization constant, it is necessary to apply appropriate boundary conditions on $\mathbb{T}^4$. Applying twists in both $12$ and $34$ planes in the path integral defining the normalization factor would result in fermion zero modes, causing $\mathcal{N}$ to vanish. To prevent the occurrence of zero modes, we apply twists of $-k$ along only one of the planes of $\mathbb{T}^4$:
 \begin{eqnarray}
 {\cal N} = \sum_{\nu \in \mathbb Z}\int [D A_\mu][D\lambda][D\bar \lambda] [D D]\; e^{-S_{SYM}}\bigg\vert_{n_{12}=-k\,,n_{34}=0}\,.\end{eqnarray}
 The quantity $\mathcal{N}$ is recognized as the path-integral formulation of the Witten index, $I_W$. In previous computations using the Hamiltonian formalism, $I_W$ was identified as $N$ for $SU(N)$ SYM, as famously calculated by Witten \cite{Witten:1982df}. This was the value we adopted for $\mathcal{N}$ in \cite{Anber:2022qsz}, leading to an additional factor of $N$ in our earlier calculations of the bilinear condensate (or an additional factor of $2$ in the case of $SU(2)$). Upon revisiting our analysis, we have discovered that the path integral computations actually yield ${\cal N}=N^2$ instead of $N$. This correction eliminates the extra factor of $N$ found in \cite{Anber:2022qsz} and provides the correct value for the condensate (\ref{our main result of the cond}).  While this resolves our earlier issue, it introduces a discrepancy in the Witten index between the Hamiltonian and path integral formalisms. We provide a possible approach toward a resolution of this issue. We point out that a similar problem, i.e., a discrepancy between the Hamiltonian and path-integral formalisms, arises in the $\mathbb Z_N$ BF theory (a topological field theory) formulated on a torus, and a careful definition of the measure by means of a triangulation or lattice formulation resolves the issue. We argue that this formulation holds lessons for the definition of the measure in the Yang-Mills theory. However, since Yang-Mills theory is not a topological theory, achieving a complete resolution of the discrepancy  is a challenging task that is left for the future.
 
This paper is organized as follows. In Section \ref{sec:reviewsolution}, we succinctly review the self-dual instanton calculations on the deformed $\mathbb T^4$, providing the reader with the necessary background to cruise smoothly into the paper. In particular, we introduce our notation and explain the nature of the lumpy structure we found in \cite{Anber:2023sjn}, the origin of the bosonic moduli space, and the gaugino zero modes in the background of these lumps. Section \ref{Shape and volume of the bosonic moduli space} is devoted to a detailed study of the bosonic moduli space, as identifying the shape and volume of this space is indispensable for studying the gaugino condensates. These results are employed in Section \ref{The gaugino condensates} to carry out the calculations of the higher-order condensates using the path integral formalism. Contrasting the results in the path integral and Hamiltonian formalisms, calculating the normalization constant ${\cal N}$, and resolving the puzzle encountered in our previous publication  \cite{Anber:2022qsz} is carried out in Section \ref{sec:hamiltonian}. We end with concluding remarks and outlook in Section \ref{Conclusions}. 

To maintain the main text at a manageable length, we have moved many important and detailed calculations to the appendices. In Appendix \ref{appx:ward}, we work out supersymmetric Ward identities on $\mathbb T^4$ in the presence of twists, showing that the condensates must be holomorphic in the strong scale $\Lambda$ and $x_i$ insertion-independent. In Appendix \ref{appx:solution}, we present the explicit order-$\sqrt\Delta$ solution of the full nonabelian instanton with topological charge $k/N$. The gaugino-zero modes' explicit form in the nonabelian solution's background is reviewed in Appendix \ref{appx:fermion}. Many important calculations needed to determine the symmetries of Wilson's lines are discussed in Appendix \ref{appx:wilson}.  The shape and volume of the bosonic moduli space are determined in Appendix \ref{Determining the shape of the moduli space}. Appendix \ref{appx:localization} contains a proposal for using the localization technique to compute the Witten index. Finally, in Appendix \ref{appx:measure}, we discuss  the $\mathbb Z_N$ BF theory on  $\mathbb T^2$ and draw lessons about defining the measure in Yang-Mills theory.

%%%%%%%%%%%%%%%%%%%%%%%%%%%%%%%%%%%%%%%%%%%%%%%%%%%%%%%%
\section{Review of self-dual instantons on the deformed $\mathbb T^4$}
\label{sec:reviewsolution}
%%%%%%%%%%%%%%%%%%%%%%%%%%%%%%%%%%%%%%%%%%%%%%%%%%%%%%%%
In this section, we introduce the notation and summarize the solution of the self-dual fractional instanton on the detuned $\mathbb T^4$. We shall be brief yet give sufficient information about the instanton backgrounds to make the exposition self-contained. For more details and derivations, see \cite{Anber:2023sjn}.

\subsection{Action, boundary conditions, and transition functions}
\label{sec:actionboundary}

We study minimal $SU(N)$ super-Yang-Mills theory in four dimensions on the four torus. Its Euclidean action is:
\begin{equation}
\label{symaction2}
S_{SYM} = {1 \over g^2} \int\limits_{\T^4} \tr_\Box \left[ {1\over 2} F_{\mu\nu} F_{\mu\nu} + 2 (\partial_\mu \bar\lambda_{\dot\alpha} + i [A_\mu, \bar\lambda_{\dot\alpha}]) \bar\sigma_\mu^{\dot\alpha \alpha} \lambda_\alpha  + D^2\right]~.
\end{equation}
Here $A_\mu = A_\mu^a T^a$ is the $SU(N)$ gauge field with hermitian Lie-algebra generators obeying $\tr_\Box \left(T^a T^b\right) =  \delta^{ab}$, $\lambda_\alpha = \lambda_\alpha^a T^a$ is the adjoint fermion (gaugino), and the field strength is $F_{\mu\nu}=\partial_\mu A_\nu-\partial_\nu A_\mu +i[A_\mu, A_\nu]$.  The symbol $\Box$ denotes the defining (fundamental) representation, with the normalization $\tr_\Box \left(T^a T^b\right) =  \delta^{ab}$ chosen to ensure that the simple roots satisfy $\bm \alpha^2=2$. Under this normalization, the root and co-root lattices, as well as the weight and co-weight lattices, become identical, significantly simplifying the analysis. From now on, we remove the symbol $\Box$ from the traces, remembering that the trace is always taken in the defining representation.   The adjoint gaugino field is represented by $\bar\lambda_{\dot\alpha} =\bar\lambda_{\dot\alpha}^a T^a$ and $\lambda_\alpha = \lambda_\alpha^a T^a$, independent complex Grassmann variables, and $D= D^a T^a$ is the scalar auxiliary field of the vector supermultiplet of minimal 4d supersymmetry.\footnote{\label{footnote:notation0}Here, $\sigma_\mu \equiv(i\vec\sigma,1)$, $\bar\sigma_\mu \equiv(-i\vec\sigma,1)$, $\vec \sigma$ are the Pauli matrices which determine the $\mu={1,2,3}$ components of the four-vectors $\sigma_\mu, \bar\sigma_\mu$. In addition, for any spinor, $\eta^\alpha = \epsilon^{\alpha \beta} \eta_\beta$, with $\epsilon^{12} = \epsilon_{21} = 1$, and likewise for the dotted ones. In addition, 
$
\bar\sigma_\mu^{\dot\alpha \alpha} = \epsilon^{\dot\alpha\dot\beta} \epsilon^{\alpha\beta} \sigma_{\mu \; \beta \dot\beta}$, $\sigma_{\mu \; \beta \dot\beta} = \epsilon_{\beta\alpha} \epsilon_{\dot\beta \dot\alpha} \bar\sigma_\mu^{\dot\alpha \alpha}$. All our notation is that of \cite{Dorey:2002ik}, except that we use Hermitean gauge fields.} As the theory only has adjoint fields, it has a $1$-form $\Z_N^{(1)}$  (in the modern terminology \cite{Gaiotto:2014kfa}) global center symmetry acting on Wilson line operators.
It   also  has a $0$-form $\Z_{2N}^{\chi}$ global chiral symmetry acting on the gaugino as $\lambda \rightarrow e^{i {2 \pi \over  2 N}} \lambda$. These symmetries have a mixed anomaly \cite{Gaiotto:2014kfa,Gaiotto:2017yup}, which will play a role in our discussion in section \ref{sec:hamiltonian}.

We take the torus to have periods of length $L_\mu$, $\mu=1,2,3,4$, where  $\mu,\nu$ runs over the spacetime dimensions. The gauge fields $A_\mu$  obey the boundary conditions
\begin{eqnarray}
A_\nu(x+L_\mu \hat e_\mu)=\Omega_\mu(x) A_\nu(x) \Omega_\mu^{-1}(x)-i \Omega_\mu(x) \partial_\nu \Omega_\mu^{-1}(x)\,,
\label{conditions on gauge field}
\end{eqnarray}
as we traverse $\mathbb T^4$ in each direction. The boundary conditions ensure that local gauge invariant quantities are periodic functions of $x$, with periods equal to the periods of $\T^4$. The fermions $\lambda$, $\bar\lambda$, and the auxiliary field $D$ from (\ref{symaction2}) obey identical boundary conditions, but without the inhomogeneous term in (\ref{conditions on gauge field}). The action (\ref{symaction2}) is invariant under supersymmetry transforms,\footnote{These are given in Appendix \ref{appx:localization}, eqn.~(\ref{susydelta}).} with supersymmetry consistent with the twisted boundary conditions on $\T^4$.

 Here, $\Omega_\mu$ are
the transition functions (or twist matrices), $N\times N$ unitary matrices, and  $\hat e_\nu$ are unit vectors in the $x_\nu$ direction. The transition functions satisfy the cocycle conditions:
\begin{eqnarray}\label{cocycle}
\Omega_\mu (x + \hat{e}_\nu L_\nu) \; \Omega_\nu (x) = e^{i {2 \pi \over N} n_{\mu\nu}}\; \Omega_\nu (x+ \hat{e}_\mu L_\mu) \; \Omega_\mu (x)\,,
\end{eqnarray}
where the exponent $e^{i {2 \pi \over N} n_{\mu\nu}}$, with integers $n_{\mu\nu}=-n_{\nu\mu}$, is in the $\mathbb Z_N$ center of $SU(N)$.
The nonvanishing twists that we shall consider in this paper are of the form\footnote{In ref.~\cite{Anber:2023sjn}, the more general case with $n_{12}=-r$ (instead of (\ref{twists1})) is studied. Thus, (\ref{twists1}) corresponds to taking $r=k$. This case is  singled out for reasons discussed there, also mentioned in Appendix \ref{appx:solution}.}
\begin{equation}\label{twists1}
n_{12} = - n_{21} = - k,~ n_{34} = - n_{43}= 1,
\end{equation}
and are chosen so   a Yang-Mills configuration  obeying (\ref{conditions on gauge field})  carries   fractional topological charge \cite{tHooft:1979rtg,tHooft:1981sps,vanBaal:1982ag}:
\begin{eqnarray} \label{Q of n}
Q= -\frac{n_{12}n_{34}+n_{13}n_{42}+n_{14}n_{23}}{N} ~({\rm mod} \; 1) = {k \over N} ~({\rm mod} \; 1)~. 
\end{eqnarray}
We study the entire range of  possible values $k \in {1,\ldots , N-1}$.

't Hooft \cite{tHooft:1981nnx} found a solution to the cocycle conditions (\ref{cocycle}), giving rise to the fractional $Q$ in (\ref{Q of n}). This was achieved by embedding the $SU(N)$ transition functions $\Omega_\mu(x)$ in $SU(k)\times SU(\ell)\times U(1)\subset SU(N)$, such that $N=k+\ell$.  To present the solution, we use the same notation followed in \cite{Anber:2023sjn}: we take primed upper-case Latin letters to denote elements of $k\times k$ matrices: $C', D'=1,2,...,k$, and the unprimed upper-case Latin letters to denote $\ell\times \ell$ matrices: $C,D=1,2,..,\ell$.   We also introduce the matrices $P_k$ and $Q_k$ (similarly the matrices $P_\ell$ and $Q_\ell$), the $k\times k$ (similarly $\ell\times\ell$) shift and clock matrices satisfying the relation 
\begin{equation}
\label{clockshift}
P_kQ_k=e^{i\frac{2\pi}{k}}Q_kP_k .
\end{equation} Explicitly, we have that  $(P_k)_{B'C'}=\gamma_k\delta_{B',C'-1 \; (\text{mod}\, k)}$ and $(Q_k)_{C'B'}=\gamma_k \; e^{i2\pi \frac{C'-1}{k}}\delta_{C'B'}$, for the matrix elements of $P_k$ and $Q_k$, where the coefficient $\gamma_k=e^{i\frac{\pi(1-k)}{k}}$ is chosen to ensure that $\mbox{Det}(P_k)=\mbox{Det}(Q_k)=1$. The matrix $\omega$ is the $U(1)$ generator:
\begin{eqnarray}\label{omega}
\omega=2\pi\mbox{diag}(\underbrace{\ell, \ell,...,\ell}_{k\, \mbox{times}},\underbrace{ -k,-k,...,-k}_{\ell\,\mbox{times}})\,,
\end{eqnarray}
commuting with $P_k,P_\ell, Q_k,Q_\ell$. 

Without much ado, we give the explicit form of the transition functions $\Omega_\mu$ obeying (\ref{cocycle}) with $n_{\mu\nu}$ of (\ref{twists1}):
 \begin{eqnarray}
\nonumber
\Omega_1&=& (-1)^{k-1}I_k \oplus I_\ell e^{i \omega \frac{ x_2}{N  L_2}} = \left[\begin{array}{cc}(-1)^{k-1}I_k e^{i2\pi \ell   \frac{x_2}{N  L_2}}&0\\0& e^{-i 2\pi k\frac{x_2}{NL_2}}I_\ell\end{array}\right]\,,\quad
\Omega_2=Q_k\oplus I_\ell = \left[\begin{array}{cc}Q_k&0\\0& I_\ell\end{array}\right],\\
\Omega_3&=&I_k\oplus P_\ell e^{i \omega \frac{x_4}{N\ell L_4}} = \left[\begin{array}{cc} e^{i2\pi  \frac{x_4}{N L_4}} I_k&0\\0& e^{-i 2\pi k\frac{x_4}{N \ell L_4}}P_\ell\end{array}\right]\,,\quad
\Omega_4=I_k\oplus Q_\ell = \left[\begin{array}{cc}I_k&0\\0& Q_\ell\end{array}\right],
\label{the set of transition functions for Q equal r over N, general solution}
\end{eqnarray}
and $I_k$ ($I_\ell$) is the $k\times k$ ($\ell\times \ell$) unit matrix, reminding the reader that $\ell = N-k$. The reader can easily check that they obey the correct cocycle conditions, eqns.~(\ref{cocycle}, \ref{twists1}).
%%%%%%%%%%%%%%%%%%%%%%%%%%%%%%%%%%%%%
\subsection{The abelian self-dual solution on the self-dual $\mathbb T^4$}
%%%%%%%%%%%%%%%%%%%%%%%%%%%%%%%%%%%%%
\label{sec:abeliansoltn}

It was also shown by 't Hooft  \cite{tHooft:1981nnx} that an abelian gauge field configuration along the $U(1)$ generator $\omega$ of (\ref{omega}) exists, which obeys the boundary conditions (\ref{conditions on gauge field}) specified by the $\Omega_\mu$  given in (\ref{the set of transition functions for Q equal r over N, general solution}) and which satisfies the vacuum Yang-Mills equations of motion. Our choice of $n_{\mu\nu}$ (\ref{twists1}) and the transition functions gives the abelian solution 
 \begin{equation}\label{gaugewithholonomies}
\hat A_\mu = A_\mu+\delta A_\mu= A_\mu + \left[\begin{array}{cc}|| \delta A_{\mu \; C' B'}||&0\\0& ||\delta A_{\mu \; C B}|| \end{array}\right]~,
 \end{equation}
 where we recall that $C',B' = 1, ...k$, while $C,B = 1,...\ell$ and $N=k+\ell$. We have split the solution into two parts. The first term ($A_\mu$) is the moduli-independent part and the second ($\delta A_\mu$) contains the dependence on the moduli. We start by discussing the first term.  The moduli-independent part of the solution $A_\mu$ is given in terms of $\omega$ of (\ref{omega}) by
 \begin{eqnarray}\label{naked expressions}
 A_1&=&0\,, \quad A_2=-\omega\frac{   x_1}{N  L_1L_2}\,, \quad  A_3=0\,, ~\quad A_4=-\omega \frac{x_3}{N\ell L_3L_4}\,.
 \end{eqnarray}
 The corresponding field strength  is constant  on $\mathbb T^4$:
\begin{eqnarray}\label{abelian F to leading}
F_{12}=\hat F_{12}&=&-\omega\frac{1}{N  L_1L_2}\,,\quad F_{34}=\hat F_{34}=-\omega\frac{1}{N\ell L_3L_4}\,.
\end{eqnarray}
 The reader can verify that the topological charge of this solution is $Q={k \over N}$.
A self-dual fractional instanton must satisfy the relation $F_{12}=F_{34}$, from which we find that  the ratio
of the torus sides has to be tuned to
\begin{eqnarray}\label{selfdualtorus1}
\frac{L_1L_2}{L_3L_4}=N-k \,.
\end{eqnarray}
A torus with periods that satisfy the above relation is said to be a self-dual torus. The action of the self-dual solution is
\begin{eqnarray}\label{action of fractional instanton}
S_{0}=\frac{1}{2g^2}\int_{\mathbb T^4}\mbox{tr}\left[F_{\mu\nu}F_{\mu\nu}\right]=\frac{8\pi^2|Q|}{g^2}=\frac{8\pi^2 k}{N g^2}\,.
\end{eqnarray}

Next, we discuss the moduli term in (\ref{gaugewithholonomies}).  $|| \delta A_{\mu \; C' B'}||$ is a $k\times k$ matrix with components $\delta A_{\mu \; C' B'}$, while  $|| \delta A_{\mu \; C B}||$ is a $\ell \times \ell$ matrix with components $\delta A_{\mu \; C B}$. From the index theorem, one expects that there are a total of $4 k$ bosonic moduli, as appropriate for a self-dual solution of topological charge ${k \over N}$.
As the moduli space plays a crucial role in the calculation of the gaugino condensates, we now give two equivalent parameterizations of the moduli, both of which are used at various stages later in the paper.  
 
 Among these, 
there are $4$ translational moduli denoted by $z_\mu$. In addition, there are $4(k-1)$ moduli, denoted by $\phi_\mu^{C'}$. These are the holonomies along the $SU(k)$ Cartan generators in each spacetime direction.  The matrix components  $\delta A_{\mu \; C B}$ and $ \delta A_{\mu \; C' D'}$ are given in terms of $z_\mu$ and $\phi_\mu^{C'}$  by
 \begin{eqnarray}\label{deltaA1}
 \delta A_{\mu \; C D} &=& \delta_{CD} \; 2 \pi k {z_\mu \over L_\mu}\,,\quad
 \delta A_{\mu \; C' D'} = \delta_{C'D'}(- 2 \pi \ell {z_\mu \over L_\mu} + \phi_\mu^{C'})\,,
\end{eqnarray}
where $\phi_\mu^{C'} = \phi_\mu^{C'-k ({\rm mod} \; k)} \equiv \phi_\mu^{\[C'-k\]_k} \; {\rm and} \; \sum_{C'=1}^{k}\phi_\mu^{C'} = 0$ and we use the notation $\[x\]_q \equiv x ({\rm mod} \; q)$.\footnote{\label{footnotewrap}The notation $[C'-k]_k$, allowing the ``wrapping" of the index $C'$ past $k$ is only used when displaying the explicit form of the nonabelian solution.} Clearly, for the special case of $k=1$, there are only $4$ translational moduli $z_\mu$, and we set the holonomies $\phi_\mu^{C'}=0$. 

For the general case of $k>1$,   we can equivalently write $\delta A_\mu$ of (\ref{deltaA1}) using the Cartan generators $\bm H_k$ of $SU(k)$, embedded in $SU(N)$ by adding zeros in their lower $\ell \times \ell$ block,  as
\begin{eqnarray}
\nonumber
\delta A_1&=&-\omega \frac{z_1}{L_1}+{2\pi \over L_1} \bm a_1\cdot \bm H_k\,,\quad \delta A_2=-\omega\frac{z_2}{L_2}+{2\pi \over L_2} \bm a_2\cdot \bm H_k\,,\\
\delta A_3&=&-\omega \frac{z_3}{L_3}+{2\pi \over L_3} \bm a_3\cdot \bm H_k\,,\quad  \delta A_4=-\omega \frac{z_4}{L_4}+{2\pi \over L_4} \bm a_4\cdot \bm H_k\,,
\label{r over N abelian sol}
\end{eqnarray}
where, e.g., $\bm a_\mu=(a^1_\mu,a^2_\mu,..,a^{k-1}_\mu)$. 
Here ${\bm H}_k\equiv (H_k^1,...,H_k^{k-1})$ are the $SU(k)$ Cartan generators obeying $\tr { H}_k^{a}  { H}_k^{b} = \delta^{ab}$, $a,b=1,...,k-1$. Recall also that these can be expressed via the weights of the fundamental representation, ${ H}_k^b$=
diag$({ \nu}^b_1,  { \nu}^b_2,..., { \nu}^b_k)$, where $\bm\nu_1,...,\bm\nu_k$ are the weights of the fundamental representation of $SU(k)$. These are $(k-1)$-dimensional vectors that obey $\bm \nu_{B'} \cdot \bm \nu_{C'} = \delta_{B'C'} - {1 \over k}$, where $B',C'=1,..,k$. 

%In the following, we shall use both forms of the moduli $\phi^{C'}_\mu$ and $\bm a_\mu$, as each of them proves useful in some context. While the former is easier to use when writing explicit expressions of general solutions of non-abelian fractional instantons, as we shall see in the next section,  the latter is more convenient when integrating over the moduli space in the path integral.

%%%%%%%%%%%%%%%%%%%%%%%%%%%%%%%%%%%%%
\subsection{The  nonabelian self-dual solution on the deformed $\mathbb T^4$ and its (multi-) lump structure.}
%%%%%%%%%%%%%%%%%%%%%%%%%%%%%%%%%%%%%
\label{sec:lumpy}

The self-dual abelian solution in the background of a self-dual $\mathbb T^4$ has a simple form, and it is tempting to use it to compute the gaugino condensates. However, as was shown in \cite{Anber:2022qsz}, the trouble with the abelian solution is that it yields more fermion zero modes than needed. In particular, the Dirac equation of both the dotted and the undotted spinors have normalizable solutions,\footnote{In other words, the Dirac operators $D$ and $\bar D$ have non-empty kernels. This, however, does not contradict the index theorem since the index is given the by difference  ${\cal I}=\mbox{ker} D-\mbox{ker}\bar D$.} while only the undotted (or dotted) spinors are expected to have zero modes. To make things worse, the additional zero modes source the Yang-Mills equations of motion, rendering the self-dual abelian solution inconsistent in the presence of adjoint matter. 

Ensuring the solution's self-duality is crucial for maintaining stability, as it prevents the presence of negative modes in the background. To lift the extra fermion-zero modes, we consider a non-self-dual (deformed) $\mathbb{T}^4$, while still requiring that the Yang-Mills solution itself remains self-dual. This approach necessitates exploring nonabelian solutions. There are no known exact self-dual nonabelian solutions on non-self-dual $\mathbb T^4$.  Yet, one can devise a method to find an approximate solution using perturbation analysis. The brief discussion below and in Appendices \ref{appx:solution}, \ref{appx:fermion}  is intended to  give an idea of the method, originated in \cite{GarciaPerez:2000aiw,Gonzalez-Arroyo:2019wpu} and further developed in \cite{Anber:2022qsz,Anber:2023sjn}, and discuss the properties of the solutions relevant for our calculation of the gaugino condensate.

One begins by introducing the detuning parameter $\Delta$,  parameterizing the deviation from the self-dual torus:
\begin{eqnarray}\label{deltadef}
\Delta\equiv \frac{k \ell L_3L_4-k L_1 L_2}{\sqrt{V}}\,,
\end{eqnarray}
and $V=\prod_{\mu=1}^4L_\mu$ is the volume of $\mathbb T^4$.
We assume, without loss of generality, $\Delta\ge0$. We search for a self-dual instanton solution with topological charge $Q=\frac{k}{N}$ on a deformed $\mathbb T^4$, following the strategy of  \cite{GarciaPerez:2000aiw,Gonzalez-Arroyo:2019wpu}. We write the general gauge field  on the non-self-dual torus in the form
\begin{eqnarray}
A_\mu(x)= \hat A_\mu +{\cal S}_\mu^\omega(x) \;\omega +\delta_\mu(x)\,.
\label{gauge field in general}
\end{eqnarray}
Here, $\hat A_\mu$ is the abelian gauge field with constant field strength defined previously in  (\ref{gaugewithholonomies}) and ${\cal S}_\mu^\omega(x)$ is the nonconstant field component along the $U(1)$ generator. The non-abelian part $\delta_\mu(x)$ is given by an $N\times N$ matrix, which is decomposed in a block form:
\begin{eqnarray}\label{fields1}
\delta_\mu=\left[\begin{array}{cc} {\cal S}_{\mu }^k & {\cal W}_\mu^{k\times \ell}\\ {\cal W}_\mu^{\dagger \ell \times k}& {\cal S}_\mu^\ell\end{array} \right] \; \; \;\;  \equiv \left[\begin{array}{cc} ||{\cal S}_{\mu \; B' C'}^k|| & ||{\cal W}_{\mu \; B' C}|| \\ ||{\cal (W}_{\mu}^\dagger)_{C B'}||& ||{\cal S}_{\mu \; B C}^\ell||\end{array} \right] \,.
\end{eqnarray}
Next, we write the various functions as series expansions in $\Delta$:
\begin{eqnarray}
{\cal W}^{k\times \ell}_\mu=\sqrt{\Delta} \sum_{a=0}^\infty \Delta^a {\cal W}^{(a)k\times \ell}_\mu\,,\quad
{\cal S}_\mu= \Delta \sum_{a=0}^\infty \Delta^a{\cal S}^{(a)}_\mu\,, \label{expansion1}
\end{eqnarray}
where ${\cal S}_\mu$ accounts for ${\cal S}^\omega_\mu$, ${\cal S}^k_\mu$, and ${\cal S}^\ell_\mu$.
The field strength $F_{\mu\nu}$ of the instanton configuration is 
\begin{eqnarray}\nonumber
F_{\mu\nu}&=&\partial_\mu A_\nu-\partial_\nu A_\mu +i[A_\mu, A_\nu]\\
&=&\hat F_{\mu\nu}+F_{\mu\nu}^s\omega+\left[\begin{array}{cc} F_{\mu\nu}^k & {\cal F}_{\mu\nu}^{k\times\ell}\\ {\cal F}_{\mu\nu}^{\dagger \ell\times k} & F_{\mu\nu}^{\ell} \end{array}\right]\,,
\label{feild strength}
\end{eqnarray}
where $\hat F_{\mu\nu}$ is given by (\ref{abelian F to leading}), 
while the explicit expressions of $F_{\mu\nu}^s$, etc. in terms of ${\cal W}^{k\times \ell}_\mu$ and  ${\cal S}_\mu$  are given in \cite{Anber:2023sjn}, and we refrain from repeating these expressions here as they do not serve us any later convenience. Since we are looking solely for a self-dual solution, we impose the self-duality constraint
 \begin{eqnarray}
\bar \sigma_{\mu\nu}F_{\mu\nu}=0\,,\label{selfduality1}
\end{eqnarray}
where
  $\bar\sigma_{\mu\nu}=\frac{1}{4}(\bar\sigma_\mu\sigma_\nu-\bar\sigma_\nu\sigma_\mu)$. 

One proceeds by imposing the constraint (\ref{selfduality1}) to the leading order in $\Delta$ by considering solutions of ${\cal W}^{k\times \ell}_\mu$ to order $\sqrt {\Delta}$ and ${\cal S}_\mu$ to order $\Delta$, thus keeping only the terms ${\cal S}^{(0)}_\mu$ and ${\cal W}^{(0)}_\mu$ in (\ref{expansion1}). The solution of the resulting equations that satisfy the boundary conditions (\ref{conditions on gauge field}) was found in \cite{Anber:2023sjn}. To  ${\cal{O}}(\sqrt{\Delta})$, the solution for ${\cal W}^{(0)}_\mu$ in (\ref{expansion1}) is given in eqns.~(\ref{expressions of W2 and W4 with holonomies}--\ref{form of Phi}) of Appendix \ref{appx:solution}.

One of the main results of \cite{Anber:2023sjn} was that a solution with $Q={k \over N}$ consists of $k$ strongly overlapping lumps.  This can be envisaged by studying the $x$-dependence of gauge-invariant densities, e.g., $\mbox{tr}\left[F_{12}F_{12} \right]$.  From the formulae given in Appendix \ref{appx:solution}, one finds that the $x$-dependent part of this gauge invariant density has the form
\begin{eqnarray}
&&\mbox{tr}\left[F_{12}F_{12} \right] \label{multilump1} \\
&\sim &\sum\limits_{C'=1}^k  F \left(x_1 - {L_1 L_2 \over 2 \pi} \hat\phi_2^{C'} - {L_1 C' \over k},\; 
x_2 +  {L_1 L_2 \over 2 \pi} \hat\phi_1^{C'},\; x_3 -  {\ell L_3 L_4 \over 2 \pi} \hat\phi_4^{C'},\; x_4+  { \ell L_3 L_4 \over 2 \pi} \hat\phi_3^{C'} \right),  \nonumber 
\end{eqnarray}
where $
\hat \phi_\mu^{C'} \equiv - 2 \pi N {z_\mu \over L_\mu} + {2 \pi \over L_\mu} \bm{a}_\mu \cdot \bm{\nu}_{C'} 
$ is related to the moduli (\ref{r over N abelian sol}) and the function $F$ is explicitly defined in (\ref{fractional1}). 

The important point is that,  
for every $C'=1,2,..,k$, the summand is given by the same function $F(x_1,x_2, x_3,x_4)$ defined above, but centered (lumped) at a different point $x_\mu$ on $\mathbb T^4$. The location of these lumps is specified by the moduli $\hat{\phi}_\mu^{C'}$ for $\mu = 1, 2, 3, 4 $ and $C' = 1, \ldots, k$. The sizes of these lumps are inherently tied to the size of $\mathbb{T}^4$, which serves as the sole scale in this scenario. Consequently, these lumps are not distinctly separate but significantly overlap, resembling a liquid's behavior more than a dilute gas. In Figure \ref{visual of liquid}, we give a visual representation of this lumpy structure.

%%%%%%%%%%%%%%%%%%%%%%%%%%%%%%%%%%%%%%%%%%%
\subsection{Fermion zero modes on the  deformed $\mathbb T^4$ and their localization on the ``lumps'' of the multi-fractional instanton}
%%%%%%%%%%%%%%%%%%%%%%%%%%%%%%%%%%%%%%%%%%%
\label{sec:fermzero}

The fermion zero modes are found by solving the Dirac equation $\sigma^\mu D_\mu \lambda=0$ in the self-dual background (\ref{gauge field in general}). 
To simplify the treatment,  we cast the $\lambda$ matrix in the form
\begin{eqnarray}
\lambda=\left[\begin{array}{cc}||\lambda_{C'B'}|| & ||\lambda_{C'C}|| \\ ||\lambda_{CC'}|| & ||\lambda_{CB}||  \end{array}\right]\,.
\end{eqnarray}
The solution of the Dirac equation to ${\cal O}(\Delta^0)$ yields the diagonal zero mode solutions
\begin{eqnarray} \label{fermion1}
\lambda_{\alpha \; B'C'} = \delta_{B'C'}\; \theta_\alpha^{C'}\,,\quad
\lambda_{\alpha \; BC} = - \delta_{BC} \;{1 \over \ell} \sum_{C'=1}^k \theta_\alpha^{C'}, 
\end{eqnarray}
where  $\alpha=1,2$ is the spinor index and $C',B'=1,2..,k$. There are $2k$ zero modes in (\ref{fermion1}), in accordance with the index theorem for the charge $Q={k\over N}$ instanton. 

The leading order zero modes (\ref{fermion1}) get deformed at order $\sqrt{\Delta}$ \cite{Anber:2023sjn}. 
The off-diagonal matrices to ${\cal O}\left(\sqrt{\Delta}\right)$ are given in Appendix \ref{appx:fermion}.  
The important point, discussed there and in \cite{Anber:2023sjn}, is that one can construct order-$\Delta$ gauge invariants from the fermion zero modes that display a pattern similar to the bosonic invariants and are characterized by a lumpy structure. 

One finds that each of the $k$ lumps of (\ref{multilump1}) hosts two zero modes, with their positions determined by the moduli $\hat\phi^{C'}_\mu$.  
Specifically, see \cite{Anber:2023sjn} and Appendix \ref{appx:fermion}, the order-$\Delta$ contribution to the gauge-invariant $\tr (\lambda \lambda)$  formed from the fermion zero modes include terms such as
\begin{eqnarray} \label{fermionzeromodedensitytext} 
&& \sum_{C'=1}^k\sum_{D=1}^\ell \lambda_{1 \; C'D} \lambda_{2 \; DC'}  \sim  \\
&& \sum_{C'=1}^k \bar\eta_1^{C'} \bar\eta_2^{C'}  \bigg| \sum\limits_{m} e^{i\frac{2 \pi m}{L_2} (x_2 + {L_1 L_2 \over 2 \pi} \hat\phi_1^{C'})   -\frac{  \pi  }{  L_1 L_2}\left[x_1-\frac{  L_1 L_2}{2\pi  }\hat \phi_2^{C'}  -\frac{L_1 C'}{k} + L_1 \frac{1+k}{2k}- L_1 m \right]^2} \bigg|^2\times \nonumber \\
\nonumber
&&   \bigg| \sum\limits_{n\in \mathbb Z}   \left( x_3-\frac{\ell L_3 L_4}{2\pi } \hat \phi_4^{C'} -L_3 \ell  n - L_3 \frac{  1+\ell}{2} \right) e^{i \frac{2\pi n}{\ell L_4} (x_4 + {\ell L_3 L_4 \over 2 \pi} \hat\phi_3^{C'}) -\frac{\pi }{\ell L_3 L_4}\left[x_3-\frac{\ell L_3 L_4}{2\pi } \hat \phi_4^{C'} -L_3\left( \ell n  +\frac{1+\ell}{2}\right)\right]^2  } \bigg|^2, \nonumber \end{eqnarray}
where $\eta_\alpha^{C'}$ are linear functions of $\theta_\alpha^{C'}$ (defined in  Appendix \ref{appx:fermion}, see (\ref{etavstheta})).
The expression  (\ref{fermionzeromodedensitytext}) for the order-$\Delta$ contribution  highlights the localization properties of the fermion zero modes, as dictated by the holonomies $\hat\phi^{C'}_\mu$, that were evident in the bosonic solution described in (\ref{fractional1}). From (\ref{fermionzeromodedensitytext}), we see that every one of the $k$ lumps in the sum over $C'$ in (\ref{fermionzeromodedensitytext}) hosts $2$ fermion zero modes.

In our calculation of the (multi-)gaugino condensates, order-$\Delta$ contributions to the gaugino bilinear will not be included. The reason is that  in order to compute to order $\Delta$, one needs the full\footnote{On the other hand,  the $\tr F_{12} F_{12}$ invariant is fully determined to ${\cal{O}}(\Delta)$, see \cite{Anber:2023sjn}.}  order-$\Delta$ solution and thus requires knowledge of the order-$\Delta$ contribution to ${\cal{S}}_\mu$ of the deformed solution (\ref{expansion1}). While it is uniquely determined by solving a recursive relation, its explicit form is very complicated and has not yet been found.
For the $k=1$ case, we have an explicit reason to expect $\Delta$-independence of the gaugino condensate since all fermion zero modes are related by supersymmetry to the bosonic background \cite{Anber:2022qsz}. For $k>1$, supersymmetric Ward identities (which hold on $\mathbb{T}^4$ as well) lead us to expect that expectation values of products of $\tr \lambda\lambda(x) \tr \lambda\lambda(y)...$ do not depend on the coordinates $x, y$, etc. Based on holomorphy,  one also does not expect a volume  dependence on the result.\footnote{For a derivation of the supersymmetric Ward identities for the $\T^4$ case, see Appendix \ref{appx:ward}. In particular, it is shown there that a  dependence of the gaugino condensate on $L |\Lambda|$, where $L$ is any measure of the torus size,  is not allowed by holomorphy.}
Verifying this using the explicit form of the  $x$-dependent solution would be a highly nontrivial check on the instanton calculation, but, while desirable, it is not feasible given our current state of knowledge of the multifractional instantons on the twisted $\mathbb{T}^4$. We now make the following comments:
\begin{enumerate}
\item We note that mathematically, independence of the volume, as outlined above, does not preclude a dependence on $\Delta$, a dimensionless parameter which characterizes the shape of $\T^4$. However, a dependence on $\Delta$, combined with the volume-independence, would imply non-uniqueness of the infinite-volume limit. As such non-uniqueness is not expected, especially in a theory with mass gap, we expect that the ${\cal{O}}(\Delta^0)$ result is, in fact, exact.
\item Sometimes, the issue is raised of whether one should expect that the condensate, calculated in the background of the 't Hooft twist of the boundary conditions, should agree with the $\R^4$ value in the infinite volume limit, due to the insertion of topological $2$-form background field gauging the $\Z_N^{(1)}$ symmetry (the 't Hooft twist). Our reply to such concerns is that one does not expect boundary conditions to affect the thermodynamic limit. 
In support of this, in nonsupersymmetric Yang-Mills theory, lattice studies \cite{Stephenson:1989pu,Gonzalez-Arroyo:1995ynx} have found that, as  the volume becomes larger than the confinement scale, physical quantities---glueball masses and string tensions---agree between calculations done with or without 't Hooft twists.
\end{enumerate}

With these remarks in mind, we now introduce a parameterization of the order-$\Delta^0$ zero modes (\ref{fermion1}), which is more in line with the parameterization of the bosonic moduli space of eqn.~(\ref{r over N abelian sol}) and which we shall use in our calculations. Thus, we define, instead of (\ref{fermion1}), the ${\cal O}(\Delta^0)$ fermion zero modes:
\begin{equation}
\label{fermionzero1}
\lambda_\alpha = \frac{\omega}{2\pi \sqrt{kN(N-k)}} \; \zeta^k_\alpha + \bm{\zeta}_\alpha \cdot \bm{H}_k,
\end{equation}
i.e. we take the $2k$ zero modes to be parameterized by the $k$ two-spinor Grassmann variables $\bm{\zeta}_\alpha = (\zeta_\alpha^1,..., \zeta_\alpha^{k-1})$ and $\zeta_\alpha^k$. Here, as in (\ref{r over N abelian sol}), $\bm{H}_k$ are the $SU(k)$ Cartan generators embedded in $SU(N)$ by taking zeros in the $\ell \times \ell$ part of the $N\times N$ matrix. The factor $1/2\pi\sqrt{kN(N-k)}$ that accompanies the first term is introduced for convenience.

%%%%%%%%%%%%%%%%%%%%%%%%%%%%%%
\section{Shape and volume of the bosonic moduli space}
\label{Shape and volume of the bosonic moduli space}
%%%%%%%%%%%%%%%%%%%%%%%%%%%%%%

We do not know of a systematic approach to determine the shape of the moduli space. However, as we noticed in our previous work, one important condition that helps us in this endeavor is that in pure $SU(N)$ Yang-Mills theory, the expectation value of Wilson lines that wrap around any of the $4$ directions must vanish identically on a small $\mathbb T^4$ (as we review below).
 
Let $q_\mu$ be an integer. Then, a general Wilson  line wrapping $q_\mu$ times in the direction $x_\mu$ is given by 
\begin{equation}\label{wilsondef}
 W_\mu^{q_\mu}[A]=\mbox {tr}\left[e^{i q_\mu\oint A_\mu (x)}\Omega_\mu^{q_\mu}\right] .
 \end{equation}
The statement of the vanishing of their expectation values is:  
\begin{eqnarray}\label{condition of vanishing W}
\left\langle \prod_{\mu=1}^4W_\mu^{q_\mu} \right\rangle=0, ~ \text{or} ~~ \int\limits_{n_{12}, n_{34}} [D A_\mu]  \left(\prod_{\mu=1}^4W_\mu^{q_\mu}[A]\right) e^{-S_{YM}}=0\,,
\end{eqnarray}
where $S_{YM}$ is the Yang-Mills action, and the path integral is over fields obeying the boundary conditions (\ref{conditions on gauge field}, \ref{cocycle}, \ref{twists1}).  For the twists given in (\ref{twists1}), this path integral sums over gauge field configurations with topological charges ${k \over N} + \nu$, for all $\nu \in \mathbb{Z}$.

There are two ways to argue that (\ref{condition of vanishing W}) must be true in pure Yang-Mills theory on $\mathbb{T}^4$. First, rather generally, the Wilson lines are charged under the center $\mathbb Z_N^{(1)}$ $1$-form symmetry. This symmetry must be preserved, i.e., $\left\langle \prod_{\mu=1}^4W_\mu^{q_\mu} \right\rangle=0$, on a small $\mathbb T^4$ since breaking the  symmetry makes sense only in the thermodynamic limit. Second, let us consider the theory in the Hamiltonian approach\footnote{See Section \ref{sec:anomalyHamilt} for a quick review of the Hamiltonian quantization on $\T^3$.} by taking space to be the three-torus $\mathbb T^3$,  say, with spatial twist $n_{12}$, treating $x_4$ as Euclidean time. Then,  the eigenstates are simultaneous eigenstates of both the Hamiltonian and the $1$-form symmetry in $\mathbb T^3$: $|\psi\rangle=|E(\vec e), \vec e\rangle$, where $\vec e\equiv (e_1,e_2,e_3)$ designates $N$ distinct eigenvalues of the $1$-form center symmetry operator in each of the $3$ spatial directions, $e_{1,2,3}\in\{0,1,..,N-1\}$. These are the electric fluxes in each of the three spatial directions \cite{tHooft:1979rtg}.  We also impose the normalization condition $\left \langle E(\vec e_{ a}), \vec e_{ a}|E(\vec e_{b}), \vec e_{ b} \right\rangle=\delta_{a,b}$. Then, for example, $
\langle W_1 \rangle=\sum_{E,\vec e_{a}}  \langle E(\vec e_{ a}), \vec e_{ a}|e^{-L_4E(\vec e_{a})} W_1| E(\vec e_{ a}), \vec e_{ a}  \rangle= 0$, where the vanishing is due to the fact $W_1$ changes the electric flux $e_1$ by unity, hence it has no diagonal matrix elements between flux eigenstates.
Thus, using the Hamiltonian approach with $x_4$ as time, we can argue that $\langle W_{1,2,3}\rangle$ should vanish. However, the Euclidean time direction can also be chosen as $x_3$, allowing us to argue that $\langle W_{4}\rangle$ should also vanish.

In the following, we shall use the path-integral approach along with the condition (\ref{condition of vanishing W}) to determine the shape and volume of the moduli space $\Gamma$ of a fractional instanton with topological charge $Q=k/N$, $1\leq k \leq N-1$. The point is that if we take pure Yang-Mills theory on a small $\mathbb{T}^4$, with twists as in (\ref{twists1}), the semiclassical approximation to (\ref{condition of vanishing W})  is expected to hold. Since instantons of all
  fractional topological charge $Q = {k \over N} (\text{mod} \; 1)$ contribute to (\ref{condition of vanishing W}), all their contributions should vanish. 
  
  In the semiclassical approximation, the path integral determining the expectation value of the Wilson loop includes an integral over the instanton moduli of the Wilson loop evaluated in the instanton background. 
  Thus, for any given Wilson  line $W_\mu^q$, evaluated in the instanton background $\hat A(\{z_\nu, \bm{a}_\nu\})$, of eqns.~(\ref{gaugewithholonomies}, \ref{r over N abelian sol}), with the solution of charge $Q={k \over N}$,
the condition  (\ref{condition of vanishing W}) reads
\begin{eqnarray}\label{explicit moduli dependence}
\int_{\Gamma} \left(\prod_{\nu=1}^4 dz_\nu d \bm a_\nu \right) W_\mu^q[\hat A(\{z_\nu, \bm{a}_\nu\})]=0,~~ \text{where} ~d \bm a_\nu = \prod\limits_{b=1}^{k-1} da_\nu^b,
\end{eqnarray}
where $\Gamma$ denotes the moduli space.
In Appendices \ref{appx:wilson} and \ref{Determining the shape of the moduli space}, we describe in detail the use of this condition and the symmetries of the Wilson loop and the local gauge invariants characterizing the solution to determine the range of the moduli $z_\mu$ and $\bm{a}_\mu$.
Here, we summarize our findings. 

The gauge invariants we consider, evaluated in the background of the solution with moduli $z_\mu$ and $\bm{a}_\mu$,  are the winding Wilson loops (\ref{wilsondef}) and the local gauge invariant densities (\ref{multilump1}). The Wilson loops, evaluated in the ${\cal{O}}(\Delta^0)$ background, are\footnote{We note that for the purposes of studying the range of the moduli, it suffices to consider $W_\mu^q$ in the ${\cal{O}}(\Delta^0)$ background since the ${\cal{O}}(\Delta)$  corrections to the Wilson loops have the same symmetry properties as the local gauge invariants, see \cite{Anber:2022qsz}.} \begin{eqnarray}\label{Wilson 1 main}
\nonumber
W_1^q &=&(-1)^{q(k-1)} e^{-i2\pi q(N-k)\left(z_1-\frac{x_2}{N L_2}\right)}\left[\sum\limits_{C'=1}^k e^{i 2\pi q \bm a_1\cdot \bm \nu_{C'}}\right]+(N-k)e^{i2\pi qk \left(z_1-\frac{x_2}{N L_2}\right)}\,, \\\nonumber
W_2^q&=& e^{-i2\pi q(N-k)\left(z_2+\frac{x_1}{N L_1}\right)} \; \left[\sum\limits_{C'=1}^k  e^{i 2\pi q (\bm a_2 - {\bm \rho \over k})\cdot \bm\nu_{C'} }\right] + (N-k)e^{i2\pi q k \left(z_2+\frac{x_1}{N L_1}\right)}\,,\\\nonumber
W_3^q&=& e^{-i2\pi q(N-k)\left(z_3-\frac{x_4}{N (N-k) L_4}\right)}\left[\sum\limits_{C'=1}^k  e^{i 2\pi q  \bm a_3 \cdot \bm\nu_{C'} }\right] + (N-k)\; e^{i2\pi q k \left(z_3-\frac{x_4}{N \ell L_4}\right)}\,\gamma_\ell^q \, \delta_{{q \over \ell}, \Z} ,\\
W_4^q&=& e^{-i2\pi q(N-k) \left(z_4+\frac{x_3}{N (N-k) L_3}\right)}\left[\sum\limits_{C'=1}^k  e^{i 2\pi q  \bm a_4 \cdot \bm\nu_{C'} }\right]+ (N-k) e^{i2\pi qk \left(z_4+\frac{x_3}{N \ell L_3}\right)} \gamma_\ell^q \;   \delta_{{q \over \ell}, \Z}\,.\end{eqnarray}
Here, $\bm \rho$ is the $SU(k)$ Weyl vector, while  $\delta_{{q \over \ell}, \Z}$ indicates that the terms do not vanish only if $q$ is an integer times $\ell = N-k$.
  
  The other invariants we study are the local gauge invariant densities, with an  ${\cal{O}}(\Delta)$ contribution  
\begin{eqnarray}\label{local 1 main}\nonumber
\tr F_{12} F_{12} && \\\nonumber
&&\sim  \sum\limits_{C'=1}^k F\left(x_1 + L_1 N z_2 - L_1 \left(\bm{a}_2-{\bm \rho \over k}\right) \cdot \bm{\nu}^{C'} -{L_1 \over 2} -{L_1 \over 2 k},  x_2 - L_2 N z_1+ L_2 \bm{a}_1 \cdot \bm{\nu}^{C'},\right.   \\
&& \left.  ~~~~~~ x_3 + L_3 \ell N z_4  -L_3 \ell \bm{a}_4 \cdot \bm{\nu}^{C'}, x_4 - L_4 \ell N z_3+ L_4 \ell \bm{a}_3 \cdot \bm{\nu}^{C'}\right)\,.
 \end{eqnarray}

{\flushleft{S}}tudying the moduli dependence of the above gauge invariants, we find (for details see Appendices \ref{appx:wilson} and \ref{Determining the shape of the moduli space}): 
 \begin{enumerate}
 \item We begin  with the local gauge density $F$ (\ref{local 1 main}). Recalling Figure \ref{visual of liquid}, we  note that $z_\mu$ can be interpreted as ``center of mass'' coordinate of the $k$-lump instanton, while $\bm{a}_\mu\cdot \bm{\nu}^{C'}$, for $C'=1,...,k$, parameterize the deviation of each lump's position from the center of mass.  It is easy to see, from the discussion below, that each of the lumps can be located anywhere on the torus.
 
 Further, $F$, like any local gauge invariant quantity,   is periodic with respect to each argument, with period given by the appropriate torus period $L_\mu$. Thus, it would appear natural to consider the variables $z_\mu$ to have periods $1\over N$ for $\mu=1,2$ and $1\over N(N-k)$ for $\mu=3,4$.

 \item However, 
 the Wilson lines (\ref{Wilson 1 main}) are not periodic functions of $z_\mu$ with these periods.
For $\mu=1,2$, a shift of $z_\mu$ by $1\over N$ performs  a $\mathbb{Z}_N^{(1)}$ 1-form symmetry transformation of the instanton background, in the respective $x_{1,2}$ direction, as it multiplies the winding Wilson lines by an $e^{i 2 \pi q {k \over N}}$ factor (recall that gcd($k,N) = 1$). Likewise,  
shifting $z_{3}$ or $z_{4}$ by $1\over N (N-k)$ corresponds to a center symmetry transformation in the respective $x_3,x_4$ directions. This follows from inspecting the traces of the Wilson lines given above. 

This, of course, reflects the fact that, in the presence of 't Hooft twists, shifting the coordinates (equivalently, the moduli $z_\mu$) on torus periods is equivalent to global $\Z_N^{(1)}$ center transformations in the corresponding direction. Since center is a global symmetry, one should include the images of an instanton under these transformations (see \cite{GarciaPerez:1992fj}, where this was studied on $\R \times \T^3$). Thus, we take the ranges of $z_\mu$:
\begin{eqnarray}\label{range of z 1}
z_\mu &\in& [0,1], \; \text{for} \;  \mu=1,2~,~ \\ \nonumber
z_\mu &\in& [0, {1 \over N-k}],  \; \text{for} \;  \mu = 3,4~, ~ \text{or} \; z_\mu \in \S^1_{\mu}~.
\end{eqnarray}
For use below, in the last line, we denoted the range of each $z_\mu$ by $\S^1_{\mu}$ of circumference as shown. 
  \item Next, we observe that the product $\bm{a}_\mu \cdot \bm{\nu}^{C'}$ shifts by an integer under $SU(k)$ root-lattice vector translations (because the product of a root with a fundamental weight is integer). Thus,  root lattice shifts are an invariance of both the Wilson loops and the local density, in view of the latter's periodicity (in fact, it is easily seen, see Appendix \ref{appx:wilson}, that these shifts are due to $\Omega$-periodic gauge transformations). Thus, we have:
  \begin{eqnarray} \label{range of a 1}
  \bm{a}_\mu \in \Gamma_{r}^{SU(k)}~, ~ \forall \mu,
  \end{eqnarray}
 where $\Gamma_r^{SU(k)}$ denotes the fundamental cell of the root lattice of $SU(k)$, which can be mapped to the torus $(\S^1)^{k-1}$.

\item Another identification on the moduli space consists of $SU(k)$ weight-lattice shifts of $\bm{a}_\mu$, compensated by shifts of $z_\mu$. These transformations, as shown in Appendix \ref{appx:weight}, leave invariant the gauge invariant Wilson loops $W_\mu^q$ (\ref{Wilson 1 main}) and the local densities (\ref{local 1 main}). Explicitly, 
\begin{eqnarray}\label{weight shift 1}
\bm{a}_\mu \rightarrow \bm{a}_\mu + \bm{w}_a,&& ~ z_\mu \rightarrow z_\mu - {{\cal{C}}_a \over k}\,, ~a = 1, 2,..., k-1,  \nonumber \\
 \text{where} \;&& N {\cal{C}}_a = a \; (\text{mod} \; k), ~ {\cal{C}}_a \in \mathbb{Z}_+ .
\end{eqnarray}
The nonneggative integer ${\cal{C}}_a$ exists because of the gcd($N,k)=1$ condition.  These shifts are also due to $\Omega$-periodic gauge transformations, as shown in Appendix \ref{appx:weight}. Furthermore, these transformations form a freely-acting $\Z_k$ group when acting on the moduli.

\item Finally, both the Wilson lines (\ref{Wilson 1 main}) and local gauge invariants (\ref{local 1 main}) do not change, as shown in Appendix \ref{appx:weyl}, upon  $SU(k)$ Weyl reflection with respect to the root $\bm \alpha_{ij}$, $i \ne j \in \{1,...,k\}$, performed simultaneously on all four moduli $\bm{a}_\mu$: \begin{eqnarray}\nonumber
\label{weyl reflection 1}
  \bm a_{\mu}&\rightarrow& \mu_{\bm\alpha_{ij}}(\bm a_\mu)\equiv\bm a_\mu-(\bm a_\mu\cdot\bm\alpha_{ij})\bm\alpha_{ij}\,,\quad \mu=1,3,4\,,\\
\bm a_{2}&\rightarrow&\bm a_2-\left[(\bm a_2-\frac{\bm \rho}{k})\cdot \bm \alpha_{ij}\right]\bm\alpha_{ij}\,.
\end{eqnarray}
It is shown in Appendix \ref{appx:weyl} that the transformations are also due to $\Omega$-periodic gauge transformations. The Weyl transformations are isomorphic to the permutation group of $k$ objects, $S_k$, of order $k!$.

A useful pictorial interpretation of (\ref{weyl reflection 1}) is that the Weyl transformation permutes the identical $k$ lump constituents of the multi-fractional instanton, described by the terms appearing in the sum (\ref{local 1 main}). 

\item We conclude (see Appendix \ref{Determining the shape of the moduli space} for details) that the moduli space is the product space of the $SU(k)$ root cell $\Gamma_r^{SU(k)}$ and the circle  $\mathbb S_\mu^1$, in each spacetime direction, modded by the action of the discrete symmetry $\mathbb Z_k$:
 \begin{eqnarray}\label{moduli space bulk 1}
 \Gamma = \prod\limits_{\mu=1}^4 {\S^1_{{\mu}} \times \Gamma_r^{SU(k)} \over \Z_k} \simeq \prod\limits_{\mu=1}^4 {(\S^1)^k \over \Z_k} \,,
 \end{eqnarray}
 and an overall action of the $S_k$ group (\ref{weyl reflection 1}) permuting the $k$ lumps.
 \item  The volume of the space ${(\S^1)^k / \Z_k}$ is  $1/k$ times the volume of  $(\S^1)^k$. In addition, one can show that (see Appx. E)  the fundamental domain of the moduli space can always be chosen to be the weight lattice of $SU(k)$, i.e., $\Gamma_w^{SU(k)}$, times the entire range of the $z_\mu$ variables given by (\ref{range of z 1}), in each spacetime direction. Thus, we can write
 \begin{eqnarray}\label{range of z and a bulk}
 \Gamma=\left\{\begin{array}{l} z_{1,2} \in [0,1)\,,\\ 
z_{3,4} \in [0, {1 \over N-k})\,,\\
\bm a_\mu  \in \Gamma_{w}^{{SU(k)}}\; \text{for} \;  \mu=1,2,3,4\,,\end{array}\right.
\end{eqnarray}
modulo the action of the $S_k$ group on $\bm a_\mu$.
 
To reinforce this conclusion, we employ a different approach in Appendix \ref{Determining the shape of the moduli space} to demonstrate that the fundamental domain of $\bm a$ is the weight lattice, given that the range of the $z_\mu$ variables is as in (\ref{range of z 1}). We examine a fractional instanton with a topological charge of $Q=(N-1)/N$, corresponding to setting $k=N-1$ and $\ell=1$. We show that, in this specific case, both the transition functions and gauge fields are completely abelian. Additionally, the holonomies $\bm a_\mu=(a_\mu^1,..,a_\mu^{N-2})$ and the four translations $z_\mu$ can be organized into a more symmetric set of moduli $\bm {\tilde \Phi}_\mu\equiv (\Phi_\mu^1,..,\Phi_\mu^{N-1})$ that lives in the Cartan subalgebra of $SU(N)$. The vanishing of the Wilson line expectation values will then be used to argue that $\bm {\tilde \Phi}_\mu$ lies in the root lattice of $SU(N)$. This finding will be shown to imply that the fundamental domain of $\bm a$ is the weight lattice of $SU(N-1)$, provided that the range of the $z_\mu$ variables is given by (\ref{range of z 1}).
\end{enumerate}

The measure on the moduli space $d\mu_B$ is 
\begin{eqnarray}\label{bosonic measure bulk}
d\mu_B=\frac{\prod_{\mu=1}^4 \prod_{b=1}^{k-1} d a_\mu^b dz_\mu \sqrt{\mbox{Det}\, {\cal U}_B}}{k!(\sqrt{2\pi})^{4k}}\,.
\end{eqnarray}
 The factor $k!$ that appears in the dominator of (\ref{bosonic measure bulk}) is the result of the fact that the lumpy solution, as well as Wilson's lines, are invariant under the Weyl group (the transformations (\ref{weyl reflection 1}) simultaneously acting on all four $\bm{a}_\mu$), which is isomorphic to the permutation group $S_k$ of order $k!$. The matrix $ {\cal U}_B$ is the metric on the moduli space, with matrix elements given by (summation over $\nu$ is implied)
 \begin{eqnarray}\label{metric on moduli in a and z 2}\nonumber
 {\cal U}_{B\,ab}^{\mu\mu'}&=&\frac{2}{g^2}\int_{\mathbb T^4}\mbox{tr}\left[\frac{\partial A_\nu}{\partial a_\mu^a}\frac{\partial A_\nu}{\partial a_{\mu'}^b}\right]\,, \quad a,b=1,..,k-1\,,\\\nonumber
  {\cal U}_{B\,zz}^{\mu\mu'}&=& \frac{2}{g^2}\int_{\mathbb T^4}\mbox{tr}\left[\frac{\partial A_\nu}{\partial z_\mu}\frac{\partial A_\nu}{\partial z_{\mu'}}\right]\,, \\
  {\cal U}_{B\,zb}^{\mu\mu'}&=& \frac{2}{g^2}\int_{\mathbb T^4}\mbox{tr}\left[\frac{\partial A_\nu}{\partial z_\mu}\frac{\partial A_\nu}{\partial a_{\mu'}^j}\right]\,, \quad b=1,..,k-1\,.
 \end{eqnarray} 
Using $\mbox{tr} (H_{k}^a H_{k}^b)=\delta_{ab}$ (remember that $\bm H_{k}=(H_{k}^1,..,H_{k}^{k-1})$ are embedded in $SU(N)$ by putting zeros in the $\ell\times \ell$ lower-right matrix), and $\mbox {tr} (\omega^2)=4\pi^2Nk(N-k)$, along with $\mbox{tr}[H_k^b\omega]=0$, we find that the metric on the moduli space in each spacetime direction $\mu$ is given by the $k\times k$ diagonal matrix
\begin{eqnarray}
{\cal U}^{\mu\mu'}_B= \frac{8\pi^2 V}{g^2 L_\mu^2}\delta^{\mu\mu'}\mbox{diag}\left(\underbrace{1,1,..,1}_{k-1}, k\ell N\right)\,,
\end{eqnarray}
and the square root of the determinant of ${\cal U}_B$  is
\begin{eqnarray}
\sqrt{\mbox{Det}\, {\cal U}_B }=\left(\sqrt{k (N-k) N}\right)^4\left(\frac{8\pi^2 \sqrt V}{g^2}\right)^{2k}\,.
\end{eqnarray}
The volume of the bosonic moduli space is obtained by integrating $d\mu_B$ over $\Gamma$. As described above, we choose the fundamental domain to be the weight lattice of $SU(k)$ times the range of $z_\mu$ as in (\ref{range of z and a bulk}).

 Collecting the above results, recalling the fact that the volume of the weight lattice of $SU(k)$ is $1/\sqrt{k}$, and performing the integral over the collective coordinates, we readily find (see Appendix \ref{Determining the shape of the moduli space} for details)
\begin{eqnarray}\label{volume of mub}
\mu_B=\int_{\Gamma}\frac{\prod_{\mu=1}^4 \prod_{b=1}^{k-1} d a_\mu^b dz_\mu \sqrt{\mbox{Det}\, {\cal U}_B}}{k!(\sqrt{2\pi})^{4k}}
&=&\frac{N^2}{k!} \left(\frac{4\pi \sqrt V}{g^2}\right)^{2k}\,.
\end{eqnarray}
Here, we integrated over the bosonic moduli space $\Gamma$ because, as we show in the next section, to leading order in $\Delta$, the integrand $(\tr \lambda\lambda)^k$ of the path integral does not depend on the bosonic moduli.

%%%%%%%%%%%%%%%%%%%%
\section{The gaugino condensates}
\label{The gaugino condensates}
%%%%%%%%%%%%%%%%%%%%%

In this section, we combine the above information to compute the higher-order condensate ${\cal C}(x_1,...,x_k)\equiv\langle \prod_{i=1}^k\mbox{tr}(\lambda\lambda)(x_i) \rangle$ in $SU(N)$ super Yang-Mills theory on a small deformed $\mathbb T^4$. As we discussed above, there are $2k$ fermion zero modes in the background of a fractional instanton carrying a topological charge $Q=k/N$. Therefore, we expect that these zero modes will saturate the condensate. We start by expressing ${\cal C}(x_1,...,x_k)$ in the path integral formalism, with action (\ref{symaction2}) (taking $D=0$):
\begin{eqnarray}\label{k correlator}\nonumber
{\cal C}(x_1,...,x_k)={\cal N}^{-1}\; \sum_{\nu \in \mathbb Z}\int [D A_\mu][D\lambda][D\bar \lambda] \left[\prod_{i=1}^k\mbox{tr}(\lambda\lambda)(x_i)\right]e^{-S_{SYM}-i\theta\left(\nu+\frac{k}{N}\right)}\bigg\vert_{n_{12}=-k\,,n_{34}=1}\,.\\
\end{eqnarray}
Here, we have emphasized that the computations are performed in the presence of the twists imposed by the transition functions (\ref{the set of transition functions for Q equal r over N, general solution}). The sum is over topological charges $\nu + {k\over N}$, $\nu \in \mathbb Z$, keeping in mind that it is only the sector $\nu=0$ (of topological charge $k/N$) that contributes to ${\cal C}(x_1,..,x_k)$ on a small $\mathbb T^4$ in the semi-classical regime. The pre-coefficient ${\cal N}^{-1}$ is a normalization constant we shall return to. We also set the vacuum angle $\theta=0$ from here on. 

One proceeds with the calculations of (\ref{k correlator}) by gauge-fixing and using the Faddeev-Popov method and finding the one-loop determinants of the bosonic and fermionic fluctuations in the background of the fractional instanton. As we elaborated previously, there are both bosonic and fermion zero modes (moduli), in addition to higher mode fluctuations. Taking the contribution from each of these sectors is a standard procedure. The upshot is that the correlator ${\cal C}(x_1,...,x_k)$ is given by:
\begin{eqnarray}\label{multicorrelator}
{\cal C}(x_1,...,x_k)= {\cal N}^{-1} \;M_{\scriptsize \mbox{PV}}^{3k}e^{-\frac{8\pi^2 k }{N g^2}}\int_{\Gamma} d\mu_B\int  d\mu_F\left[\prod_{i=1}^k\mbox{tr}(\lambda\lambda)(x_i)\right]\,.
\end{eqnarray}
The pre-factor $M_{\scriptsize \mbox{PV}}^{3k}e^{-\frac{8\pi^2 k }{N g^2}}$ arises from the bosonic and fermionic determinants of the non-zero modes after employing the Pauli-Villars regularization technique, and $M_{\scriptsize \mbox{PV}}$ is the Pauli-Villars mass.\footnote{The reader can consult the reviews in \cite{Vandoren:2008xg,Dorey:2002ik} for details. We  note that, due to supersymmetry, only the zero modes contribute in the self-dual instanton background. The power of $M_{PV}$ in (\ref{multicorrelator}) equals $n_B - {1 \over 2} n_F$, where $n_B = 4 k$ and $n_F=2k$ is the number of bosonic and fermionic zero modes.} Additionally, we note that $S_0=\frac{8\pi ^2 k}{g^2 N}$ is the action of a fractional instanton with a topological charge $Q=k/N$, see eqn. (\ref{action of fractional instanton}).

The measure of the bosonic moduli $d\mu_B$ was introduced and discussed in the previous section, with the result for its volume $\mu_B = \int_{\Gamma} d\mu_B$ given in (\ref{volume of mub}). The measure on the fermionic moduli space $d\mu_F$ is determined as in e.g.~\cite{Vandoren:2008xg}. It is given by
\begin{eqnarray}\label{dmuf}
d\mu_F=\frac{\prod_{C'=1}^k d\zeta_1^{C'}d\zeta_2^{C'}}{\sqrt{\mbox{Det}\, {\cal U}_F}}\,,
\end{eqnarray}
where ${\cal U}_F$ is the metric on the fermionic moduli space, with components 
\begin{eqnarray}\label{fermionic metric}\nonumber
({\cal U}_{F})_{ B'  C' }^{~\beta\; \gamma}&=&\frac{2}{g^2}\int_{\mathbb T^4}\mbox{tr}\left[\frac{\partial \lambda^\alpha}{\partial\zeta_\beta^{B'}}\frac{\partial \lambda_\alpha}{\partial\zeta_\gamma^{C'}}\right]\,,\quad B',C'=1,..,k\,,\quad \beta, \gamma =1,2\,,\\
&=&\frac{2 V}{g^2}\delta_{B'C'} \,\epsilon^{\gamma\beta}.
\end{eqnarray}
To go from the first to the second line, we used the parameterization of the zero modes of eqn.~(\ref{fermionzero1}) and employed the identities $\mbox{tr}\left[H_k^{a}H_k^{b}\right]=\delta_{ab}$, where $a,b=1,..,k-1$, $\mbox{tr}\left[\omega^2\right]=4\pi^2 kN(N-k)$, and $\mbox{tr}\left[\omega H_k^{a}\right]=0$. 
From (\ref{fermionic metric}),  we immediately find
\begin{eqnarray}\label{DetFermion}
\sqrt{\mbox{Det}\, {\cal U}_F}=\left(\frac{2 V}{g^2}\right)^{k}\,.
\end{eqnarray}

The last piece of computation involves the fermion multi-linear $\prod_{i=1}^k\mbox{tr}(\lambda\lambda)(x_i)$, recalling that we are only interested in the result to ${\cal O}\left(\Delta^0\right)$, as per the discussion at the end of section \ref{sec:fermzero} (this ensures that this multi-linear is position- and bosonic moduli-independent).  Using  (\ref{fermionzero1}),  we obtain for the gauge-invariant bilinear:
\begin{eqnarray}
\mbox{tr}\left(\lambda\lambda\right)=2\mbox{tr}\left(\lambda_2\lambda_1 \right)\big\vert_{\text{zero modes}}=2\sum_{C'=1}^k\zeta_2^{C'}\zeta_1^{C'}\,,
\end{eqnarray}
from which we find
\begin{eqnarray}\label{multilinear fermion}
\prod_{i=1}^k\mbox{tr}(\lambda\lambda)(x_i)=2^kk!\prod_{C'=1}^k \zeta_2^{C'}\zeta_1^{C'}\,.
\end{eqnarray}
Substituting (\ref{volume of mub}, \ref{dmuf}, \ref{DetFermion}, \ref{multilinear fermion}) into (\ref{multicorrelator}), and using\footnote{This definition of $\Lambda$ is standard in supersymmetric instanton calculations (e.g.~\cite{Davies:2000nw,Vandoren:2008xg, Dorey:2002ik}) and is written here in terms of the canonical coupling.  $\Lambda$ is also the holomorphic scale and can equivalently be expressed in terms of the holomorphic coupling $g_h(\mu)$, which only runs to one loop (see \cite{Shifman:1986zi,Arkani-Hamed:1997qui}), as $\Lambda^3 = \mu^3 e^{-\frac{8\pi^2 }{N g_h^2(\mu)}}$.}  the strong scale $\Lambda$
\begin{equation}\label{coupling1}
\Lambda^3\equiv {\mu^3 \over g^2(\mu)} e^{-\frac{8\pi^2 }{N g^2(\mu)}},
\end{equation}
where the energy scale $\mu$ is taken to be the inverse size of $\mathbb T^4$, we finally obtain
\begin{eqnarray}\nonumber \label{gauginocondensates}
{\cal C}(x_1,...,x_k)=\left\langle \prod_{i=1}^k\mbox{tr}(\lambda\lambda)(x_i) \right\rangle&=& {\cal N}^{-1} \;N^2\left(\frac{16 \pi^2 M_{\scriptsize \mbox{PV}}^3}{g^2}e^{-\frac{8\pi^2 }{N g^2}}\right)^k \int \prod_{C'=1}^k d\zeta_1^{C'} d\zeta_2^{C'} \zeta_2^{C'}\zeta_1^{C'}\\
&=&{\cal N}^{-1} \; N^2\left(16\pi^2 \Lambda^3\right)^k\,.
\end{eqnarray}
In conclusion, our result for $\left\langle \prod_{i=1}^k\mbox{tr}(\lambda\lambda)(x_i) \right\rangle$ shown in (\ref{gauginocondensates}), momentarily ignoring the normalization factor ${\cal N}^{-1}$, is $N^2$ times the known result from the weakly coupled (multi)-instanton calculations on $\R^4$. We next turn to a discussion of the subtleties involved.

%%%%%%%%%%%%%%%%%%%%%%%%%%%%%%%%%%%%%%%%%%%%%%%%%%%%%%%%%%%%%%%%%%%%%%%
\section{The Hamiltonian on $\mathbb{T}^3$ with a twist, the path integral, the normalization ${\cal{N}}$, and the gaugino condensate}
\label{sec:hamiltonian}
%%%%%%%%%%%%%%%%%%%%%%%%%%%%%%%%%%%%%%%%%%%%%%%%%%%%%%%%%%%%%%%%%%%%%%%%

So far in this paper, we performed a computation of the Euclidean path integral (\ref{k correlator}) with 't Hooft twists $n_{12}=-k$ and $n_{34}=1$, leading to the result  (\ref{gauginocondensates}). We notice the factor of $N^2$    obtained in the calculation of $\langle (\tr \lambda\lambda)^k \rangle$ on the twisted torus. In order to discuss the normalization factor ${\cal{N}}^{-1}$ and facilitate comparison to the $\R^4$ result, here we reinterpret the calculation using the Hamiltonian formalism on a spatial $\mathbb{T}^3$.

We first recall that supersymmetric Ward identities lead, via the holomorphy argument,  reviewed in Appendix \ref{appx:ward}, to the requirement that  the gaugino condensates on the four torus  be independent of the volume and thus coincide with the  $\R^4$ result. That this should be so has been the expectation at least since \cite{Shifman:1986zi} (and probably the original toron calculation of \cite{Cohen:1983fd}; we stress again that the numerical coefficient was not computed until our previous work \cite{Anber:2022qsz} and its extension here). 

Now, we  interpret  our calculation in the Hamiltonian formalism. The exposition below may look familiar since the Hamiltonian formalism was also an essential part of the discussion in \cite{Anber:2022qsz}. However, 
apart from the more general focus of this paper (e.g., going beyond $N=2$, $k=1$), there are  a few subtleties that were missed there and that point toward the understanding of the mismatch pointed out in our earlier work.

\subsection{Mixed anomaly, degeneracies, and $\langle (\tr \lambda^2)^k \rangle$}
\label{sec:anomalyHamilt}

We begin by taking, for definiteness, space to be comprised of the  $x_{1,2,3}$ directions and interpret $x_4$ as Euclidean time. In view of $n_{12}=-k$, there is 't Hooft ``magnetic flux'' $m_3 = n_{12} = -k$ on the spatial torus.\footnote{Since gcd$(k,N)=1$, a completely equivalent (to eqn.~(\ref{gauginohamiltonian}) below) result is obtained if we consider, say $x_{3,4,1}$ (or $x_{3,4,2}$) to be the spatial torus coordinates with unit twist $n_{34}$.} The quantization of $SU(N)$ super-Yang-Mills theory on a three-torus with twists is already familiar from the calculation of the Witten index \cite{Witten:1982df,Witten:2000nv}; a more recent introduction, also discussing generalized anomalies in this framework, is in  \cite{Cox:2021vsa}.

Briefly, upon quantizing (super) Yang-Mills theory on $\mathbb{T}^3$, the energy eigenstates (with eigenvalues $E$) can also be labeled by ``electric flux,'' the eigenvalues of the $1$-form center symmetry generators $\hat T_i$ in the $x_i$, $i=1,2,3$, directions. Thus, let $ |E, \vec{e} \rangle_{m_3}$ be the simultaneous eigenstates of $\hat T_i$ and the Hamiltonian $\hat H$ in the Hilbert space of states on $\mathbb{T}^3$ with spatial twist $m_3 = n_{12} = -k$ (further below, we denote this Hilbert space by ${\cal{H}}_{m_3}$).
Here $e_j$ $(\vec{e}=(e_1, e_2, e_3)$) are the (mod $N$) integer electric fluxes, labeling the eigenvalues of the $\Z_N^{(1)}$ generators, 
$\hat T_j | E, \vec{e} \rangle_{m_3} = |E, \vec{e} \rangle_{m_3} e^{i {2\pi \over N} e_j}$. 

It is well known that super Yang-Mills theory has a discrete $\Z_{2N}^{(0)}$ $0$-form chiral symmetry, generated by the operator $\hat{X}_{2N}$. In the presence of 't Hooft twists, the generators of the center symmetry along the magnetic flux do not commute with the chiral symmetry,  reflecting the mixed chiral/center anomaly \cite{Gaiotto:2014kfa,Gaiotto:2017yup}. Here, we write the commutation relation for our choice $m_3 = -k$, see  \cite{Cox:2021vsa} for derivation:
\begin{eqnarray}
\label{anomaly 1}
\hat T_3 \; \hat X_{2N} \; \hat T_3^{-1} = e^{ i {2 \pi \over N} k} \hat X_{2N}~.
\end{eqnarray}
This relation implies that $\hat X_{2N} |E, \vec{e}\rangle$ is an eigenstate of $\hat T_3$ with eigenvalue $e_3 + k$. But since $\hat X_{2N}$ is a symmetry, $\hat X_{2N} |E, \vec{e}\rangle$ has the same energy as $|E, \vec{e}\rangle$. Since gcd$(N,k)=1$, we conclude that there are  $N$ degenerate eigenstates of the same energy, labeled by the $N$ different values of $e_3$. This is an exact degeneracy (in addition to the degeneracy due to supersymmetry) of all states in the Hilbert space on $\mathbb{T}^3$ with 't Hooft twist $m_3=k$, with gcd$(N,k)=1$.

In the Hamiltonian formalism, we consider expectation values of operators $\hat {\cal{O}}$, evaluated using the twisted partition function, a trace over the Hilbert space ${{\cal{H}}_{m_3}}$:
 \begin{eqnarray}\label{defoftrace main}
 \langle \hat{ \cal{O}} \rangle \equiv {\cal{N}}^{-1} \; \tr_{\small{{\cal{H}}_{m_3}}}\left[ \hat{\cal{O}} e^{- \beta H} \hat{T}_3   (-1)^F\right]~.
 \end{eqnarray}
Here, $\beta$ ($=L_4$) is the extent of the Euclidean time direction, and the fermion number operator $(-1)^F$ is inserted to impose periodic boundary conditions on the fermions. The insertion of the center symmetry generator $\hat T_3$ (along the direction of the magnetic flux $m_3$)  implements twisted boundary conditions in the 34 plane.  A normalization factor ${\cal{N}}$ is inserted for a later convenience.
 
 For $\hat {\cal{O}}=   \prod_{i=1}^k\mbox{tr}(\lambda\lambda)(x_i) $, eqn.~(\ref{defoftrace main}) is precisely the path integral (\ref{k correlator}) computed semiclassically in this paper. For brevity, in what follows we denote $\hat {\cal{O}} =  (\tr \lambda^2)^k$ and write (\ref{defoftrace main}) as
\begin{eqnarray} \label{gauginos 1}
\langle (\tr \lambda^2)^k \rangle &=&  {\cal{N}}^{-1}\; \sum_{E, \vec{e}} e^{- \beta E} (-1)^F \langle E, \vec{e}| (\tr \lambda^2)^k\; \hat T_3 |E, \vec{e} \rangle \nonumber \\
&=&   {\cal{N}}^{-1} \;\sum_{E, \vec{e}} e^{- \beta E} (-1)^F \langle E, \vec{e}| (\tr \lambda^2)^k |E, \vec{e} \rangle e^{i {2 \pi \over N} e_3}~.\end{eqnarray}
The sum is over all energy and center symmetry eigenstates  $|E, \vec{e}\rangle_{m_3}$ (we omit the subscript $m_3$ for brevity).

Next, we use $\hat X_{2N}^{-1} \; (\tr \lambda^2)^k  \; \hat X_{2N}  = e^{- i {2\pi \over N} k} \; (\tr \lambda^2)^k$ to argue that the expectation values of $(\tr \lambda^2)^k$ in degenerate flux states differing by $k$ units  of $e_3$ flux differ  by a $\Z_N$ phase:
\begin{equation}\label{lambdavev}
\langle E, \vec{e} + \delta_{i3} k| (\tr \lambda^2)^k | E, \vec{e} + \delta_{i3} k\rangle = e^{- i{2 \pi \over N} k} \langle E, \vec{e}  | (\tr \lambda^2)^k | E, \vec{e} \rangle~.
\end{equation}
 Consider now the contribution to the sum in (\ref{gauginos 1}) of the $N$ degenerate states of energy $E$, skipping the energy eigenvalue and the other flux labels. Using the facts that $e_3$ is a (mod $N$) integer, that gcd$(N,k)=1$, and using (\ref{lambdavev}), we obtain (omitting $e_1$, $e_2$ for brevity):
 \begin{eqnarray}\nonumber
 \sum_{e_3=0}^{N-1} \langle e_3 | (\tr \lambda^2)^k | e_3\rangle e^{i {2 \pi \over N} e_3} &=& \sum_{q=0}^{N-1} \langle q k |  (\tr \lambda^2)^k | q k\rangle e^{i {2 \pi \over N} qk}  \\\nonumber
 &=& \sum_{q=0}^{N-1} \langle e_3=0 |  (\tr \lambda^2)^k | e_3=0\rangle e^{- i {2 \pi \over N} qk} e^{i {2 \pi \over N} qk} = N \langle e_3 =0| (\tr \lambda^2)^k | e_3=0\rangle\,.\\
 \end{eqnarray}
 Thus, returning to (\ref{gauginos 1}), we obtain
 \begin{eqnarray}\label{gauginohamiltonian}
\langle (\tr \lambda^2)^k \rangle &=&  {\cal{N}}^{-1} \; N \sum_{E, e_1, e_2} e^{- \beta E} (-1)^F \langle E, \vec{e}| (\tr \lambda^2)^k |E, \vec{e} \rangle\big\vert_{e_3=0}\,.
\end{eqnarray}

The conclusion from the above discussion is that the twist by $\hat T_3$ in (\ref{defoftrace main}) compensates the phases of the gaugino condensate in the different degenerate $e_3$ states. In effect, this makes the twisted torus partition function (\ref{defoftrace main}) sum up the gaugino condensates in the $N$ degenerate states by absolute value, as per (\ref{gauginohamiltonian}), instead of of weighting them by their different $\Z_N$ phases (the ones in (\ref{lambdavev}), which would make the result add to zero).

In order to obtain the gaugino condensate in one of the $N$ vacua, it is natural to take the normalization factor ${\cal{N}}$ in (\ref{gauginohamiltonian}) to be equal to the Witten index $I_W$ (equal to $N$). We continue by recalling the computation of the $I_W$ using the Hamiltonian on $\T^3$ with a twist and discussing the comparison with the path integral calculation (\ref{gauginocondensates}).

\subsection{A review of the Witten index on $\T^3$ with twist}
\label{sec:witten}

We recall that in \cite{Anber:2022qsz} we normalized the gaugino condensate by dividing by the Witten index, thus removing the factor of $N$ found in (\ref{gauginohamiltonian}) by taking ${\cal{N}}= I_W = N$. 
The Witten index is the partition function (\ref{defoftrace main}) with $\hat {\cal{O}}=1$, ${\cal{N}}=1$, and without the $\hat T_3$ insertion:\footnote{Inserting $\hat T_3$ makes the partition function with $\hat {\cal{O}}=1$ vanish, because of the degeneracy of flux states due to (\ref{anomaly 1}) and  the summation over $e_3$ in (\ref{gauginos 1}).}
 \begin{eqnarray}\label{windex main}
I_W  \equiv \tr_{\small{{\cal{H}}_{m_3}}}\left[  e^{- \beta H} \    (-1)^F\right] = N.
 \end{eqnarray}
The calculation of $I_W$, giving the result shown, $I_W=N$, was done in the Hamiltonian formalism on $\T^3$ with magnetic flux $m_3$ in the original paper \cite{Witten:1982df}. We recall, see also \cite{Witten:2000nv}, that $I_W$ is independent of $\beta$ as well as on the volume of $\T^3$ or the spatial twist of the boundary conditions.
We now briefly review the calculation with $m_3 \ne 0$. Even though it is well-known, the steps leading to the result will be useful in Section \ref{sec:wittenpath}.

We begin by recalling the advantage of using a twist $m_3 = n_{12} = -k$ with gcd$(N,k)=1$. It is simply that the twist removes the zero modes of all fields and fully gaps the excitation spectrum above a discrete set of zero energy states.\footnote{The excitation spectrum in pure Yang-Mills theory on a small $\T^3$ with a twist has been studied in \cite{GonzalezArroyo:1987ycm,Daniel:1989kj}.}
 Only these zero energy states contribute to $I_W$. For small $\T^3$, where semiclassical quantization should hold, these are simply quantum states obtained from quantizing around the classical configurations with zero field strength. 
 In the Hilbert space ${\cal{H}}_{m_3 = -k}$, there is a discrete set of precisely $N$ such gauge non-equivalent zero energy configurations. These are most easily written in an appropriately chosen gauge\footnote{Usually a gauge where $\Omega_{1,2,3}$ are taken constant; for other gauges see \cite{Poppitz:2022rxv} and references therein.} for the transition functions on $\T^3$, in the $A_4=0$ gauge. The $N$ classical gauge nonequivalent  zero energy configurations are:
\begin{eqnarray}\label{zeroenergy1}
\sum\limits_{i=1}^3 A_i^{(p)} d x_i = - i \hat T_3^{p}(x_1,x_2,x_3) d \; \hat T_3^{- p}(x_1,x_2,x_3),~p = 0,1,...,N-1.
\end{eqnarray}
Here $\hat T_3(x_1,x_2,x_3)$ is the generator of center symmetry in the direction of the magnetic flux. We recall that in Hamiltonian quantization, see  \cite{GonzalezArroyo:1987ycm,Daniel:1989kj,Cox:2021vsa,Poppitz:2022rxv}, center symmetry is generated by an ``improper'' gauge transformation,  $\hat T_3(x_1,x_2,x_3)$, which is not $\Omega$-periodic but changes by a $\Z_N$ center element upon traversing the $x_3$ direction, i.e. one that obeys
\begin{equation}\label{t3}
\hat T_3(\vec{x} + \vec{e}_i L_i) =e^{i {2\pi \over N} \delta_{i 3}} \; \Omega_i \;\hat T_3 (\vec{x})\; \Omega_i^{-1} ,
~\text{where} ~ \Omega_{3}=1.\end{equation}
An explicit expression for $\hat T_3$ obeying these boundary conditions is possible to find (e.g.~\cite{Poppitz:2022rxv}), but we only need its property (\ref{t3}). 

That (\ref{zeroenergy1}), with $\hat T_3$ obeying (\ref{t3}),  are zero-energy field configurations, is clear from the fact that they are locally pure gauge. That they are gauge inequivalent follows from the fact that they are distinguished by the different values of the holonomy in the $x_3$ direction, $W_3$ defined in (\ref{wilsondef}):
\begin{equation}\label{wilson witten}
W_3[A^{(p)}] = e^{i {2 \pi \over N} p} , ~ p = 0,...,N-1,
\end{equation}
 which  takes  $N$ different   values in the $N$ vacua (\ref{zeroenergy1}). The result (\ref{wilson witten})  follows directly from the boundary condition (\ref{t3}) on $\hat T_3$ and the definition of $W_3$. 

Equivalently stated, the $N$ nonequivalent ground states are the trivial vacuum $A^{(0)}=0$ and its $N-1$ images under  the global $\Z_N^{(1)}$ 1-form symmetry in the $x_3$ direction.\footnote{As explained in \cite{Witten:1982df}, acting with  center symmetry transformations in $x_1$ and $x_2$ leaves $A^{(0)}=0$ invariant.} 
The fact that there are precisely $N$ zero-energy nonequivalent classical field configurations (\ref{zeroenergy1}) means, at small $\T^3$, that there are $N$ quantum states of zero energy. Since these are the only states that contribute to $I_W$ and since $I_W$ does not depend on the volume of $\T^3$ and on $\beta$, one concludes that $I_W=N$.

\subsection{The gaugino condensate: Hamiltonian vs. path integral and ${\cal{N}}=I_W = N$}

Returning to our result (\ref{gauginohamiltonian}) for the gaugino condensate from the Hamiltonian trace and taking
 ${\cal{N}}=N$, the Witten index, we obtain, after taking $\beta \rightarrow \infty$ (so that only zero energy states contribute):
 \begin{eqnarray}\label{gauginohamiltonian 2}
\langle (\tr \lambda^2)^k \rangle\big\vert_{\beta \rightarrow \infty}&=&    \langle E=0, \vec{e}=0| (\tr \lambda^2)^k |0, \vec{e}=E=0 \rangle~.
\end{eqnarray}
Here, we took into account that, as explained in section \ref{sec:anomalyHamilt},  eqn.~(\ref{gauginohamiltonian}) is proportional to the contribution of one of $N$ degenerate zero energy states built over the classical states (\ref{zeroenergy1}), the one with $e_3=0$ (and $e_1=e_2=0$); we also took $(-1)^F=1$.\footnote{A slight technical remark is that $e_3$ flux states ($\hat T_3$  eigenstates) are a discrete $\Z_N$ Fourier transform of the states defined in (\ref{zeroenergy1}). The latter map into each other upon the $\hat T_3$ action.} 
Further taking the infinite $\T^3$-volume limit, $V_3 \rightarrow \infty$, as per the remarks at the end of Section \ref{sec:fermzero}, eqn.~(\ref{gauginohamiltonian 2}) becomes the gaugino condensate in one of the $N$ vacua on $\R^4$. We denote the large $V_3$ limit of (\ref{gauginohamiltonian 2}) by  $\langle (\tr \lambda^2)^k \rangle\big\vert_{\beta, V_3 \rightarrow \infty}$.

We now recall that our semiclassical path-integral calculation of (\ref{defoftrace main})  yielded eqn.~(\ref{gauginocondensates}). Upon taking ${\cal{N}}=N$ (as  done above and in \cite{Anber:2022qsz}) and using the volume independence, we arrive from (\ref{gauginocondensates})  at the result 
\begin{eqnarray}
\label{gaugino 22}
\langle (\tr \lambda^2)^k \rangle\big\vert_{\beta, V_3 \rightarrow \infty}&=&  {\cal{N}}^{-1} \; N^2 \left(16\pi^2 \Lambda^3\right)^k\, = N \left(16\pi^2 \Lambda^3\right)^k\,.
\end{eqnarray}
Thus, assuming that the path integral we calculated, with ${\cal{N}}=N$, matches the Hilbert space expression (\ref{gauginohamiltonian 2}) above, gives the expected result for $\langle (\tr \lambda^2)^k \rangle$ in one of the $\R^4$ vacua, albeit with a factor of $N$ discrepancy. This discrepancy was already observed for $k=1$, $N=2$ in \cite{Anber:2022qsz}. The calculation of this paper, valid for general values of $k, N$ (with gcd($k,N)=1$), yields the same discrepancy. We take this to imply that  the discrepancy has a common origin, as we now describe.

\subsection{The gaugino condensate: ${\cal{N}}$ as a semiclassical path integral}
\label{sec:wittenpath}

We go back to the path integral  (\ref{k correlator}) and the normalization factor ${\cal{N}}$.  Sticking entirely with the path integral formalism, it is natural to take it as given by the path integral with the SYM action (\ref{symaction2}):\footnote{The integral over the auxiliary field is denoted by $[D D]$.}
\begin{eqnarray}\label{normalization 1}
 {\cal N} = \sum_{\nu \in \mathbb Z}\int [D A_\mu][D\lambda][D\bar \lambda] [D D]\; e^{-S_{SYM} }\bigg\vert_{n_{12}=-k\,,n_{34}=0}\,. \end{eqnarray}
We note that the sum here, as opposed to (\ref{k correlator}), is over integer topological charges $\nu$, since with the twists indicated, the topological charge  (\ref{Q of n}) is integer.

 We consider two lines of thought on the expected value of $ {\cal N}$. The first,  more intuitive and based on the expected validity of semiclassics at small $\T^4$, is presented below. The second, more formal argument (which, however, we think is worthy of further development), is based on supersymmetric localization; it appears to lead to a similar result and is presented in Appendix \ref{appx:localization}.

{\flushleft{\bf Semiclassics at small $\T^4$:}} This is the limit where all our calculations were done. The gauge coupling is weak and we expect that a semiclassical calculation of the path integral (\ref{normalization 1}) holds. Sectors with $\nu >0$ require fermion insertions and so should not contribute to  (\ref{normalization 1}). At small volume, the contribution of the sector with $\nu=0$ can be evaluated perturbatively, by expanding around minimum action configurations. The minimum bosonic action in the $\nu=0$ sector is zero. Clearly, the zero action configuration $A=0$ is a classical saddle point (recall that it is an isolated saddle point, as the twists remove the  continuous degeneracy, with a massive supersymmetric spectrum of excitations). Thus, one finds that due to supersymmetry all bosonic and fermionic fluctuations  cancel and the contribution of this saddle point to ${\cal N}$ is simply unity. 

We expect that in this semiclasssical small-$\T^4$ limit the path integral (\ref{normalization 1}) sums over the contributions of all possible gauge-nonequivalent zero-action classical configurations.  The point now is that, as shown in \cite{Gonzalez-Arroyo:1997ugn},  the Euclidean action on $\T^4$ with nonvanishing twists $n_{12}=-k, n_{34}=0$
 has exactly $N^2$ gauge nonequivalent zero action configurations. Since the twists with  gcd$(N,k)=1$ lift all the continuous zero modes, the only zero action configurations  are  discrete holonomies.
 These zero-action   Euclidean configurations on $\T^4$ obey   boundary conditions appropriate to the given twists,   are locally pure gauge, and   can be enumerated by mapping the problem to the study of the irreducible representations of the ``twist group,'' as described in \cite{Gonzalez-Arroyo:1997ugn}.
 
 We refer to the more abstract derivation in the cited reference. Here, we  describe these configurations explicitly, using the language already employed in eqns.~(\ref{zeroenergy1}, \ref{t3}, \ref{wilson witten}), in the Hamiltonian calculation of the Witten index. We begin by stating the general result \cite{Gonzalez-Arroyo:1997ugn}:
the $N^2$ inequivalent configurations, which we label $A^{(p,q)}$, with  $p$ and $q$ taking $N$ values each, are distinguished by the values of the winding Wilson loops in $x_3$ and $x_4$:\footnote{With $W_1=W_2=0$. Recently, these zero action Euclidean minima were also found numerically and characterized, as in (\ref{zeroaction many}), as part of the lattice study of fractional instantons  \cite{Wandler:2024hsq}.}
 \begin{eqnarray}\label{zeroaction many}
  W_3[A^{(p,q)}] &=& e^{i {2 \pi\over N}q}, ~q=0,...,N-1, \nonumber \\
 W_4[A^{(p,q)}] &=& e^{i {2 \pi \over N}p}, ~p=0,...,N-1.
 \end{eqnarray}
 A qualitative explanation of the existence of the zero action configurations characterized by (\ref{zeroaction many}) is that the path integral (\ref{normalization 1}) with only nonzero $n_{12}$ twist allows either $x_3$ or $x_4$ to be taken as the Euclidean time direction. Thus, one can consider center symmetry transformations of the trivial $A=0$ zero-action saddle point in each of these two directions---and, in fact, in both directions, as shown below. 
 
 With this in mind,  we can write an explicit expression for the $N^2$ gauge-inequivalent zero action configurations $A^{(p,q)}$:
 \begin{eqnarray}\label{zeroenergy2}
\sum\limits_{\mu=1}^4 A^{(p,q)}_\mu dx_\mu = - i \hat T_3^{q}(x_1,x_2,x_3) \hat T_3^{p}(x_1,x_2,x_4) d \left(\hat T_3^{q}(x_1,x_2,x_3) \hat T_3^{p}(x_1,x_2,x_4)\right)^{-1}~,
 \end{eqnarray}
 where $\hat T_3$ is the same function of three arguments as appeared in (\ref{zeroenergy1}), a function obeying (\ref{t3}) with $i=3$ denoting the last argument (supplemented by $\Omega_4=1$). However, notice that $\hat T_3$ above is taken to have different arguments: the function $\hat T_3(x_1,x_2,x_3)$ performs a center symmetry transform along $x_3$ while 
 $\hat T_3 (x_1,x_2,x_4)$ is mathematically the same expression, but performing a center symmetry transform along $x_4$ (center symmetry transforms in $x_3$ and $x_4$ commute). The fact that  (\ref{zeroenergy2}) are characterized by the Wilson loop traces of eqn.~(\ref{zeroaction many}) follows simply from the boundary condition (\ref{t3}) obeyed by $\hat T_3$ with respect to its last argument.\footnote{This calculation is easiest done upon taking $\Omega_1$ and $\Omega_2$ to be constant, given as appropriate powers of the clock and shift matrices; recall also that $\Omega_3=\Omega_4=1$.}
 
 Thus, on $\T^4$, there are $N^2$ zero-action saddle points that are gauge nonequivalent (rather than $N$, as assumed in our previous work). Due to supersymmetry, the contribution of each saddle point to the path integral (\ref{normalization 1}) is unity\footnote{The massive spectra of excitations around the $A^{(p,q)}$ saddle points are identical to the ones at the trivial $A^{(0,0)}=0$ one, and the corresponding eigenfunctions are obtained by applying appropriate powers of, the matrices $\hat T_3^{q} \hat T_3^{p}$ (schematically), the ones relating the  backgrounds (\ref{zeroenergy2}) to the trivial one.} just like the contribution of the $A^{(0,0)}=0$ trivial vacuum.
In conclusion, the above chain of arguments, based on the small-$\T^4$ semiclassical evaluation of (\ref{normalization 1}),  makes us declare that 
\begin{equation}
\label{normalization 2} {\cal{N}}=N^2.
\end{equation} 
 Another  formal  argument, presented in Appendix \ref{appx:localization},  which also appears to lead to the result (\ref{normalization 2}), and is worthy of pursuit (see the arguments there) is based on supersymmetric localization of the path integral (\ref{normalization 1}).

Thus, accepting the value for ${\cal{N}}$ given by the above semiclassical reasoning, eqn.~(\ref{normalization 2}), and returning to the result of our calculation (\ref{gauginocondensates}), we now obtain 
\begin{eqnarray}
\label{gaugino 33}
\langle (\tr \lambda^2)^k \rangle\big\vert_{\beta, V_3 \rightarrow \infty}&=&  {\cal{N}}^{-1} \; N^2 \left(16\pi^2 \Lambda^3\right)^k\, =  \left(16\pi^2 \Lambda^3\right)^k\,,
\end{eqnarray}
a result agreeing for all $k$ and $N$, assuming gcd$(k,N)=1$, with the $\R^4$ result.

 It goes without saying that the discrepancy between the Witten index calculated via the path integral and Hamiltonian approaches clearly calls for resolution. In this work, we do not aim to provide a complete solution to this puzzle. Nonetheless, we outline a potential path forward toward addressing this issue, as we discuss below and in Appendix \ref{appx:measure}.

%%%%%%%%%%%%%%%%%%%%%%%%%%%%%%%%%%%%%%%%%
\section{Conclusions and outlook}
\label{Conclusions}
 %%%%%%%%%%%%%%%%%%%%%%%%%%%%%%%%%%%%%%
 
 The focus of this paper is the calculation of the higher-order gaugino condensate on the twisted $\T^4$, leading to the result (\ref{gaugino 33}), agreeing with the weakly coupled instanton calculations  \cite{Dorey:2002ik}  on $\R^4$ and the recent lattice results \cite{Bonanno:2024bqg}. While initially attempted 40 years ago \cite{Cohen:1983fd},  the $\T^4$ calculation could be completed only after the recent understanding of generalized anomalies involving center symmetry \cite{Gaiotto:2014kfa,Gaiotto:2017yup}, including in the Hamiltonian formalism \cite{Cox:2021vsa}, and the  construction of spatially-dependent fractional instanton solutions on the torus, pioneered in \cite{GarciaPerez:2000aiw} and developed in \cite{Gonzalez-Arroyo:2019wpu,Anber:2023sjn}.
 
{\flushleft{L}}et us now summarize our main results:
 
\begin{enumerate}
\item
The result (\ref{gaugino 33}) for the higher-order condensate is comforting. It explicitly demonstrates that the small-$\T^4$ semiclassical setup directly computes quantities relevant to the $\R^4$ limit, thanks to the protection afforded by supersymmetry. It emphasizes the role of the $k$-lump multi-fractional instanton ``liquid''-like self-dual configurations (pictured on Figure~\ref{visual of liquid}) in determining the value of the multi-gaugino condensate. 

At the technical level, the determination of the shape and size of the moduli space of the $Q=k/N$ instanton is crucial in obtaining  (\ref{gaugino 33}). We also note that, for $k>1$, the calculation using the $Q=k/N$ solution is technically significantly simpler compared to the one using the ADHM background in super QCD (SQCD) on $\R^4$ \cite{Dorey:2002ik}. 
 
\item As opposed to the  weakly coupled calculation via SQCD on $\R^4$, the objects contributing to the higher-order gaugino condensate are closely related to the ones causing chiral symmetry breaking and confinement, as on $\R^3 \times \S^1$ \cite{Unsal:2007jx}. Fractional instantons, monopole-instantons, and center vortices are all objects, which in different geometries can be argued  to lead to semiclassical confinement  \cite{RTN:1993ilw,Gonzalez-Arroyo:1998hjb,Unsal:2007jx,Unsal:2008ch,Tanizaki:2022ngt}. The fractional instantons used in our calculation are continuously connected, upon taking various limits of $\T^4$ sizes and twists, to both the monopole-instantons (responsible for semiclassical chiral symmetry breaking and confinement on $\R^3 \times \S^1$) and center vortices (which accomplish this on $\R^2 \times \T^2$). This continuity  has been suggested earlier  and shown recently (via analytical or numerical tools, see \cite{GarciaPerez:1999hs,Unsal:2020yeh,Poppitz:2022rxv,Hayashi:2024yjc,Guvendik:2024umd,Wandler:2024hsq}), demonstrating that the space of multi-fractional instantons on $\T^4$ contains  the topological objects arising in the  semi-infinite volume  $\R^3 \times \S^1$ or $\R^2 \times \T^2$   limits.
 \end{enumerate}
As we already mentioned, there are aspects of our calculation that need better understanding. We end by listing some of the  issues left for future studies:
 \begin{enumerate}
\item \label{wittenconfusion}
There is one unsettling element left in our determination of (\ref{gaugino 33}). Based on the usual relation between Hamiltonian and path integral formalisms, one expects that the path integral (\ref{normalization 1})  equals the Hamiltonian trace (\ref{windex main}). As is clear from our discussion, the reason for the discrepancy are the saddle points with $p \ne 0$ in (\ref{zeroenergy2}). They contribute due to the fact that the Euclidean path integral (\ref{normalization 1}) can be  time-sliced so that time is either  $x_3$ or $x_4$. 
In order for ${\cal{N}}$ of (\ref{normalization 1}) to yield the same result as the trace $I_W$ of (\ref{windex main}), the contribution of the saddle points representing center symmetry images of $A=0$ in the $x_4$ direction has to be omitted. Currently, we do not know how to justify this. Presumably, an appropriate choice of the (infinite dimensional) complex contour of integration would be required (see Appendix \ref{appx:localization}).

\item  We note, however, that a similar story unfolds in the $\mathbb{Z}_N$ BF theory on a torus, as we discuss in Appendix \ref{appx:znlattice}. For example, consider this theory on $\mathbb{T}^2$. In a Euclidean setup, this theory has $N^2$ saddle points, naively giving the partition function $Z = N^2$. However, in a Hamiltonian formalism, one finds that there are $N$ ground states, contributing $N$ to the partition function. In Appendix \ref{appx:znlattice}, we provide a detailed explanation of this finding and show how a careful treatment of the measure yields the correct result, $Z = N$. In Appendix \ref{appx:g2}, we employ a similar construction in lattice Yang-Mills theory to argue that a proper formulation of the Yang-Mills measure can potentially resolve the puzzle of the Witten index.\footnote{We are grateful to an anonymous referee for pointing out an analogous issue arising in $\Z_N$ topological gauge theories. We also thank Tin Sulejmanpasic for discussions and Theodore Jacobson for pointing out the related refs.~\cite{Choi:2021kmx,Hellerman:2010fv}.}

\item
 We also note that the same procedure as suggested in point \ref{wittenconfusion}. above---ignoring center symmetry images in the $x_4$ time direction---can be applied directly to the calculation of the gaugino condensate (\ref{k correlator}). Recall from (\ref{multilump1}) that $z_4$ shifts by ${1\over N(N-k)}$ perform center symmetry transforms in $x_4$. Thus, ignoring center transforms in $x_4$ would make the range of $z_4$ $N$ times smaller than indicated in (\ref{range of z 1}), i.e. would have the range of $z_4 \in [0, {1 \over N(N-k)}]$. This restriction would  produce, instead of (\ref{gauginocondensates}), an answer $N$ times smaller, agreeing with the $\R^4$ answer upon taking  ${\cal{N}}=N$.
 
  % As in the previous point, at the moment, we do not know in great details how the Euclidean path integral imposes this restriction. We suggest that, perhaps, answering the same question regarding the Witten index, as discussed above, is a simpler first step
   
  \item The calculation of the higher-order condensates has yet to be generalized for the cases with gcd$(k,N)>1$. That one expects subtleties is  already clear from the fact that the finite volume degeneracies discussed in section \ref{sec:anomalyHamilt}  depend  on whether one takes $x_{1,2}$ or $x_{3,4}$ to be the spatial directions. In addition, the moduli space of our multifractional instantons has not been fully analyzed for this case. 
  
  \item Finally, the relation between gaugino condensates and fractional instantons on $\T^4$ found here  is specific for $SU(N)$ gauge groups. For other gauge groups,  with a smaller center or  without a center, the gaugino condensate can be seen to arise due to monopole-instantons on $\R^3\times \S^1$ \cite{Davies:2000nw}, but these objects do not appear  related to fractional topological charge objects on $\T^4$. The physics of the gaugino condensate on $\T^4$ is then likely to be more complicated and remains to be uncovered. 
\end{enumerate}

{\bf {\flushleft{Acknowledgments:}}}  E.P. is grateful to Margarita Garc\' ia P\' erez and Antonio Gonz\' alez-Arroyo for  many conversations on fractional instantons and the gaugino condensate, and especially for sharing their understanding of  zero-action Euclidean configurations on the torus. E.P. also acknowledges the hospitality of the Instituto de F\' isica Te\' orica, UAM-CSIC, Madrid. We are grateful to an anonymous referee for pointing  the analogy with $\Z_N$ topological gauge theories. We also thank Tin Sulejmanpasic  and Theodore Jacobson for discussions and Andrew Cox for help with the numerical minimization discussed in Appendix \ref{appx:g2}.  M.A. is supported by STFC through grant ST/T000708/1.   E.P. is supported by a Discovery Grant from NSERC.

 \appendix
 
 \section{Supersymmetric Ward identities on $\mathbb{T}^4$}
  \label{appx:ward}
  
 Here, we argue that the supersymmetric Ward identities, usually discussed on $\R^4$,  also hold when expectation values are computed via path integral on $\mathbb{T}^4$. The Ward identities that we use are:
 \begin{eqnarray} \label{ward1}
 \Lambda^* {\partial \over \partial \Lambda^*} \langle \phi_1(x^1)...\phi_s(x^s) \rangle &=&0,   \\
 \bar\sigma_\mu^{\dot\alpha \alpha }{\partial \over \partial x^i_\mu} \langle \phi_1(x^1)...\phi_i(x^i)... \phi_s(x^s)\rangle &=&0, \label{ward2}
 \end{eqnarray}
 where $\phi_i$ are lowest components of chiral superfields and $\Lambda^*$ is the antiholomoprhic scale. On $\mathbb{T}^4$, the brackets $\langle ... \rangle$ mean
 \begin{eqnarray}\label{defoftrace}
 \langle \hat{\cal{O}} \rangle \equiv\tr_{{\cal{H}}_{m_3}} \left[ \hat{\cal{O}} e^{- \beta H} \hat{T}_3   (-1)^F\right],
 \end{eqnarray}
 where the trace is over the Hilbert space on $\mathbb{T}^3$ (spanned by $x_{1,2,3}$, with a spatial twist $n_{12}=-k$) and the insertion of the center symmetry generator, $\hat{T}_3$, assures that $n_{34}=1$.   Usually, supersymmetric Ward identities are considered in the limit $\beta \rightarrow \infty$ (in fact, also in the infinite $\mathbb{T}^3$ limit), where only the ground state contributes. The proof uses the fact that the supercharges annihilate the ground state; see the review \cite{Amati:1988ft}. In contrast, at finite $\beta$, all excited states contribute, and hence, the proof of the Ward identities requires some modifications. 
 
 Before we consider these, we stress that (\ref{ward1}) is important since it shows that gaugino condensates can not depend on the antiholomorphic scale $\Lambda^*$, excluding a dependence on the size of $\mathbb{T}^4$ (on dimensional grounds, the size can only enter through dependence on  $L | \Lambda|$, where $L$ is any linear dimension of the torus). Thus, holomorphy implies, for example, that $\langle \tr (\lambda\lambda)  \rangle  = c \Lambda^{3}$ for some constant $c$.
The second Ward identity (\ref{ward2}) states that expectation values of products of lowest components of chiral superfields are $x_i$-independent. It has been used in attempts to relate results obtained for small $|x_i-x_j|$ (e.g., in strongly-coupled instanton calculations, whose validity has been questioned by many, see \cite{Dorey:2002ik}) to those at arbitrary separations, allowing the use of cluster decomposition in the infinite volume limit. 

The fact that the Ward identities also hold on $\mathbb{T}^4$ appears to have been known (or obvious) to the authors of \cite{Shifman:1986zi}.    The required modification in the proof for the $\mathbb{T}^4$ case is minimal, and we present it for completeness.  The point is that the $\Lambda^*$ derivative is proportional to an insertion of the highest component of an antichiral superfield,\footnote{This is because $\Lambda^*$ couples as $\ln \Lambda^* \int d^4x d^2 \bar \theta\, \tr \bar W^2$, with $\int d^2 \bar \theta\, \tr \bar W^2 = F^*$, see  \cite{Shifman:1986zi}.} which obeys $F^* \sim \{ \bar Q_{\dot\alpha}, \bar\psi^{\dot\alpha}\}$,  where $\bar\psi^{\dot\alpha}$ is the middle component of an antichiral superfield.\footnote{We only consider (\ref{ward1}), noting that the modification of the proof for (\ref{ward2}) is identical:  using the fact that $\bar\sigma_\mu^{\dot\alpha \alpha} \partial_\mu \phi(x)  = \{\bar Q^{\dot\alpha}, \psi^\alpha\}$, one proceeds via steps identical to   (\ref{ward3}-\ref{ward8})  followed in the proof of (\ref{ward1}).}
 Thus, denoting $\phi_1(x^1)...\phi_s(x^s) = \hat{\cal{O}}$, we have
\begin{eqnarray} \label{ward3}
 \Lambda^* {\partial \over \partial \Lambda^*}\langle\hat  {\cal{O}} \rangle \sim \langle  \hat{\cal{O}}     \{ \bar Q_{\dot\alpha}, \bar\psi^{\dot\alpha}\}\rangle \sim \langle  \hat{\cal{O}} ( \bar Q_{\dot 1}  \bar\psi_{\dot 2} +\bar\psi_{\dot 2}  \bar Q_{\dot 1} )\rangle - ( \dot{1} \leftrightarrow \dot{2}).
\end{eqnarray}
Then, we use $[ \bar Q_{\dot\alpha},\hat {\cal{O}}]=0$ to cast the r.h.s. of (\ref{ward3}) in the form
\begin{eqnarray}\label{ward4}
 \Lambda^* {\partial \over \partial \Lambda^*}\langle\hat {\cal{O}} \rangle \sim \langle  \bar Q_{\dot 1} ({\cal{O}} \bar\psi_{\dot 2}) + ({\cal{O}} \bar\psi_{\dot 2}) \bar Q_{\dot 1} \rangle - ( \dot{1} \leftrightarrow \dot{2})
 \end{eqnarray}
In the $\beta \rightarrow \infty$ limit where $\langle ...\rangle$ means vacuum (zero-energy) expectation value, (\ref{ward4}) vanishes since the supercharges annihilate the ground state, completing the proof. At finite $\beta$, using (\ref{defoftrace}), denoting $X_{\dot\alpha} \equiv \hat {\cal{O}} \bar\psi_{\dot\alpha}$, we have instead:\footnote{Keeping only   the first term on the r.h.s. in (\ref{ward4}); the vanishing of the second term follows similarly to what we show below.}
\begin{eqnarray}\label{ward5}
 \langle  \bar Q_{\dot 1}   X_{\dot 2} + X_{\dot 2}  \bar Q_{\dot 1} \rangle = \sum\limits_E (-1)^F e^{-\beta E} \langle E|  \bar Q_{\dot 1}   X_{\dot 2} + X_{\dot 2}  \bar Q_{\dot 1} |E\rangle~.
\end{eqnarray}
The states with $E=0$ are annihilated by the supercharges, just like in $\R^4$. For the $E>0$ states, we use a representation of the supersymmetry generators in terms of fermionic creation and annihilation operators
$\bar Q_{\dot \alpha} \sim a_\alpha^\dagger, Q_\alpha \sim a_\alpha$, with $\{ a_\alpha^\dagger, a_\beta\} = \delta_{\alpha \beta}$. Then for each energy eigenstate,\footnote{There can be extra degeneracies of the energy levels, in addition to the one due to supersymmetry, as seen in the Hamiltonian formalism on $\mathbb{T}^3$ in the presence of twists, section \ref{sec:anomalyHamilt}. The proof of the Ward identities given holds whether or not twists are present.} we have four degenerate states:
\begin{eqnarray}\label{fermionstates}
|0\rangle, \; |1\rangle = a_1^\dagger |0\rangle, \; |2 \rangle = a_2^\dagger |0\rangle, \; |0'\rangle = a_1^\dagger a_2^\dagger |0\rangle, ~\text{with} ~ a_\alpha |0\rangle = 0 ~ \text{and} ~ a_\alpha^\dagger |0'\rangle = 0\,,
\end{eqnarray}
where, without loss of generality, $|0\rangle$ and $|0'\rangle$ are taken bosonic and $|1 \rangle, |2\rangle$ are fermionic. 
Thus, the contribution of any given $E>0$ supermultiplet of states to (\ref{ward5}) is proportional to:
\begin{eqnarray}
&&\langle 0 | \{\bar Q_{\dot 1}, X_{\dot 2} \} |0\rangle + \langle 0' | \{\bar Q_{\dot 1}, X_{\dot 2} \} |0'\rangle - \langle 1 | \{\bar Q_{\dot 1}, X_{\dot 2} \} |1\rangle -  \langle 2 | \{\bar Q_{\dot 1}, X_{\dot 2} \} |2\rangle \nonumber \\
&\sim& \langle 0 | \{ a_1^\dagger, X_{\dot 2} \} |0\rangle + \langle 0' | \{a_1^\dagger, X_{\dot 2} \} |0'\rangle - \langle 1 | \{a_1^\dagger, X_{\dot 2} \} |1\rangle -  \langle 2 | \{a_1^\dagger, X_{\dot 2} \} |2\rangle \label{ward7}
\end{eqnarray}
Then using (\ref{fermionstates}), we find that the last line in (\ref{ward7}) equals
\begin{eqnarray}\label{ward8}
&& \langle 0 |X_{\dot 2}  |1\rangle + \langle 2 |  X_{\dot 2}  |0'\rangle - \langle 0|   X_{\dot 2}  |1\rangle -  \langle 2 |  X_{\dot 2}  |0'\rangle =0.
\end{eqnarray}
This shows that the contribution of each nonzero energy state  to the sum in (\ref{ward5}) cancels out and the holomorphy, eqn.~(\ref{ward1}), holds as on $\R^4$.   As noted above, the $x_\mu$ derivative in the other Ward identity (\ref{ward2}) reduces, using the supersymmetry transformations of chiral superfields, to an expression similar to (\ref{ward5}) and leads to a similar conclusion.

%%%%%%%%%%%%%%%%%%%%%%%%%%%%%%%%%%%%
\section{The nonabelian part  of the solution to leading order in $\Delta$}
\label{appx:solution}
%%%%%%%%%%%%%%%%%%%%%%%%%%%%%%%%%%%%%

Here, we give the explicit expressions of ${\cal W}_{\mu}^{(0)k\times \ell}$  that correspond to a fractional instanton carrying a topological charge  $Q={r\over N}$. We set $n_{12}=-r$, $n_{34}=1$, and employ the embedding $SU(N)\supset SU(k)\times SU(\ell)\times U(1)$. This is a more general case than  $n_{12} = - k$ of (\ref{twists1}) (and with the same $n_{34}=1$) considered in the bulk of this paper. In our discussion of the properties of this general solution, it will become apparent why we chose to take $k=r$ through the bulk of the paper.  This general solution was explicitly constructed in \cite{Anber:2023sjn}.

To order $\sqrt{\Delta}$, the solution (\ref{expansion1}) is determined by two functions,\footnote{The remaining functions determining the nonabelian solution (\ref{gauge field in general}), e.g. ${\cal{S}}_\mu$ from  (\ref{expansion1}) appear first at order-$\Delta$. They are uniquely determined in terms of the order-$\sqrt{\Delta}$ functions given below by solving a recursion relation, see \cite{Anber:2023sjn}, but we do not have  an explicit form.}
  ${\cal W}_4^{(0)k\times \ell}$ and ${\cal W}_2^{(0)k\times \ell}$, given by:\footnote{For the notation, recall footnote \ref{footnotewrap}.}
\begin{eqnarray}
\left({\cal W}_{2,4}^{(0)k\times \ell}\right)_{C'C}=V^{-1/4} \sum_{p=0}^{\scriptsize \frac{r}{\mbox{gcd}(k,r)}-1}{\cal C}^{[C'+pk]_r}_{2,4}\Phi^{(p)}_{C'C}(x,\hat\phi)  \,.
\label{expressions of W2 and W4 with holonomies}
\end{eqnarray}
The remaining two functions ${\cal W}_1^{(0)k\times \ell}$ and ${\cal W}_3^{(0)k\times \ell}$ appearing in the leading-order expansion (\ref{expansion1}) are determined by ${\cal W}_2^{(0)k\times \ell}$ and ${\cal W}_4^{(0)k\times \ell}$ via
\begin{eqnarray}\label{assertion1}
 {\cal W}_1^{(0)k\times \ell}=-i{\cal W}_2^{(0)k\times \ell}\,,\quad {\cal W}_3^{(0)k\times \ell}=-i{\cal W}_4^{(0)k\times \ell}\,.
\end{eqnarray}
In (\ref{expressions of W2 and W4 with holonomies}), $C_{2,4}^{[C'+pk]_r}$ are complex constants whose significance is discussed in the last paragraph of this section, while 
  $\Phi^{(p)}_{C'C}(x,\hat\phi)$ are functions of $x$ and the moduli $\hat\phi_\mu^{C'} $. The latter are  defined in terms of $\phi_\mu^{C'}$ from eqn.~(\ref{deltaA1}) via
 \begin{equation}\label{holonomy1}
\hat\phi_\mu^{C'} \equiv \phi_\mu^{C'} - 2 \pi N {z_\mu \over L_\mu}, ~ {\rm{with}}~ \hat\phi_\mu^{C'} =  \hat\phi_\mu^{[C'-r]_k}\,.
\end{equation} 
To complete the explicit form of the order-$\sqrt{\Delta}$ solution, we now give the form of $\Phi^{(p)}_{C'C}(x,\hat\phi)$:
\begin{eqnarray}
\nonumber
\Phi^{(p)}_{C' B}(x,\hat\phi)&=& \sum_{\scriptsize m=p+\frac{rm'}{\mbox{gcd}(k,r)},\, m'\in \mathbb Z}~~\sum_{n'\in \mathbb Z}e^{\frac{i2\pi x_2 }{L_2}(m+\frac{2	C'-1-k}{2k})}e^{\frac{i2\pi x_4 }{L_4}(n'-\frac{2	B-1-\ell}{2\ell})}\\
\nonumber
&&~~\times e^{-i\frac{\pi (1-k)}{k}\left(C'-\frac{1+k(1-2m)}{2}\right)} e^{i\frac{\pi(1-\ell)}{\ell}\left(B-\frac{1+\ell(2n'+1)}{2}\right)}\\
\nonumber
&&~~\times \; e^{-\frac{\pi r}{k L_1 L_2}\left[x_1-\frac{k L_1 L_2}{2\pi r}(\hat \phi_2^{[C']_r}-i\hat \phi_1^{[C']_r})-\frac{L_1}{r}\left(km +\frac{2C'-1-k}{2}\right)\right]^2}\\
&&~~\times \; e^{-\frac{\pi }{\ell L_3 L_4}\left[x_3-\frac{\ell L_3 L_4}{2\pi }(\hat \phi_4^{[C']_r}-i\hat \phi_3^{[C']_r})-L_3\left(\ell n' -\frac{2B-1-\ell}{2}\right)\right]^2}\,.
\label{form of Phi}
\end{eqnarray}

Finally, the complex coefficients ${\cal C}^{[C'+pk]_r}_2$ and ${\cal C}^{[C'+pk]_r}_4$ are $4r$ arbitrary parameters, and a subset of these parameters serve as additional moduli. A careful analysis, see \cite{Anber:2023sjn}, reveals that ${\cal C}^{[C'+pk]_r}_2$, ${\cal C}^{[C'+pk]_r}_4$, in addition to the holonomies $\phi^{C'}_\mu$ and translations $z_\mu$, comprise in total $4r$ independent bosonic moduli, as per the index theorem.

In the limiting case $r=1$, one finds ${\cal C}^{[C'+pk]_r}_4=0$, while ${\cal C}^{[C'+pk]_r}_2$ remains as arbitrary unphysical $U(1)$ phase. Here, we can set $\phi_\mu^{C'}=0$, and thus, we are left with the $4$ translational moduli $z_\mu$. 
In the general case $1<r<N$, the moduli ${\cal C}^{[C'+pk]_r}_2$ and ${\cal C}^{[C'+pk]_r}_4$ are non-compact, resulting in infinities when integrated over in the path integral. This issue is resolved by setting $k=r$, which corresponds to choosing the embedded groups $SU(k)=SU(r)$ and $SU(\ell)=SU(N-r)$ within $SU(N)$. This specific choice eliminates ${\cal C}^{[C'+pk]_r}_4$ and sets ${\cal C}^{[C'+pk]_r}_2$ to an arbitrary nonphysical $U(1)$ phase, leaving only the compact moduli $\phi^{C'}_\mu$ and $z_\mu$ as the relevant moduli in the problem. Notice that here $C'=1,..,r$ and that $\phi^{C'}_\mu$ is subject to the constraint $\sum_{C'=1}^r \phi^{C'}_\mu=0$. Thus, there are $4(r-1)$ holonomies. Adding the $4$ translations $z_\mu$ gives a total of $4r$ bosonic moduli. This $r=k$ choice is the one assumed throughout this work. 

One can study the gauge-invariant densities of the solution, e.g., $\mbox{tr}\left[F_{12}F_{12} \right]$.  Using (\ref{gauge field in general}, \ref{fields1}, \ref{expansion1}, \ref{feild strength}), it is tedious but straightforward to show that  to ${\cal O}(\Delta)$:
\begin{eqnarray}\nonumber
\mbox{tr}\left[F_{12}F_{12} \right]=\mbox{tr}[\omega^2]\left\{ \hat F_{12}^\omega \hat F_{12}^\omega+2\Delta \hat F_{12}^\omega\left(\partial_{1}{\cal S}^{(0)\omega }_{2}-\partial_{2}{\cal S}^{(0)\omega }_{1}\right) \right\}+ 8\pi N \Delta \hat F_{12}^\omega\mbox{tr}_k\left[{\cal W}_{2}^{(0)}{\cal W}_{2}^{\dagger(0)}\right]\,,\\
\label{field strength F12F12}
\end{eqnarray}
where $\mbox{tr}_k$ denotes taking the trace of the respective $k\times k$ matrix. 
The first term $\hat F_{12}^\omega \hat F_{12}^\omega$ is constant on the deformed $\mathbb T^4$, while it can be shown that the second term $
\left(\partial_1{\cal S}_2^{(0)\omega}-\partial_2{\cal S}_1^{(0)\omega} \right)=O\, \mbox{tr}_k\left[{\cal W}_2^{(0)}{\cal W}_2^{\dagger(0)}\right]$ for some differential operator $O$ whose explicit form can be found in \cite{Anber:2023sjn}. Then, using  (\ref{expressions of W2 and W4 with holonomies}, \ref{form of Phi}) we find for the $x$-dependent part of the gauge-invariant density (\ref{field strength F12F12})
\begin{eqnarray}\label{fractional1}
&&\mbox{tr}_k\left[{\cal W}_2^{(0)}{\cal W}_2^{\dagger(0)}\right]  \sim \nonumber \\
&  & \sum\limits_{C'=1}^k\bigg|  \sum_{m'\in \mathbb Z}\;\;e^{i \left(\frac{ 2\pi x_2 }{L_2}+L_1\hat \phi_1^{C'}\right) m'  -\frac{\pi  }{  L_1 L_2}\left[x_1-\frac{L_1 L_2}{2\pi }\hat \phi_2^{C'}- \frac{L_1 C'}{k} - L_1 (m' -\frac{1+k}{2k}) \right]^2}\bigg|^2 \nonumber \\
 &&\;\;\; \times \;
\bigg|  \sum_{n'\in \mathbb Z} \;e^{i\left(\frac{2\pi x_4 }{\ell L_4}+ L_3\hat \phi_3^{C'}\right) n'   -\frac{\pi }{\ell L_3 L_4}\left[x_3-\frac{\ell L_3 L_4}{2\pi }\hat \phi_4^{C'}-L_3\left(\ell n' +\frac{ 1+\ell}{2}\right)\right]^2}\, \bigg|^2\nonumber  \\
&&\equiv \sum\limits_{C'=1}^k  F \left(x_1 - {L_1 L_2 \over 2 \pi} \hat\phi_2^{C'} - {L_1 C' \over k},\; 
x_2 +  {L_1 L_2 \over 2 \pi} \hat\phi_1^{C'},\; x_3 -  {\ell L_3 L_4 \over 2 \pi} \hat\phi_4^{C'},\; x_4+  { \ell L_3 L_4 \over 2 \pi} \hat\phi_3^{C'} \right),  \nonumber \\
\end{eqnarray}
where the last equality defines the function $F$. The lumpy structure of the $Q={k \over N}$ solution seen in the above formula is discussed at the end of section \ref{sec:lumpy}, see Figure \ref{visual of liquid}. We also stress that local gauge invariant densities, as opposed to winding Wilson loops, are periodic functions on $\mathbb{T}^4$ with periods equal to the torus periods $L_\mu$. 

%%%%%%%%%%%%%%%%%%%%%%%%%%%%%%%%%%%%%%%%
\section{The nonabelian part of the fermion zero modes and their localization}
 \label{appx:fermion}
%%%%%%%%%%%%%%%%%%%%%%%%%%%%%%%%%%%%%%%

 The solutions of the undotted Dirac equation, the fermion zero modes $\lambda$, are given as a matrix in the form
\begin{eqnarray}
\lambda=\left[\begin{array}{cc}||\lambda_{C'B'}|| & ||\lambda_{C'C}|| \\ ||\lambda_{CC'}|| & ||\lambda_{CB}||  \end{array}\right]\,.
\end{eqnarray}
Here, we only consider the case $r=k$ discussed in the bulk of the paper. 
The solution of the Dirac equation to ${\cal O}(\Delta^0)$ yields the diagonal zero mode solutions
\begin{eqnarray} \label{fermion1apx}
\lambda_{\alpha \; B'C'} = \delta_{B'C'}\; \theta_\alpha^{C'}\,,\quad
\lambda_{\alpha \; BC} = - \delta_{BC} \;{1 \over \ell} \sum_{C'=1}^k \theta_\alpha^{C'}, 
\end{eqnarray}
where  $\alpha=1,2$ is the spinor index and $C',B'=1,2..,k$. 
The order $\sqrt{\Delta}$ off-diagonal solutions are, see \cite{Anber:2023sjn}:
\begin{eqnarray}\label{lambdazeromode11apx}\nonumber
\lambda_{1 \; C'D} &=& -i V^{1/4}\sqrt{\Delta} \eta_2^{C'} {\cal F}_{13 \; C'D}^{(0)}\,, \quad
\lambda_{2 \; C'D} = 0~\,\\
\lambda_{1 \; CD'} &= & 0\,,\quad
\lambda_{2 \; CD'} = i V^{1/4}\sqrt{\Delta}\eta_1^{D'}   {\cal F}_{13 \; D'C}^{* \; (0)}\,,
\end{eqnarray}
where we introduced the spinor $ \eta^{C'}_\alpha$ defined, modulo an overall multiplicative factor, as
 \begin{equation}\label{etavstheta}
 \eta^{C'}_\alpha \equiv    \theta^{C'}_\alpha + {1 \over \ell} \sum\limits_{B'=1}^{r} \theta^{B'}_\alpha,
 \end{equation}
and
\begin{eqnarray}
\nonumber
{\cal F}^{(0)}_{13,C',C}(x,\hat\phi)&=&-iV^{1/4}\frac{2\pi }{\ell L_3 L_4}\ \sum_{m'\in \mathbb Z}\sum_{n'\in \mathbb Z}e^{\frac{i2\pi x_2 }{L_2}(m'+\frac{2	C'-1-k}{2k})}e^{\frac{i2\pi x_4 }{L_4}(n'-\frac{2	C-1-\ell}{2\ell})}\\
\nonumber
&&\times e^{-i\frac{\pi (1-k)}{k}\left(C'-\frac{1+k(1-2m')}{2}\right)} e^{i\frac{\pi(1-\ell)}{\ell}\left(C-\frac{1+\ell(2n'+1)}{2}\right)}\\
\nonumber
&&\times \left( x_3-\frac{\ell L_3 L_4\hat \phi_4^{C'}}{2\pi  } -L_3\left(\ell n'-\frac{2C-1-\ell}{2}\right)\right) \\
\nonumber
&&\times e^{-\frac{\pi  }{  L_1 L_2}\left[x_1-\frac{  L_1 L_2}{2\pi   }(\hat \phi_2^{C'}-i\hat \phi_1^{C'})-\frac{L_1}{k}\left(km +\frac{2C'-1-k}{2}\right)\right]^2}\\
&&\times e^{-\frac{\pi }{\ell L_3 L_4}\left[x_3-\frac{\ell L_3 L_4}{2\pi }(\hat \phi_4^{C'}-i\hat \phi_3^{C'})-L_3\left(\ell n' -\frac{2C-1-\ell}{2}\right)\right]^2}\,.
\label{the G3 function}
\end{eqnarray}
is the off-diagonal field strength of the instanton background to ${\cal O}(\sqrt{\Delta})$, where $\hat\phi^{C'}$ are the holonomies from (\ref{holonomy1}).

As predicted by the index theorem, there are $2k$ fermion zero modes associated with the spinors $\theta^{C'}_\alpha$ for $C'=1,\ldots,k$ and $\alpha=1,2$. These modes arise in the background of a self-dual fractional instanton with a topological charge of $Q=k/N$ on the deformed $\mathbb{T}^4$.

One can construct order-$\Delta$ gauge invariants from the fermion zero modes that display a pattern similar to the bosonic invariants, characterized by a lumpy structure. Each of the $k$ lumps hosts 2 zero modes, with their positions determined by the moduli $\hat\phi^{C'}_\mu$. Specifically, the order-$\Delta$ gauge invariants formed from the fermion zero modes include terms such as
\begin{eqnarray} \label{fermionzeromodedensity}\nonumber
&& \sum_{C'=1}^k\sum_{D=1}^\ell \lambda_{1 \; C'D} \lambda_{2 \; DC'}  \sim  \\
&& \sum\limits_{C'=1}^k \bar\eta_1^{C'} \bar\eta_2^{C'}  \bigg| \sum\limits_{m} e^{i\frac{2 \pi m}{L_2} (x_2 + {L_1 L_2 \over 2 \pi} \hat\phi_1^{C'})   -\frac{  \pi  }{  L_1 L_2}\left[x_1-\frac{  L_1 L_2}{2\pi  }\hat \phi_2^{C'}  -\frac{L_1 C'}{k} + L_1 \frac{1+k}{2k}- L_1 m \right]^2} \bigg|^2\times \nonumber \\
\nonumber
&&   \bigg| \sum\limits_{n}   \left( x_3-\frac{\ell L_3 L_4}{2\pi } \hat \phi_4^{C'} -L_3 \ell  n - L_3 \frac{  1+\ell}{2} \right) e^{i \frac{2\pi n}{\ell L_4} (x_4 + {\ell L_3 L_4 \over 2 \pi} \hat\phi_3^{C'}) -\frac{\pi }{\ell L_3 L_4}\left[x_3-\frac{\ell L_3 L_4}{2\pi } \hat \phi_4^{C'} -L_3\left( \ell n  +\frac{1+\ell}{2}\right)\right]^2  } \bigg|^2\,.\\ \end{eqnarray}
This expression highlights the localization properties of the fermion zero modes, as dictated by the holonomies $\hat\phi^{C'}$, which were evident in the bosonic solution described in (\ref{fractional1}). From (\ref{fermionzeromodedensity}), we see that very one of the $k$ lumps in the sum in (\ref{fermionzeromodedensity}) hosts $2$ fermion zero modes.

%%%%%%%%%%%%%%%%%%%%%%%%%%%%%%%%%%%%
\section{Wilson  lines and symmetries of gauge-invariant observables}
\label{appx:wilson}
%%%%%%%%%%%%%%%%%%%%%%%%%%%%%%%%%%%%

Here, we study the Wilson loops' moduli dependence. We set $k=r$, $q_\mu=q$ and impose the condition 
\begin{eqnarray}\label{explicit moduli dependenceappx}
\int_{\Gamma} \left(\prod_{\nu=1}^4 dz_\nu d \bm a_\nu \right) W_\mu^q[\hat A(\{z_\nu, \bm{a}_\nu\})]=0,~~ \text{where} ~d \bm a_\nu = \prod\limits_{C'=1}^{k-1} da_\nu^{C'}~.
\end{eqnarray}
Using Eqs. (\ref{the set of transition functions for Q equal r over N, general solution}, \ref{gaugewithholonomies}, \ref{naked expressions}, \ref{r over N abelian sol}), we obtain to the zeroth order of $\Delta$:
\begin{eqnarray}\label{Wilson 1}
\nonumber
W_1^q &=&(-1)^{q(k-1)} e^{-i2\pi q(N-k)\left(z_1-\frac{x_2}{N L_2}\right)}\left[\sum\limits_{C'=1}^k e^{i 2\pi q \bm a_1\cdot \bm \nu_{C'}}\right]+(N-k)e^{i2\pi qk \left(z_1-\frac{x_2}{N L_2}\right)}\,, \\\nonumber
W_2^q&=& e^{-i2\pi q(N-k)\left(z_2+\frac{x_1}{N L_1}\right)} \; \left[\sum\limits_{C'=1}^k  e^{i 2\pi q (\bm a_2 - {\bm \rho \over k})\cdot \bm\nu_{C'} }\right] + (N-k)e^{i2\pi q k \left(z_2+\frac{x_1}{N L_1}\right)}\,,\\\nonumber
W_3^q&=& e^{-i2\pi q(N-k)\left(z_3-\frac{x_4}{N \ell L_4}\right)}\left[\sum\limits_{C'=1}^k  e^{i 2\pi q  \bm a_3 \cdot \bm\nu_{C'} }\right] + (N-k)\; e^{i2\pi q k \left(z_3-\frac{x_4}{N \ell L_4}\right)}\,\gamma_\ell^q \, \delta_{{q \over \ell}, \Z}\,,\\
W_4^q&=& e^{-i2\pi q(N-k) \left(z_4+\frac{x_3}{N \ell L_3}\right)}\left[\sum\limits_{C'=1}^k  e^{i 2\pi q  \bm a_4 \cdot \bm\nu_{C'} }\right]+(N-k)\; e^{i2\pi qk \left(z_4+\frac{x_3}{N \ell L_3}\right)} \gamma_\ell^q \; \; \delta_{{q \over \ell}, \Z}\,,
\end{eqnarray}
and $\ell=N-k$. 
Here, $\bm \rho$ is the $SU(k)$ Weyl vector, obeying\footnote{\label{Rk footnote}To derive this identity, we may use the $\mathbb R^k$ basis $\{\bm e_i\}$, $i=1,...,k$, where $\bm e_i$ is a unit vector in the $i$-th direction. The Weyl vector is given by  $\bm \rho=\sum_{a=1}^{k-1}\bm w_a$, where the weights $\bm w_a$ are expressed in terms of the basis vectors as $\bm w_a=\sum_{i=1}^a \bm e_i-\frac{a}{k}\sum_{i=1}^{k}\bm e_i$. We also use the expression of $\bm \nu_{C'}$, $C'=1,2,..,k$, in terms of the basis vectors: $\bm \nu_{C'}=\bm e_{C'}-\frac{1}{k}\sum_{i=1}^k\bm e_i$.} $\bm \rho\cdot \bm \nu_{C'}=-C'+\frac{k+1}{2}$. Also, we used $\delta_{{q \over \ell}, \Z}$ to denote unity if $q$ is divisible by $\ell \;(= N-k)$ and $0$ otherwise. In the special case $k=1$, one disregards $\bm a_\mu$ since there are solely $4$ translational moduli $z_\mu$.

We also examine the local gauge-invariant densities. 
Recall that we introduced the variable $\hat \phi^{C'}_\mu$, in terms of which we wrote the local gauge invariants. This variable is related to $z_\mu$ and $\bm a_\mu$ as follows:
\begin{equation}\label{phiofa}
\hat \phi_\mu^{C'} = - 2 \pi N {z_\mu \over L_\mu} + {2 \pi \over L_\mu} \bm{a}_\mu \cdot \bm{\nu}_{C'}.
\end{equation}
Local gauge invariant densities have the form of a sum of $k$ lumps centered at values determined by $\hat\phi^{C'}$:
\begin{equation}\label{density1}
\sum\limits_{C'=1}^k F\left(x_1 - L_1L_2{\phi_2^{C'}\over 2 \pi} - {L_1 C'\over r}, x_2 + L_2 L_1 {\phi_1^{C'} \over 2 \pi}, x_3 - \ell L_3 L_4 {\phi_4^{C'} \over 2 \pi}, x_4 +  \ell L_4 L_3 {\phi_3^{C'} \over 2 \pi}\right)~.
\end{equation}
We now use (\ref{phiofa}) and rewrite (\ref{density1}) in terms of $\bm a_\mu$, where we use again the identity  
${C'\over k} = - {\bm \rho \over k} \cdot \bm \nu_{C'} +{1 \over 2} + {1 \over 2 k}$ to replace the $C'\over k$ factor:\begin{eqnarray}\label{density2}
&& \sum\limits_{C'=1}^k F\left(x_1 + L_1 N z_2 - L_1 \left(\bm{a}_2-{\bm \rho \over k}\right) \cdot \bm{\nu}_{C'} -{L_1 \over 2} -{L_1 \over 2 k},\right.   \\
&& \left.  ~~~~~~ x_2 - L_2 N z_1+ L_2 \bm{a}_1 \cdot \bm{\nu}_{C'}, x_3 + L_3 \ell N z_4  -L_3 \ell \bm{a}_4 \cdot \bm{\nu}_{C'}, x_4 - L_4 \ell N z_3+ L_4 \ell \bm{a}_3 \cdot \bm{\nu}_{C'}\right)~.\nonumber
 \end{eqnarray}

\subsection{Symmetries of Wilson  lines and gauge-invariant densities}

\subsubsection*{Root lattice translations}

We first note that the Wilson lines are invariant under root lattice translations of $\bm a_\mu$:
\begin{eqnarray}\label{rootshift}
\bm a_\mu \rightarrow \bm a_\mu+\bm\alpha_{ij}\,.
\end{eqnarray}
 Here,  $\bm \alpha_{ij}$, where  $i,j=1,...k$ and $i\neq j$, are the $SU(k)$ roots. This easily follows from (i) any weight of the fundamental representation $\bm \nu_{C'}$ can be written as a linear superposition of the fundamental weights $\bm w_{a}$, $a=1,...,k-1$, with integer coefficients, (ii) any root $\bm \alpha_{ij}$ is written as a linear superposition of the simple roots $\bm \alpha_a$, $a=1,..,k-1$, and (iii) the fundamental weights and simple roots satisfy the identity $\bm \alpha_a\cdot \bm w_b=\delta_{ab}$. Moreover, the gauge-invariant densities, e.g. (\ref{fractional1}),  are invariant under the shifts (\ref{rootshift}). This can be easily verified using the properties (i) to (iii), substituting $\hat \phi_\mu^{C'} = - 2 \pi N {z_\mu \over L_\mu} + {2 \pi \over L_\mu} \bm{a}_\mu \cdot \bm{\nu}_{C'}$, and keeping in mind that $F(x_1,x_2,x_3,x_4)$ in (\ref{fractional1}) is a periodic function of its arguments. Then, it is natural to conclude that $a_\mu$ lives in the root space:
 \begin{eqnarray}\label{root app}
 \bm{a}_\mu \in \Gamma_r^{SU(k)}\,.
 \end{eqnarray}
where $\Gamma_r^{SU(k)}$ denotes the fundamental cell of the root lattice of $SU(k)$, which can be mapped to the torus $(\S^1)^{k-1}$.

It is important to note that the transformations (\ref{rootshift})  
are $\Omega$-periodic gauge transformations. Recall that these are gauge transformations $g(x)$ preserving the boundary conditions (fixed and determined by the chosen gauge for the transition functions $\Omega_\mu(x)$) obeyed by the fields being integrated over in the path integral. In Hamiltonian quantization, these are the transformations leaving the physical states invariant. Explicitly, $\Omega$-periodic $g(x)$  obey
\begin{equation}\label{omegagauge}
g^{-1} (x+L_\mu)\; \Omega_\mu (x) \; g(x) = \Omega_\mu(x)~, ~ \forall  \mu,
\end{equation}
with $\Omega_\mu (x)$ from (\ref{the set of transition functions for Q equal r over N, general solution}). With some abuse of notation, we use $x+L_\mu$ to denote a shift of only the $\mu$-th component of the four-vector $x$ by the corresponding $\T^4$ period $L_\mu$.

The root-lattice translations (\ref{rootshift})  can  be seen to be due to an $\Omega$-periodic gauge transformation:
\begin{eqnarray}
\label{translation g}
g_{ij,\mu}(x)= e^{- i \frac{2\pi x_\mu} { L_\mu} \bm{\alpha}_{ij} \cdot \bm{H_k}}\,.
\end{eqnarray}

\subsubsection*{Combined $SU(k)$ weight-lattice and $z_\mu$ shifts}
\label{appx:weight}
 
 Furthermore, we observe that $SU(k)$ weight-lattice shifts of $\bm{a}_\mu$ in (\ref{Wilson 1 main}) are compensated by shifts of $z_\mu$,  thus leaving $W_\mu^q$ invariant. Explicitly, 
\begin{eqnarray}\label{weylshift}
\bm{a}_\mu \rightarrow \bm{a}_\mu + \bm{w}_a,&& ~ z_\mu \rightarrow z_\mu - {{\cal{C}}_a \over k}\,, ~a = 1, 2,..., k-1,  \nonumber \\
 \text{where} \;&& N {\cal{C}}_a = a \; (\text{mod} \; k), ~ {\cal{C}}_a \in \mathbb{Z}_+ .
\end{eqnarray}
The non-negative integer ${\cal{C}}_a$ exists because of the gcd($N,k)=1$ condition, which we assume throughout the paper.  

To see that (\ref{weylshift}) leaves gauge invariants evaluated in the background of the solution invariant, notice that under the shift (\ref{weylshift}) of $\bm{a}_\mu$ by $\bm{w}_a$, $a \in {1,...,k-1}$, we have that $\bm{a}_\mu\cdot \bm{\nu}^{C'}$ changes by $- {a \over k} + \theta_{a,C'}$, where $\theta_{a, C'} = 1$ if $a \ge C'$ and zero otherwise. For $W_\mu^q$, this is compensated by the shift of $z_\mu$ indicated above.  
The shift (\ref{weylshift}) also preserves the invariance of the local density (\ref{density2}). In this case, the integer-valued shift $\theta_{a, C'}$, along with a similar contribution from the shift of $z_\mu$, is absorbed by the periodic nature of the local gauge invariants $F$. Unlike the winding Wilson lines, these invariants are periodic functions of the coordinates, with periods matching those of the torus.

Finally, it is important to note that the transformations   (\ref{weylshift})  are also $\Omega$-periodic gauge transformations.   Consider the following $x$-dependent gauge transformation:
 \begin{equation}\label{gamu}
 g_{a,\mu}(x) = e^{ - i 2 \pi {x_\mu \over L_\mu}( \bm{w}_a \cdot \bm{H}_k +{{\cal{C}}_a \over 2 \pi k}  \omega )}~,
\end{equation}
with $\omega$ from (\ref{omega}).
First, we observe that $g_{a,\mu}(x)$ commutes with all transition functions (\ref{the set of transition functions for Q equal r over N, general solution}), since $Q_k$ is diagonal while $g_{a,\mu}(x)$ is proportional to unity in the lower $\ell\times \ell$ corner. Thus, the $\Omega$-periodicity condition (\ref{omegagauge}) reduces to the condition that $g_{a,\mu}$ be periodic, or that $g^{-1}_{a,\mu}(x+L_\mu) g_{a,\mu}(x)$ be the unit matrix. From (\ref{gamu}), the remarks after (\ref{weylshift}), and the explicit form (\ref{omega}) of $\omega$,  we find 
\begin{eqnarray}
g^{-1}_{a,\mu}(x+L_\mu) g_{a,\mu}(x) = \text{diag}(e^{- i 2 \pi ({a \over k} - {N {\cal{C}}_a \over k})} I_k, I_\ell)
\end{eqnarray}
which is, indeed, the unit matrix in view of the gcd$(N,k)=1$ condition, $N {\cal{C}}_a = a \; (\text{mod} \; k)$, as per (\ref{weylshift}). On the other hand, the $x$-dependence of the $\Omega$-periodic gauge transform (\ref{gamu}) means that the instanton background  $A_\mu(x)$ shifted by 
\begin{eqnarray}
- i g_{a,\mu}(x) \partial_\mu g^{-1}_{a,\mu}(x) = { 2 \pi \over L_\mu} ( \bm{w}_a \cdot \bm{H}_k +{{\cal{C}}_a  \over 2 \pi k} \omega ).
\end{eqnarray}
Thus, recalling the definition of the moduli (\ref{r over N abelian sol}), we find that the gauge transformation (\ref{gamu}) precisely affects the shift (\ref{weylshift}) of the moduli $z_\mu$ and $\bm{a}_\mu$.

\subsubsection*{Weyl reflections}
\label{appx:weyl}
 
The Wilson lines (\ref{Wilson 1}) and the local gauge-invariant densities (\ref{density2}) are left invariant under Weyl transformations acting simultaneously on all $\bm{a}_\mu$. As discussed in the main text, they have the interpretation of permuting the $k$ identical lumps which comprise the multi-fractional instanton.

The Weyl group is the group of reflections about hyperplanes orthogonal to all roots performed simultaneously on all four $\bm{a}_\mu$:\footnote{Later, we also argue that (\ref{weylmoduli}) are due to a particular set of $\Omega$-periodic gauge transformations.}
\begin{eqnarray}\nonumber
\label{weylmoduli}
\bm a_{\mu}&\rightarrow& \mu_{\bm\alpha_{ij}}(\bm a_\mu)\equiv\bm a_\mu-(\bm a_\mu\cdot\bm\alpha_{ij})\bm\alpha_{ij}\,,\quad \mu=1,3,4\,,\\
\bm a_{2}&\rightarrow&\bm a_2-\left[(\bm a_2-\frac{\bm \rho}{k})\cdot \bm \alpha_{ij}\right]\bm\alpha_{ij}\,, 
\end{eqnarray}
where $i,j=1,2,..,k$.
The derivation, notation, and the meaning of these reflections will be discussed momentarily. 
The group generated by the reflections (\ref{weylmoduli}) is the Weyl group, which is isometric to the permutation group $S_k$ of order $k!$. 

To study the Weyl reflections, it is more convenient to use the $\mathbb R^k$ basis. To this end, let $\bm a_\mu=(a_\mu^{1},..., a_{\mu}^{k})$ be a vector that lives in $\mathbb R^k$, i.e., it has $k$ components,  such that it satisfies the constraint $\sum_{i=1}^k a_\mu^{k}=0$, which eliminates the unphysical component in $(a_\mu^{1},..., a_{\mu}^{k})$. Then, the weights of the defining representation are given by $\bm\nu_{C'}=\bm e_{C'}-\frac{1}{k}\sum_{i=1}^k \bm e_{i}$, where $\bm{e}_{i}$ are orthonormal unit vectors in $\mathbb R^k$.\footnote{Notice that the indices $i,j \in \{1,...,k\}$ label the different $\mathbb R^k$ vectors, e.g. $\bm{e}_i$ instead of $\bm{e}_{C'}$. We hope using either $C', D'$ or $i,j$ in this Appendix is not too confusing.} Using this construction,  it is easy to prove the identity \begin{eqnarray}\label{imp identity}\bm a_\mu\cdot \bm \nu_{C'}=a_{\mu}^{C'}\,,\quad C'=1,2,..,k\,.\end{eqnarray}
 
  A general positive or negative root of $SU(k)$ is given by $\bm\alpha_{ij}=\bm e_i-\bm e_j$, $i\neq j$, with $i,j \in \{1,...,k\}$, keeping in mind that $SU(k)$ possesses $k^2-k$ roots. The Weyl reflection operation acting on $\bm a_\mu$ about the root $\bm \alpha_{ij}$ is given by
\begin{eqnarray}\label{Weyl reflection def}
\mu_{\bm \alpha_{ij}}(\bm a_\mu)\equiv \bm a_\mu-\left(\bm \alpha_{ij}\cdot \bm a_\mu\right)\bm \alpha_{ij}\,,
\end{eqnarray}
and the product $\mu_{\bm \alpha_{ij}}(\bm a_\mu)\cdot \bm \nu_{C'}$ is
\begin{eqnarray}
\mu_{\bm \alpha_{ij}}(\bm a_\mu)\cdot \bm \nu_{C'}=\bm a_\mu\cdot\bm \nu_{C'}-\left(\bm \alpha_{ij}\cdot \bm a_\mu\right) \left(\bm \alpha_{ij}\cdot \bm \nu_{C'}\right)=a_\mu^{{C'}}-\left(a_\mu^{i}-a_\mu^{j}\right)\left(\delta_{i{C'}}-\delta_{j{C'}}\right)\,.
\end{eqnarray}
We write this product explicitly as
\begin{eqnarray}\label{modulipermute1}
\mu_{\bm \alpha_{ij}}(\bm a_\mu)\cdot \bm \nu_{C'}=\left\{\begin{array}{lr}a_\mu^{{C'}} &\quad {C'}\neq j\, \mbox{and}\, {C'}\neq j\\ a_\mu^{i}-(a_\mu^{i}-a_\mu^{j})=a_\mu^{j}&\quad {C'}=i\,\mbox{and}\, {C'}\neq j \\ a_\mu^{j}+(a_\mu^{i}-a_\mu^{j})=a_\mu^{i}&\quad {C'}=j\,\mbox{and}\, {C'}\neq i\,. \end{array}  \right.
\end{eqnarray}
Thus, a reflection of $\bm a$ about $\bm\alpha_{ij}$ swaps the components $a_\mu^{i}$ and $a_\mu^{j}$, leaving all other components unchanged. Recalling (\ref{imp identity}), we observe that the only effect of a Weyl reflection is to interchange the terms $e^{i2\pi q a_\mu^i}$ and $e^{i 2\pi q a_\mu^j}$ within the sum $\sum_{C'=1}^k e^{i2\pi  q \bm a_\mu\cdot \bm \nu_{C'}}$ appearing in $W_{1,2,3}^q$. This means that a Weyl reflection preserves the value of this sum. Consequently, the Wilson lines $W_{1,3,4}^q$ remain invariant under the Weyl reflection defined in (\ref{Weyl reflection def}).

The invariance of $W_2^q$ under reflections is more involved, thanks to the phase $e^{-i2\pi q\bm \rho\cdot \bm \nu_{C'}}$. Recall (\ref{imp identity}) and the identity $\bm \rho\cdot \bm \nu_{C'}=-C'+\frac{k+1}{2}$. Then, to show the invariance of $W_{2}^q$ under reflections about $\bm\alpha_{ij}$, one needs to swap the phases $e^{i 2\pi qi/k}$ and $e^{i 2\pi qj/k}$ that accompany the terms $e^{i2\pi q a_2^i}$ and $e^{i 2\pi q a_2^j}$, respectively, within the sum that appears in $W_2^q$. This is easily achieved by defining a new vector $\bm a_2'$:
\begin{eqnarray}
\bm a_2'\equiv \bm a_2-\frac{\bm \rho}{k}\,,
\end{eqnarray}
such that the reflection operation about $\bm \alpha_{ij}$ should involve the newly defined $\bm a_2'$:
\begin{eqnarray}\label{ref a2}
\mu_{\bm \alpha_{ij}}\left( \bm a_2'\right)=\bm a_2'-\left(\bm \alpha_{ij}\cdot \bm a_2'\right)\bm \alpha_{ij}\,.
\end{eqnarray}
We already found above how $\mu_{\bm \alpha_{ij}}$ acts on $\bm a_2$. What remains is to find  $\mu_{\bm \alpha_{ij}}(\bm \rho)$:
\begin{eqnarray}
\mu_{\bm \alpha_{ij}}(\bm \rho)=\bm \rho-\left(\bm \alpha_{ij}\cdot \bm \rho\right)\bm \alpha_{ij}\,.
\end{eqnarray}
Using $\bm \rho=\sum_{a=1}^{k-1}\bm w_a$ and  $\bm w_a=\sum_{i=1}^a \bm e_i-\frac{a}{k}\sum_{i=1}^{k}\bm e_i$ (see Footnote \ref{Rk footnote}), we find
\begin{eqnarray}
\nonumber
\bm \rho\cdot \bm \alpha_{ij}&=& \bm \rho\cdot\bm  e_i-\bm \rho\cdot \bm e_j=j-i\,,\\
\mu_{\bm \alpha_{ij}}(\bm \rho)&=&\bm \rho-\left(\bm \alpha_{ij}\cdot \bm \rho\right)\bm \alpha_{ij}=\bm \rho-(j-i)\bm \alpha_{ij}\,.\label{rho1}
\end{eqnarray}
Thus, we conclude
\begin{eqnarray}
\mu_{\bm \alpha_{ij}}(\bm \rho)\cdot \bm \nu_{C'}=\bm \rho\cdot \bm \nu_{C'}-(j-i)\left(\delta_{iC'}-\delta_{jC'}\right)\,.
\end{eqnarray}
Using $\bm \rho\cdot \bm \nu_{C'}=-{C'}+\frac{k+1}{2}$, we find
\begin{eqnarray}
\nonumber
&&\mu_{\bm \alpha_{ij}}(\bm \rho)\cdot \bm \nu_{C'}\\
\nonumber
&&=\left\{\begin{array}{lr} \bm \rho\cdot \bm \nu_{C'} & i\neq {C'}\, \mbox{and}\, j\neq {C'} \\
\bm \rho\cdot \bm \nu_i-(j-i)=-i+\frac{k+1}{2}-j+i=-j+\frac{k+1}{2}=\bm \rho\cdot \bm \nu_j & i={C'}\, \mbox{and}\,  j\neq {C'} \\ \bm \rho\cdot \bm \nu_i& j={C'} \, \mbox{and}\, i\neq {C'}\,.  \end{array} \right.\\
\end{eqnarray}
This shows that the terms $e^{i 2\pi q \bm \rho\cdot \bm \nu_i}$ and $e^{i 2\pi q \bm \rho\cdot \bm \nu_j}$ are swapped under a reflection about $\bm \alpha_{ij}$. Therefore, the reflection defined in (\ref{ref a2}) leaves $W_2^q$ invariant.

As for the local density (\ref{density2}), the Weyl reflection (\ref{weylmoduli}) with a given $i,j$ interchanges the $C'=i$ and $C'=j$ terms in (\ref{density2}), i.e. permutes two of the $k$ lumps, provided it is performed on all four moduli $\bm{a}_\mu$. That this is so follows immediately from the action of the shifts (\ref{weylmoduli}) on the dot products (described in the various equations above) that enter in (\ref{density2}). This provides a pictorial representation of the action of this symmetry of the moduli space (and in addition to (\ref{gworks21}) below provides a physical argument why the Weyl reflection should be performed simultaneously in all four directions).

The Weyl transformation (\ref{weylmoduli}) can be seen to be due to an $\Omega$-periodic gauge transformation. Without much ado, we simply state its form 
\begin{eqnarray}
\label{gworks21}
g_{ij}(x_2)= e^{- i 2\pi {x_2 \over L_2}{ (j-i) \over k} \bm{\alpha}_{ij} \cdot \bm{H_k}} \; P_{ij}~,
\end{eqnarray}
where $P_{ij}$ is a constant permutation matrix permuting the $i$-th and $j$-th eigenvalues  of a $k \times k$  diagonal  matrix (when acting as $P_{ij} ... P_{ij}^{-1}$; it is also embedded trivially in $SU(N)$). Below, we explain why $g_{ij}$ of (\ref{gworks21}) performs the transforms (\ref{weylmoduli}), without giving all steps, which can be easily reproduced by the reader.  

First, we immediately see that (\ref{gworks21}) obeys (\ref{omegagauge}) for  $\mu = 1,3,4$, since $\Omega_{1,3,4}$ are proportional to the unit matrix of $SU(k)$ and $g_{ij}$ is independent of $x_{1,3,4}$.
Furthermore, when it acts on the $A_{1,3,4}$ components of the solution, its effect is that of the permutation $P_{ij}$, which precisely permutes the moduli $a_\mu^{i}$ and $a_\mu^j$, in the $\R^k$ basis notation, as explained after eqn.~(\ref{modulipermute1}). 

We next consider the $\Omega$-periodicity of $g_{ij}$ for $\mu=2$ and its action on the $A_2$ component of the solution. That $g_{ij}$ is $\Omega$-periodic also for $\mu=2$, i.e. obeys all of (\ref{omegagauge})  follows from the identity 
 $g_{ij}^{-1} (x_2 = L_2) Q_k g_{ij} (x_2=0) = Q_k$, which can be checked from the explicit form of $g_{ij}$ and $Q_k$, see (\ref{clockshift}) (using the $\R^k$ basis for the roots, as used around eqn.~(\ref{Weyl reflection def}) is also helpful). That $g_{ij}(x_2)$ also affects the constant shift of $\bm{a}_2$ from (\ref{weylmoduli}) follows from its $x_2$ dependence as well as the expressions for $\bm{\rho}$ and the roots in the $\R^k$ basis.

%%%%%%%%%%%%%%%%%%%%%%%%%%%%
\section{Determining the shape and volume of the moduli space}
\label{Determining the shape of the moduli space}
%%%%%%%%%%%%%%%%%%%%%%%%%%%%

The integrals of the Wilson loops should vanish for any value of $x_\mu$. Thus, in each case, we require that the integrals over the moduli of each of the two terms appearing in each $W_\mu^q$ in (\ref{Wilson 1})  vanish. This can only be accomplished by restricting the range of $z_\mu$. 

{\flushleft{\bf The range of the $z_\mu$ moduli:}}

For $W_1^q$ and $W_2^q$, assuming that $\mbox{gcd}(k, N-k)=1$,  the integrals of 
$e^{-i2\pi q (N-k) \left(z_1-\frac{x_2}{N L_2}\right)}$ and $e^{i2\pi q k \left(z_1-\frac{x_2}{N L_2}\right)}$ over $z_1$, as well as of $e^{-i2\pi q (N-k) \left(z_2+\frac{x_1}{N L_2}\right)}$ and $e^{i2\pi q k \left(z_2+\frac{x_1}{N L_2}\right)}$ over $z_2$ vanish, for any integer $q$ and for all values of $x_\mu$, provided the limits of integration are taken $z_{1,2} \in [0,1)$.\footnote{If $\mbox{gcd}(k, N-k)>1$, shorter periods, $1/{\rm gcd}(k,N-k)$, might be expected. However, this case presents additional complexities and is left for future investigation.}

For the case of $W_3^q$, we require that, for integer $q \neq \ell$, the integral of  $e^{-i2\pi q (N-k) z_3}$ over $z_3$ vanishes (we do not show the $x_4$ dependence, as the vanishing must hold for all $x_4$). This leads to the condition $z_3 \in [0, \frac{1}{N-k})$. The second term in $W_3^q$ is only nonzero when $q = \ell = N-k$. To ensure that the integral of $e^{i2\pi (N-k) k \left(z_3 - \frac{x_4}{N (N-k) L_4}\right)}$ over $z_3$ vanishes, we again find that $z_3 \in [0, \frac{1}{N-k})$ is the appropriate minimal range.

Similarly, for $W_4^q$, the condition for the first term, with $q \neq \ell \mathbb{Z}$, requires $z_4 \in [0, \frac{1}{N-k})$. For $q = \ell\mathbb Z$, as in the case of $W_3^\ell$, we demand that the integral of $e^{i2\pi (N-k) k  z_4}$ over $z_4$ vanishes, establishing the same minimal range, $z_4 \in [0, \frac{1}{N-k})$.

In conclusion, the ranges of $z_\mu$:
\begin{eqnarray}\label{range of z 1 app}
z_\mu &\in& [0,1], \; \text{for} \;  \mu=1,2~,~ \\ \nonumber
z_\mu &\in& [0, {1 \over N-k}],  \; \text{for} \;  \mu = 3,4~, ~ \text{or} \; z_\mu \in \S^1_{\mu}~.
\end{eqnarray}

{\flushleft{\bf The range of the ${\bm a}_\mu$  moduli:}} 

 We first parameterize $\bm{a}_\mu \in \Gamma_r^{SU(k)}$ by defining (skipping the index $\mu$ for brevity):
\begin{equation}\label{alpharoot}
\bm{a} = \sum\limits_{a=1}^{k-1} t_a \bm{\alpha}_a, ~\text{where} ~ t_a \in [0,1], \; \forall a=1,..., k-1.
\end{equation}
Thus the unit cell of the root lattice is seen to be equivalent to $(\S^1)^{k-1}$ parameterized by the $t_a$'s, all defined modulo $1$. Further, the action of the weight lattice shifts are found to be
\begin{equation}\label{weightshift1}
\bm{w}_b: \; t_a \rightarrow t_a - {a b\over k}~,~ a, b = 1,..., k-1, 
\end{equation}
where we used the (mod $1$) property of $t_a$. The action of the weight lattice shift on $\Gamma_r^{SU(k)}$ is immediately seen to be a $\Z_k$ transformation. Thus it is enough to consider the action (\ref{weightshift1}) with $b=1$, which generates all other transformations.

The weight-lattice shifts (\ref{weylshift}) also act on the variable $z$. For $\mu = 1,2$, as we discussed, the  range of $z_{1,2}$ is $[0,1]$, while for $\mu = 3,4$ we had the range $[0, {1 \over N-k}]$. To discuss all $z_\mu$ uniformly (and hence omit the label $\mu$), we define $, \hat z_{1,2} \equiv z_{1,2}, \; \hat z_{3,4} \equiv (N-k) z_{3,4}$. The range of all $\hat z_\mu$ is now $[0,1]$. Then consider the shifts (\ref{weylshift}) with $b=1$ for $z_{\mu}$. Recalling that $C_1 N=1 (\text{mod}\, k)$ we find  $\hat z_{1,2} \rightarrow \hat z_{1,2} - {C_1 \over k}$, and $\hat z_{3,4} \rightarrow \hat z_{3,4} - {C_1 (N-k) \over k} = \hat z_{3,4} - {1 \over k}$, where we recall modulo $1$ property of all $\hat z_\mu$. The action of the weight lattice shifts on $\hat z_\mu$ is also a $\Z_k$ transformation. 
Thus, for now we have the space $(\S^1)^k$ with coordinates $(t_1,...,t_{k-1}, \hat z)$ all  modulo $1$. This space is subject to the identifications (one should take $C_1=1$ for $\hat{z}_{3,4}$, although the precise value is irrelevant):
\begin{eqnarray}
\label{zkidentification}
\Z_k: ~ t_a &\rightarrow& t_a - {a \over k},~ a = 1,..., k-1, \nonumber \\
z &\rightarrow& z - {C_1 \over k}, ~ N C_1 = 1 (\text{mod} k), ~ \text{gcd}(N,k)=1~.
\end{eqnarray}
The transformation is a freely-acting $\Z_k$ transformation (one which has no fixed points, as there is no solution to $t_a = t_a - {a \over k}$ modulo $1$, for $a = 1,..., k-1$). 

Let us now characterize the fundamental domain of (\ref{zkidentification}). The easiest way to do this is to split each of the $[0,1]$ circles of $t_a$ or $\hat z$ into $k$ intervals, each of length $1/k$. We label these intervals by modulo $k$ integers $p_a$, $a=1,...,k-1$, labeling the $k$ intervals for each $t_a$, and $p_k$, labeling the $k$ intervals for $z$.  Each $p_a$ and $p_k$ can take the integer values $1,...,k$, defined modulo $k$. Thus, we have $k^k$ integers label the $k^k$ cubes into which we split  $(\S^1)^k$, a $k$-dimensional cube (with opposite sides identified).

 The utility of this parameterization is that it allows us to figure out how to parameterize the fundamental domain of $\Z_k$. This is because the cubes get identified under the transformation (\ref{zkidentification}) as follows:
\begin{eqnarray}\label{zk 2}
\Z_k: ~(p_1, p_2,...,p_{k-1}, p_k) \equiv (p_1 -1, p_2 - 2, ... , p_{k-1}-k+1, p_k - C_1) ~(\text{mod}\, k)
\end{eqnarray}
There are $k^k$ cubes labeled by the sets of integers (each defined modulo $k$) $(p_1, p_2,...,p_{k-1}, p_k)$. The identification among the $k^k$ cubes given by eqn.~(\ref{zk 2}) has orbits of size $k$. Thus, there are ${k^k \over k} = k^{k-1}$ independent orbits. The fundamental domain of the $\Z_k$ action is formed by taking one representative (it does not matter which) of each orbit.\footnote{\label{footnote:ex}A simple example is $k=2$, with say $N=3$ where $C_1=1$ and we have the $2^2$ cubes $(1,1)$, $(1,2)$, $(2,1)$, $(2,2)$. The identification (\ref{zk 2}) gives the two orbits $(1,1) \equiv (2,2)$ and $(1,2) \equiv (2,1)$. Thus, there are four choices of a fundamental domain. Notice that it can be taken $(1,1)$ and $(1,2)$, which corresponds to taking the weight lattice range for $t_1$ and the entire range of $\hat z$. }

{\flushleft{\bf Volume of fundamental domain:}} 
Now, the volume of the $(t_1,...,t_{k-1}, \hat z)$-space equals the sum of the volumes of the $k^k$ cubes labeled by the $k^k$ different choices of $(p_1, p_2,...,p_{k-1}, p_k)$. After the identification of these cubes into $k$-dimensional $\Z_k$ orbits, we are left with a fundamental domain consisting of $k^{k-1}$ cubes (since, to describe the fundamental domain, we take one cube from each orbit). The volume of the fundamental domain of the combined weight-lattice/$\hat z$-shifts (\ref{zkidentification}) equals the sum of the volumes of these $k^{k-1}$ cubes and is, therefore, $k$ times smaller than the volume of the $(t_1,...t_{k-1}, \hat z)$-space prior to the identification

{\flushleft{\bf Choice of fundamental domain:}} The fundamental domain of (\ref{zk 2}) can always be chosen to be the weight lattice of $SU(k)$ times the entire range of the $\hat z$ variable, generalizing the example of Footnote \ref{footnote:ex}. 

To see this, note that the weight lattice fundamental domain is given by applying the identification (\ref{zk 2}) on all cubes labeled by $(p_1,...,p_{k-1})$, with $p_k$ (the one representing $\hat z$) omitted. The set of $k^{k-1}$ cubes labeled by the possible choices of $(p_1, p_2,...,p_{k-1})$ gives the root lattice fundamental domain, which is then identified under the weight lattice shifts:
\begin{eqnarray}\label{zk 3}
\Z_k: ~(p_1, p_2,...,p_{k-1}) \equiv (p_1 -1, p_2 - 2, ... , p_{k-1}-k+1) ~(\text{mod}\, k)~.
\end{eqnarray}
As the orbits of (\ref{zk 3}) are $k$ dimensional, the weight lattice fundamental domain is given by the choice of $k^{k-2}$ cubes (of the $k^{k-1}$ total), one from each orbit. This gives the well-known result that the volume of the weight lattice fundamental domain is $1/k$ times the volume of the root lattice. For lack of a better way, we now continue by labeling the weight lattice fundamental domain by the cubes that comprise it, i.e., by choice of $(p_1^*, p_2^*, ... p_{k-1}^*)$, where $p_i^*$ are sets of mod $k$ integers {\it not} related by (\ref{zk 3}). As described above, there are $k^{k-2}$ such cubes. 
We now consider a $k^{k-1}$ dimensional set $(p_1^*, p_2^*, ..., p_{k-1}^*, p_k)$, with $p_k$ unrestricted. In words,  to each of the cubes in the weight lattice fundamental domain, we add the $k$ cubes describing the $\hat z$ circle). This set of $k^{k-1}$ cubes is, by construction, not related by (\ref{zk 2})---as the $p_{1...k-1}^*$ are not---hence it forms a fundamental domain of (\ref{zk 2}). We stress that this argument is illustrated simply for $k=2$ in footnote \ref{footnote:ex} and that the reader can construct examples with $k>2$. 
 
{\flushleft{\bf Vanishing of the integrals of Wilson loops, eqn.~(\ref{explicit moduli dependence}) over the fundamental domain: }} To begin, we note that the integrals over the entire range of $z_\mu$ and $\bm{a}_\mu$, (\ref{range of z 1 app}), (\ref{root app}), respectively, vanish trivially. As we showed, this full range splits into $k$ copies of the fundamental domain of the $\Z_k$ action (\ref{weylshift}). 
The integrand (the Wilson loops) is invariant under (\ref{weylshift}); thus, its integral restricted to the fundamental domain also vanishes.

 {\flushleft{\bf Collecting everything}:} We arrive at the following description of the moduli space. It is the product space of the $SU(k)$ root cell $\Gamma_r^{SU(k)}$ and the circle  $\mathbb S_\mu^1$, in each spacetime direction, modded by the action of the discrete symmetry $\mathbb Z_k$:
 \begin{eqnarray}\label{moduli space bulk}
 \Gamma = \prod\limits_{\mu=1}^4 {\S^1_{{\mu}} \times \Gamma_r^{SU(k)} \over \Z_k}  \simeq \prod\limits_{\mu=1}^4 {(\S^1)^k \over \Z_k}\,,
 \end{eqnarray}
additionally modded by the permutation of the $k$ identical lumps.
 As shown above, the volume of the space ${(\S^1)^k \over \Z_k}$ is  $1/k$ times the volume of  $(\S^1)^k$. Further,  dividing by the Weyl group introduces an extra $1/k!$ factor. In addition, the fundamental domain of the moduli space can always be chosen to be the weight lattice of $SU(k)$, i.e., $\Gamma_w^{SU(k)}$, times the entire range of the $z_\mu$ variables given by (\ref{range of z 1 app}).

\subsection{The case $k=N-1$}

In this section, we study the special case of a fractional instanton carrying a topological charge $Q=(N-1)/N$, i.e., taking $k=N-1$ and $\ell=1$. We shall show that the transition functions and gauge fields are fully abelian in this special case. Also, the holonomies $\bm a_\mu=(a_\mu^1,..,a_\mu^{N-2})$ and the four translations $z_\mu$ that appear in (\ref{Wilson 1}) can be grouped into more symmetric moduli $\bm {\tilde \Phi}_\mu\equiv (\Phi_\mu^1,..,\Phi_\mu^{N-1})$ that lives in the Cartan generators of $SU(N)$. Here, we use a tilde over the boldface letter $\bm {\tilde \Phi}_\mu$ to emphasize that this is a $(N-1)$-dimensional vector, in contrast to the $(N-2)$-dimensional vector $\bm a$.  Condition (\ref{explicit moduli dependenceappx}), then, will be used to argue that $\bm {\tilde \Phi}_\mu$ lives in the root lattice of $SU(N)$. This statement will be shown to be equivalent to that $\bm a$ lives in the weight lattice of $SU(N-1)$ provided that the range of the $z_\mu$ variables are given by (\ref{range of z 1 app}). These results work as a check on our above identification of the shape of the moduli space in the general case $1 \leq k \leq N-1$.

Specializing to the case $k=N-1$, the transition functions (\ref{the set of transition functions for Q equal r over N, general solution}) are
 \begin{eqnarray}
\nonumber
\Omega_1&=&  \left[\begin{array}{cc}e^{i2\pi    \frac{x_2}{N  L_2}}I_{N-1}&0\\0& e^{-i 2\pi (N-1)\frac{x_2}{NL_2}}\end{array}\right]\,,\quad
\Omega_2=\left[\begin{array}{cc}Q_{N-1}&0\\0& I_1\end{array}\right],\\
\Omega_3&=& \left[\begin{array}{cc} e^{i2\pi  \frac{x_4}{N L_4}} I_{N-1}&0\\0& e^{-i 2\pi (N-1)\frac{x_4}{N  L_4}}\end{array}\right]\,,\quad
\Omega_4=I_{N-1}\oplus Q_\ell = \left[\begin{array}{cc}I_{N-1}&0\\0& Q_\ell\end{array}\right].
\label{the set of transition functions k is N-1}
\end{eqnarray}
Notice that, here, unlike in (\ref{the set of transition functions for Q equal r over N, general solution}),  we replace the factor $(-1)^{N}$ that accompanies the unit matrix $I_{N-1}$ in $\Omega_1$ with $1$  (without changing cocycle conditions with $n_{12} = 1-N$), greatly simplifying the treatment. Notice that the factor $(-1)^{k-1}$ in (\ref{the set of transition functions for Q equal r over N, general solution}) is traced back to our construction in  \cite{Anber:2023sjn}, where we considered the general case $r\neq k$. There, it was crucial to use $P_k^{-r}$ in building $\Omega_{1}$, where $P_k$ is the shift matrix defined before (\ref{the set of transition functions for Q equal r over N, general solution}). In the special case we are considering in this paper, we set $k=r$. Under this condition, we obtain $P_k^{-r(=-k)} = \gamma_k^{-k} = (-1)^{k-1}$.

To proceed, it proves easier to apply a gauge transformation on the transition functions $\Omega_{\mu}$, $\mu=1,2,3,4$. Let us consider the gauge transformation by the diagonal $SU(N)$ matrix $U(x)$ defined by
\begin{eqnarray}
U(x)\equiv \left[\begin{array}{cc} \underbrace{e^{-i2\pi \left(\frac{C'-1}{N-1}-\frac{N}{2(N-1)}\right) \frac{x_2}{L_2}}\delta_{B'C'}}_{(N-1)\times (N-1)} &0 \\ 0& 1 \end{array}\right]\,.
\end{eqnarray}
It is easy to check that $\mbox{Det}\,U(x)=1$. 
Under a gauge transformation by $U(x)$, the transition functions transform as
\begin{eqnarray}
\Omega_\mu'(x)=U(x_\mu=L_\mu)\Omega_\mu (x) U^{-1}(x_\mu=0)\,,
\end{eqnarray}
and thus, we find
\begin{eqnarray}
\nonumber
\Omega_1'&=&\Omega_1=I_{N}e^{i \frac{2\pi x_2}{NL_2}\scriptsize\mbox{diag}\left(\underbrace{1,1,...,1}_{N-1}, -(N-1)\right)}\,,\quad
\Omega_2'=I_N\,,\\
\Omega_3'&=&\Omega_3=I_{N}e^{i \frac{2\pi x_4}{NL_4}\scriptsize\mbox{diag}\left(\underbrace{1,1,...,1}_{N-1}, -(N-1)\right)}\,,\quad
\Omega_4'=\Omega_4=I_N\,.
\label{the transformed transition}
\end{eqnarray}
These transition functions can be cast in terms of the Cartan generators $\bm {\tilde H}$ of $SU(N)$, where the use of the tilde is to emphasize that these are $(N-1)$-dimensional vectors:
\begin{eqnarray}
\Omega_1'=e^{-i2\pi \bm {\tilde H} \cdot \bm {\tilde \nu}_N \frac{x_2}{L_2}}\,, \quad \Omega_2'=1\,,\quad \Omega_3'=e^{-i2\pi \bm {\tilde H} \cdot \bm {\tilde \nu}_N \frac{x_4}{L_4}}\,,\quad \Omega_4'=1\,,
\label{nice transition functions}
\end{eqnarray}
where $\bm {\tilde H}=\mbox{diag}\left(\bm {\tilde \nu}_1,...,\bm {\tilde \nu}_{N}\right)$. Using the identity $\bm{\tilde \nu}_{a'}\cdot  \bm{\tilde \nu}_{b'}=\delta_{a'b'}-\frac{1}{N}$, where $a',b'=1,2,..,N$, one can immediately see the equivalence between  (\ref{the transformed transition}) and (\ref{nice transition functions}). This shows that the transition functions fully abelianize in the special case $k=N-1$. 

The abelian gauge field that satisfies the boundary conditions (\ref{conditions on gauge field}) using the transition functions (\ref{nice transition functions}) is given by
\begin{eqnarray}
\label{the fully abelian solution}
\nonumber
A_1'&=&\frac{2\pi \bm {\tilde \Phi}_1\cdot \bm {\tilde H}}{L_1}\,,\quad A_2'=-\frac{2\pi x_1}{L_1L_2}\bm {\tilde H}\cdot\bm {\tilde\nu}_N+\frac{2\pi \bm {\tilde \Phi}_2\cdot \bm {\tilde H}}{L_2}\,,\\
A_3'&=&\frac{2\pi \bm {\tilde \Phi}_3\cdot \bm {\tilde H}}{L_3}\,,\quad A_4'=-\frac{2\pi x_3}{L_3L_4}\bm {\tilde H}\cdot\bm {\tilde \nu}_N+\frac{2\pi \bm {\tilde \Phi}_4\cdot \bm {\tilde H}}{L_4}\,,
\end{eqnarray}
and we used the $(N-1)$-dimensional vectors $ \bm {\tilde \Phi}_\mu$ to label the moduli space; here we have $4(N-1)$ independent moduli, as per the index theorem. The corresponding field strength is
\begin{eqnarray}
F_{12}=-\frac{2\pi}{L_1L_2}\bm {\tilde H}\cdot\bm {\tilde\nu}_N\,, \quad F_{34}=-\frac{2\pi}{L_3L_4}\bm {\tilde H}\cdot\bm {\tilde\nu}_N\,,
\end{eqnarray}
keeping in mind that the relation $L_1L_2=L_3L_4$ is satisfied to ${\cal O}(\Delta^0)$.  These expressions of the field strength exactly match those appearing in (\ref{abelian F to leading}) upon setting $k=N-1$ and $\ell=1$ in $\omega$.

As usual, the Wilson lines are given by
\begin{eqnarray}
W_\mu^q(x)=\mbox{tr}\left[e^{i q\oint A_\mu'}\Omega_\mu'\right]\,,
\end{eqnarray}
and thus
\begin{eqnarray}
W_\mu^q(x)=\mbox{tr}\left[e^{\frac{-i2\pi q x_\mu}{L_\mu}\bm {\tilde H}\cdot\bm {\tilde \nu}_{N}+i2\pi q \bm {\tilde \Phi}_\mu\cdot \bm {\tilde H}}\right]=\sum_{\bm {\tilde \nu}_{a'}} e^{\frac{-i2\pi q x_\mu}{L_\mu}\bm {\tilde \nu}_{a'}\cdot\bm {\tilde \nu}_N+i2\pi q \bm {\tilde \Phi}_\mu\cdot \bm {\tilde \nu}_{a'}}\,.
\end{eqnarray}
As before, we demand that $\langle W_\mu^q \rangle=0$, and thus, we need to integrate over the moduli space region $\Gamma$ that yields a vanishing result, i.e., we demand
\begin{eqnarray}
\int_{\Gamma}\left[\prod_{j=1}^{N-1}d\Phi_{\mu}^j\right] \sum_{\bm {\tilde \nu}_{a'}} e^{\frac{-i2\pi q x_\mu}{L_\mu}\bm {\tilde \nu}_{a'}\cdot\bm {\tilde \nu}_N+i2\pi q \bm {\tilde \Phi}_\mu\cdot \bm {\tilde \nu}_{a'}}=0\,,
\label{vanishing integral}
\end{eqnarray}
and the sum is over all the weights $\bm {\tilde\nu}_{a'}$, $a'=1,..,N$. 
In the following, we show that $\Gamma$ coincides with the fundamental cell of $SU(N)$ root lattice. To this end, we use the $\mathbb R^N$ basis of the $SU(N)$ algebra. In this basis, we take $\bm {\tilde \Phi}_\mu=\left(\Phi^1_\mu,\Phi^2_\mu,..., \Phi^N_\mu \right)$ and impose the constraint $\sum_{i=1}^N\Phi_\mu^i=0$, which eliminates the unphysical component in $\left(\Phi^1_\mu,\Phi^2_\mu,..., \Phi^N_\mu \right)$. Also, the weights in this basis are $\nu_j^i=\delta_{ij}-\frac{1}{N}$. Then,  the integral (\ref{vanishing integral}) is written as (suppressing the $x$-dependence and sum to reduce clutter)
\begin{eqnarray}
\int_{\Gamma}\left[\prod_{j=1}^{N} d\Phi_{\mu}^j\right] \delta (\Phi_{\mu}^1+...+\Phi_{\mu}^N) e^{i 2\pi q \Phi_{\mu}^j}e^{-i\frac{2\pi q}{N}(\Phi_{\mu}^1+...+\Phi_{\mu}^N)}\,.
\end{eqnarray}
The Dirac-delta function, inserted to impose the constraint, kills the term $e^{-i\frac{2\pi }{N}(\Phi_{\mu}^1+...+\Phi_{\mu}^N)}$.  Now, it is easy to see the above integral vanishes provided that we integrate the left-over term $e^{i 2\pi \Phi_{\mu}^j}$ over a region $\Gamma$: the parallelotope bounded by the simple roots $\bm {\tilde \alpha}_{i}=\bm {\tilde e}_i-\bm {\tilde e}_{i+1}$, $i=1,..,N-1$,  where $\{\bm {\tilde e}_i\}$ is the set of unit vectors spanning $\mathbb R^N$. We conclude that every term in the sum (\ref{vanishing integral}) vanishes when integrated over the fundamental cell of the $SU(N)$ root space $\Gamma$, and thus, $W^q_\mu(x)$ vanishes when integrated over $\Gamma$ . An alternative approach to reach the same result is to use the change of variables
\begin{eqnarray}
\zeta_\mu^{a'}=\bm {\tilde \Phi}_\mu\cdot \bm {\tilde w}_{a'}\,,
\end{eqnarray}
where $\bm {\tilde w}_{a'}$ are the fundamental weights of $SU(N)$ and $a'=1,2,..,N-1$. 
This transformation effectively rectifies the root lattice by leveraging the identity $\bm{\tilde{\alpha}}_{b'} \cdot \bm{\tilde{w}}_{a'} = \delta_{a'b'}$, where $\bm{\tilde{\alpha}}_{b'} $ is a simple root. As a result, the fundamental domain of the root lattice becomes the hypercubic region defined by $0 \leq \zeta_\mu^{a'} \leq 1$. Consequently, it is easy to see that the integral of $e^{i2\pi q \bm{\tilde{\Phi}}_\mu \cdot \bm{\tilde{w}}_{a'}} = e^{i2\pi q \zeta_\mu^{a'}}$ over the fundamental root lattice is zero.

In our endeavor to determine the volume of the bosonic moduli space, we need to determine the volume of the fundamental cell of the root lattice:
\begin{eqnarray}
\int_{\Gamma}\left[\prod_{j=1}^{N-1} \prod_{\mu=1}^4d\Phi_{\mu}^j\right]\,,
\end{eqnarray}
which is the fourth power (for $4$ spacetime dimensions) of the volume of a $(N-1)$-dimensional  parallelotope spanned by the simple roots $\{\bm \alpha_1,...,\bm\alpha_{N-1}\}$. This volume is given by the fourth power of the determinant of the simple roots:
\begin{eqnarray}\label{V of root space}
\int_{\Gamma}\left[\prod_{j=1}^{N-1} \prod_{\mu=1}^4d\Phi_{\mu}^j\right]=\left(\mbox{Det} \left[\begin{array}{cccc}\alpha_1^1&\alpha_1^2&...&\alpha_{1}^{N-1}\\...&...&...&....\\
\alpha_{N-1}^1&\alpha_{N-1}^2&...&\alpha_{N-1}^{N-1}\end{array}\right]\right)^4=(\sqrt N)^4=N^2\,.
\end{eqnarray}

 The metric of the bosonic moduli space spanned by $\bm{\tilde \Phi}_\mu$ is given by the matrix $U_B$, with components given by 
 \begin{eqnarray}
 U_{B\,ij}^{\mu\mu'}=\frac{2}{g^2}\int_{\mathbb T^4}\mbox{tr}\left[\frac{\partial A_\nu'}{\partial\Phi_\mu^i}\frac{\partial A_\nu'}{\partial\Phi_{\mu'}^j}\right]\,, \quad i,j=1,...,N-1\,,\quad \mu=1,2,3,4\,.
 \end{eqnarray}
 Using $\frac{\partial A_\nu'}{\partial\Phi_\mu^i}=\frac{2\pi}{L_\mu}\delta_{\mu\nu}\delta_{ik}H^k$ along with the identity $\mbox{tr}[H^i H^j]=\delta_{ij}$, for $i,j=1,2,..,N-1$, we obtain
 \begin{eqnarray}
  U_{B\,ij}^{\mu\mu'}=\frac{8\pi^2 V}{g^2 L_\mu^2}\delta_{ij}\delta_{\mu\mu'}\,,
 \end{eqnarray}
 and 
 \begin{eqnarray}
 \sqrt{\mbox{Det}\,  U_B}=\left(\frac{8\pi^2 \sqrt V}{g^2}\right)^{2(N-1)}\,.
 \end{eqnarray}
 Finally, using the collective coordinates method, one finds that the measure of the bosonic moduli space is (see \cite{Vandoren:2008xg} and Appendix B in \cite{Anber:2022qsz} for the details of these calculations)
 \begin{eqnarray}\nonumber
 d\mu_B=  \sqrt{\mbox{Det}\,  U_B}\,\frac{\prod_{j=1}^{N-1} \prod_{\mu=1}^4d\Phi_{\mu}^j}{(N-1)!(\sqrt{2\pi})^{4(N-1)}}=\left(\frac{8\pi^2 \sqrt V}{g^2}\right)^{2(N-1)}\frac{\prod_{j=1}^{N-1} \prod_{\mu=1}^4d\Phi_{\mu}^j}{(N-1)!(\sqrt{2\pi})^{4(N-1)}}\,.\\
 \end{eqnarray}
 The factor $(N-1)!$ that appears in the dominator is introduced to take into consideration the fact that the gauge-invariant observables are invariant under the Weyl group, permuting the $N-1$ moduli $\Phi_\mu^i$ in the $4$ spacetime dimensions simultaneously, which is isomorphic to the permutation group $S_{N-1}$ of order $(N-1)!$. The volume of the moduli space is obtained by performing the integral over $\prod_{j=1}^{N-1} \prod_{\mu=1}^4d\Phi_{\mu}^j$; using (\ref{V of root space}), we readily find 
 \begin{eqnarray}\label{final muB}
 \int_{\Gamma}  d\mu_B=\frac{N^2}{(N-1)!}\left(\frac{4\pi \sqrt V}{g^2}\right)^{2(N-1)}\,.
 \end{eqnarray}

Now, let us return to the moduli-space parameterization using $\bm a_\mu$ and $z_\mu$, reminding the reader that we are still treating the special case $k=N-1$. We shall find the range of $\bm a_\mu$ by demanding that the change of variables from $\bm{\tilde \Phi}_\mu$ to $z_\mu$ and $\bm a_\mu=(a_\mu^1,..,a_\mu^{N-2})$ must leave the integral $\int_{\Gamma}  d\mu_B$ invariant.  

Let ${\cal U}_{B}$ denote the metric on the moduli space spanned by $z_\mu$ and $\bm a$. Using the gauge fields $A_\mu$ given by (\ref{gaugewithholonomies}, \ref{r over N abelian sol}) we obtain the matrix elements of the metric ${\cal U}_{B}$ (summation over $\nu$ is implied):
 \begin{eqnarray}\label{metric on moduli in a and z 3}\nonumber
 {\cal U}_{B\,ab}^{\mu\mu'}&=&\frac{2}{g^2}\int_{\mathbb T^4}\mbox{tr}\left[\frac{\partial A_\nu}{\partial a_\mu^a}\frac{\partial A_\nu}{\partial a_{\mu'}^b}\right]\,, \quad a,b=1,2,..,N-2\,,\\\nonumber
  {\cal U}_{B\,zz}^{\mu\mu'}&=& \frac{2}{g^2}\int_{\mathbb T^4}\mbox{tr}\left[\frac{\partial A_\nu}{\partial z_\mu}\frac{\partial A_\nu}{\partial z_{\mu'}}\right]\,, \\
  {\cal U}_{B\,zb}^{\mu\mu'}&=& \frac{2}{g^2}\int_{\mathbb T^4}\mbox{tr}\left[\frac{\partial A_\nu}{\partial z_\mu}\frac{\partial A_\nu}{\partial a_{\mu'}^b}\right]\,, \quad b=1,..,N-2\,.
 \end{eqnarray} 
Using $\mbox{tr} (H_{N-1}^a H_{N-1}^b)=\delta_{ab}$ (remember that $\bm H_{N-1}=(H_{N-1}^1,..,H_{N-1}^{N-2})$ are embedded in $SU(N)$ by putting zeros in the $1\times 1$ lower-right element), and $\mbox {tr} (\omega^2)=4\pi^2N(N-1)$, along with $\mbox{tr}[H_{N-1}^a\omega]=0$, we find that the metric on the moduli space in each spacetime direction $\mu$ is given by the $(N-1)\times (N-1)$ diagonal matrix
\begin{eqnarray}
{\cal U}^{\mu\mu'}_B=\frac{8\pi^2 V}{g^2 L_\mu^2}\mbox{diag}\left(\underbrace{1,1,..,1}_{N-2}, N(N-1)\right)\delta^{\mu\mu'}\,,
\end{eqnarray}
and thus
\begin{eqnarray}
\sqrt{\mbox{Det}\, {\cal U}_B}=\left(\sqrt{N(N-1)}\right)^4 \left(\frac{8\pi^2 \sqrt V}{g^2}\right)^{2(N-1)}\,.
\end{eqnarray}

In terms of the moduli spanned by $z_\mu$ and $\bm a_\mu$, the differential element on the moduli space is 
\begin{eqnarray}
d\mu_B= \sqrt{\mbox{Det}\,  {\cal U}_B}\,\frac{\prod_{b=1}^{N-2} \prod_{\mu=1}^4 dz_\mu da_{\mu}^b}{(N-1)!(\sqrt{2\pi})^{4(N-1)}}\,,
\end{eqnarray}
such that the integral $\int_{\Gamma} d \mu_B$ must be given by (\ref{final muB}). Recalling that in our case, the range of $z_\mu \in [0,1)$ (see eqn. (\ref{range of z 1 app}) and recall $k=N-1$), and thus $\prod_{\mu=1}^4\int_{0}^1 dz_\mu=1$, 
 we see immediately that the result (\ref{final muB}) is obtained if and only if we demand that  $\bm a_\mu$ lives in the weight lattice of $SU(N-1)$. This confirms our assertion that appears after eqn. (\ref{moduli space bulk}) that the fundamental domain of the moduli space can always be chosen to be the weight lattice of $SU(k)$, i.e., $\Gamma_w^{SU(k)}$ (in this case $k=N-1$), times the entire range of the $z_\mu$ variables given by (\ref{range of z 1 app}).
The volume of the fundamental cell of the weight lattice is given by the volume of a $(N-2)$-dimensional parallelotope spanned by the simple weights $\{\bm w_1,...,\bm w_{N-2}\}$ in each spacetime direction. Thus, we find
 \begin{eqnarray}
 \int_{\Gamma_{w}^{SU(N-1)}} \prod_{b=1}^{N-2} \prod_{\mu=1}^4 da_{\mu}^b=\left(\mbox{Det} \left[\begin{array}{cccc}w_1^1&w_1^2&...&w_{1}^{N-2}\\...&...&...&....\\\nonumber
w_{N-2}^1&w_{N-2}^2&...&w_{N-2}^{N-1}\end{array}\right]\right)^4=\left(\frac{1}{\sqrt{N-1}}\right)^4\,,\\
 \end{eqnarray}
 where we used $\Gamma_{w}^{SU(N-1)}$ to denote the weight lattice of $SU(N-1)$.
 Collecting everything we obtain
\begin{eqnarray}\nonumber
\int_{\Gamma} d\mu_B =\frac{\left(\sqrt{N(N-1)}\times \frac{1}{\sqrt{N-1}}\right)^4}{(N-1)!}\left(\frac{4\pi \sqrt V}{g^2}\right)^{2(N-1)}=\frac{N^2}{(N-1)!}\left(\frac{4\pi \sqrt V}{g^2}\right)^{2(N-1)}\,,\\
\end{eqnarray}
matching the result in (\ref{final muB}).

%%%%%%%%%%%%%%%%%%%%%%%%%%
 \subsection{Volume of the bosonic moduli space}
%%%%%%%%%%%%%%%%%%%%%%%%%%%

Now, we use our experience from the case $k=N-1$ to determine the volume of the moduli space of the general case of $1\leq k\leq N-1$, assuming that $\mbox{gcd}(r,N-k)=1$. As we argued above,  the moduli space $\Gamma$ spanned by $z_\mu$ and $\bm a_\mu=(a_\mu^1,..,a_\mu^{k-1})$ can be taken to be:
\begin{eqnarray}\label{range of z and a}
 \Gamma=\left\{\begin{array}{l} z_{1,2} \in [0,1)\,,\\ 
z_{3,4} \in [0, {1 \over N-k})\,,\\
\bm a_\mu  \in \Gamma_{w}^{{SU(k)}}\; \text{for} \;  \mu=1,2,3,4\,,\end{array}\right.
\end{eqnarray}
where $\Gamma_{w}^{SU(k)}$ is the weight lattice of $SU(k)$, keeping in mind there is an extra identification on $\bm a_\mu$ by the Weyl group.  The volume of the weight lattice of $SU(k)$ is $1/\sqrt{k}$.  The matrix $ {\cal U}_B$ is the metric on the moduli space, with matrix elements given by (summation over $\nu$ is implied)
 \begin{eqnarray}\label{metric on moduli in a and z app 1}\nonumber
 {\cal U}_{B\,ab}^{\mu\mu'}&=&\frac{2}{g^2}\int_{\mathbb T^4}\mbox{tr}\left[\frac{\partial A_\nu}{\partial a_\mu^a}\frac{\partial A_\nu}{\partial a_{\mu'}^b}\right]\,, \quad a,b=1,..,k-1\,,\\\nonumber
  {\cal U}_{B\,zz}^{\mu\mu'}&=& \frac{2}{g^2}\int_{\mathbb T^4}\mbox{tr}\left[\frac{\partial A_\nu}{\partial z_\mu}\frac{\partial A_\nu}{\partial z_{\mu'}}\right]\,, \\
  {\cal U}_{B\,zb}^{\mu\mu'}&=& \frac{2}{g^2}\int_{\mathbb T^4}\mbox{tr}\left[\frac{\partial A_\nu}{\partial z_\mu}\frac{\partial A_\nu}{\partial a_{\mu'}^j}\right]\,, \quad b=1,..,k-1\,.
 \end{eqnarray} 
Using $\mbox{tr} (H_{k}^a H_{k}^b)=\delta_{ab}$ (remember that $\bm H_{k}=(H_{k}^1,..,H_{k}^{k-1})$ are embedded in $SU(N)$ by putting zeros in the $\ell\times \ell$ lower-right matrix), and $\mbox {tr} (\omega^2)=4\pi^2Nk(N-k)$, along with $\mbox{tr}[H_k^b\omega]=0$, we find that the metric on the moduli space in each spacetime direction $\mu$ is given by the $k\times k$ diagonal matrix:
\begin{eqnarray}
{\cal U}^{\mu\mu'}_B= \frac{8\pi^2 V}{g^2 L_\mu^2}\delta^{\mu\mu'}\mbox{diag}\left(\underbrace{1,1,..,1}_{k-1}, k\ell N\right)\,,
\end{eqnarray}
and the square root of the determinant of ${\cal U}_B$  is
\begin{eqnarray}
\sqrt{\mbox{Det}\, {\cal U}_B }=\left(\sqrt{k (N-k) N}\right)^4\left(\frac{8\pi^2 \sqrt V}{g^2}\right)^{2k}\,.
\end{eqnarray}
Collecting everything we find (performing the integral over the moduli space in all $4$ directions):
\begin{eqnarray}\nonumber
\mu_B=\int_{\Gamma}\frac{\prod_{\mu=1}^4 \prod_{b=1}^{k-1} d a_\mu^b dz_\mu \sqrt{\mbox{Det}\, {\cal U}_B}}{k!(\sqrt{2\pi})^{4k}}&=&\frac{1}{k!}\left(\frac{4\pi \sqrt V}{g^2}\right)^{2k}\underbrace{\left(\sqrt{k (N-k) N}\right)^4}_{\mbox{Det}\, {\cal U}^{\mu\mu'} }\\\nonumber&&\times
 \underbrace{\left(\frac{1}{\sqrt k}\right)^4}_{\scriptsize \mbox{volume of}\, SU(k)\, \mbox{weight}\,\mbox{in all 4 directions} }
\times \underbrace{\frac{1}{(N-k)^2}}_{\scriptsize\mbox{volume of}\, z_\mu}\\
&=&N^2 \frac{\left(\frac{4\pi \sqrt V}{g^2}\right)^{2k}}{k!}\,.
\end{eqnarray}
The factor $k!$  takes into consideration the fact that the lumpy solution and Wilson's lines are invariant under the Weyl group, given by (\ref{weylmoduli}), which is isomorphic to the permutation group $S_k$ of order $k!$.
The pre-coefficient is always $N^2$ for all values of $k$, reminding we always assume $\mbox{gcd}(k,N-k)=1$.

\section{A supersymmetric localization for ${\cal{N}}$} 
\label{appx:localization}

A more sophisticated approach to the calculation of (\ref{normalization 1}) (which needs further development, see below) is based on supersymmetric localization (see, e.g., the lecture notes \cite{KyotoSchool2016}). The point is that the super-Yang-Mills action (\ref{symaction2}) can be written as a supersymmetry variation. Explicitly, one can show by direct calculation that\footnote{A Minkowski space version of  (\ref{variation 1}) can be found in \cite{Malcha:2021ess}.}
\begin{eqnarray}\label{variation 1}
g^2 S_{SYM} &=& \delta^\alpha   {\cal{O}}_\alpha  + \delta_{\dot\alpha}{\cal{O}}^{\dot\alpha},~ \text{where} ~ {\cal{O}}_\alpha  = {1 \over 8} \int\limits_{\T^4} d^4x\; \Delta_\alpha , ~ {\cal{O}}^{\dot\alpha}  = {1 \over 8} \int\limits_{\T^4} d^4x \;\Delta^{\dot\alpha} ~. \end{eqnarray}
with $
\Delta_\alpha     \equiv  \sigma_{\mu\nu \; \alpha}^{~~~ ~~\beta}\; \lambda_\beta^a \; F_{\mu\nu}^a +   \lambda_\alpha^a D^a$, $
\Delta^{\dot\alpha}  \equiv \bar\sigma_{\mu\nu \;~~ \dot\beta}^{~~~\dot\alpha} \;\bar\lambda^{\dot\beta \; a} \; F_{\mu\nu}^a  +  \bar\lambda^{\dot\alpha \; a} D^a~$.
The supersymmetry transformations, for convenience, defined without the usual Grassmann parameters, which can be attached if desired, are
\begin{eqnarray}
\label{susydelta}
\delta_\alpha A_\mu^a &=& \sigma^\mu_{\alpha \dot\alpha} \; \bar\lambda^{\dot\alpha \; a}, ~
~~~~~~~~~~~~~~~~~\delta^{\dot\alpha} A_\mu^a = \bar\sigma^{\mu \; \dot\alpha \alpha} \lambda_\alpha^a, \nonumber \\
\delta^\beta \lambda_\alpha^a &=& -  \sigma_{\mu\nu \; \alpha}^{~~~ ~~\beta}\; F_{\mu\nu}^a + \delta^\beta_\alpha D^a, ~ ~\nonumber
\delta^{\dot\beta} \lambda_\alpha^a = 0, ~\\
\delta^\beta \bar\lambda_{\dot\alpha}^a &=& 0, ~  ~~~~~~~~~~~~~~~~~~~~~~~~~
\delta_{\dot\beta} \bar\lambda^{\dot\alpha \; a} = - \bar\sigma_{\mu\nu \;~~ \dot\beta}^{~~~\dot\alpha} \; F_{\mu\nu}^a  + \delta_{\dot\beta}^{\dot\alpha} D^a, \nonumber \nonumber \\
\delta_\alpha D^a &=& -\sigma_{\mu\;\alpha \dot\alpha} \; (D_\mu \bar\lambda^{\dot\alpha})^a, ~ ~~~~~~\delta^{\dot\alpha} D^a  =  - \bar\sigma_\mu^{\dot\alpha \beta} (D_\mu \lambda_\beta)^a~.
\end{eqnarray}
It is easy to check that $\delta^{\alpha} S_{SYM} = \delta_{\dot\alpha} S_{SYM} = 0$.

The fact that the action is a supersymmetry variation (\ref{variation 1}) and the vanishing of its supersymmetry variation imply, formally, that the path integral (\ref{normalization 1}) is coupling-independent:
 \begin{eqnarray}
{d {\cal{N}} \over d  g^{-2}} &=& - \int\limits_{\T^4\;  \text{with} \; n_{12} \ne 0\; (\text{mod} N)} {\cal{D}}A \;{\cal{D}}\lambda \; {\cal{D}}\bar\lambda\;  {\cal{D}}D \;\left[  \delta^\alpha( {\cal O}_\alpha e^{-S_{SYM}} ) + \delta_{\dot\alpha} ( {\cal O}^{\dot\alpha} e^{-S_{SYM}} ) \right]~= 0. \nonumber \\ \label{couplingindependence}
\end{eqnarray}
The vanishing follows from the  fact  that when an integrand, which is a symmetry variation, is integrated using  a  symmetry-invariant measure, one obtains zero (barring nonvanishing boundary contributions). 

The argument leading to coupling independence here is, of course, formal. One needs to gauge fix, regulate, etc.---as was done, e.g. for 4d nonconformal theories with extended supersymmetry in \cite{Pestun:2007rz}---something that we have not considered. Accepting it at face value, however, 
coupling-independence means that one can evaluate ${\cal{N}}$ at any coupling, including the $g^2 \rightarrow 0$ limit. Then, one arrives at the same conclusion as our small-$\T^4$ argument, that the computation of ${\cal{N}}$ reduces to a sum over $N^2$ zero action saddle points, thus leading us to the same expected value (\ref{normalization 2}).

While we stress the formal nature of the localization argument, we present it in the hope that it can stimulate further work. 
Based on our usual understanding of the relation between Hamiltonian formalism and path integral, we expect that the Hilbert space trace (\ref{windex main}) should equal the path integral (\ref{normalization 1}). In fact, this is what was assumed in our previous work \cite{Anber:2022qsz} (which did not take into account the existence of $N^2$ zero action configurations (\ref{zeroenergy2}) but only took the $N$ $A^{(0,q)}$ saddles into account), leading to a factor of $N$ discrepancy between the $\R^4$ and $\T^4$ calculations of the gaugino condensate.

 In order that (\ref{normalization 1}) give the answer (\ref{windex main}), instead of (\ref{normalization 2}), it is necessary to omit the contribution of zero action saddle points which are obtained via center symmetry transforms in the $x_4$ (time) direction of $A=0$, reducing thus the $N^2$ saddles (\ref{zeroenergy2}) to $N$.
 We do not yet see how the Euclidean path integral, which sees no difference between $x_3$ and $x_4$ can accomplish this. Perhaps a proper definition of the Euclidean path integral (\ref{normalization 1}), using complex contours of integration (we note that the Euclidean fields are necessarily complexified, as is already clear from the supersymmetry transformations and the fact that $\lambda$ and $\bar\lambda$ are independent) could be developed to explain this.

%%%%%%%%%%%%%%%%%%%%%%%%%%%%%%%%%%
\section{The path integral measure and the extra saddle points}
\label{appx:measure}
%%%%%%%%%%%%%%%%%%%%%%%%%%%%%%%%%%%%%%%%%%
In this appendix, we discuss the issue around the apparent disagreement we find between the path integral and Hamiltonian determination of the Witten index. 

We begin by pointing out---as suggested to us by an anonymous referee---that an analogous issue arises in topological $\Z_N$ gauge theories. Here, it is resolved by a careful definition of the measure by means of a triangulation or lattice formulation. We shall then argue that this construction holds lessons for the definition of the measure in the Yang-Mills theory of interest.

\subsection{From the Hilbert space to the path integral in the $\Z_N$ topological (lattice) gauge theory}
\label{appx:znlattice}

Without loss of generality, we now consider the simplest example of a topological $\Z_N$ theory: taking two dimensional spacetime and $N=2$. The 2-dimensional $\Z_2$ topological theory, for definiteness defined on $\T^2$, has a continuum Euclidean action:
\begin{equation}\label{z2action1}
S = i {2   \over 2 \pi} \oint_{\T^2} \phi^{(0)} d a^{(1)}, 
\end{equation}
where $\phi^{(0)}$ is a $2 \pi$ periodic scalar and $a^{(1)}$ is a compact $U(1)$ gauge field with periods $\oint d a^{(1)} = 2 \pi \Z$.
Upon canonically quantizing this theory on $\S^1$, one finds that it has two ground states, which can be described by the expectation values of the $a^{(1)}$ Wilson loops winding around $\S^1$: $e^{i \oint_{\S^1} \hat a^{(1)}} |P \rangle = |P \rangle (-1)^P$, $P=1,2$. As the Hamiltonian is zero, the $\T^2$ partition function is $Z_{\T^2} = \tr {\bf{1}} = 2$, with the trace taken over the above two-dimensional Hilbert space. 

An issue similar to our Witten index puzzle arises if we consider the partition function as an Euclidean path integral with action (\ref{z2action1}). Integrating out $\phi^{(0)}$, a Lagrange multiplier imposing flatness on the field $a^{(1)}$, we find that there are now four saddle points, the four $\Z_2$ flat connections on $\T^2$, where $\oint_i a^{(1)} \in \{0, \pi\}$ and the integral $\oint_i$, $i=1,2$, is taken along either the spatial or timelike $\S^1$. One could now argue, analogous to what we did for the case of the Witten index, that each of these zero action saddle points contributes the same amount to the partition function, leading one to expect that the path integral gives $Z_{\T^2}= 4$  instead of the Hamiltonian result $\Z_{\T^2}=2$. We now want to discuss the resolution of this apparent discrepancy upon a careful definition of the measure. 

First we review the measure defined upon a lattice discretization. It produces the correct answer, $\Z_{\T^2}=2$, for the Euclidean $\T^2$ partition function, see Appendix A of ref.~\cite{Choi:2021kmx}.\footnote{Curiously, there is also a continuum calculation of the partition function of this two dimensional $\Z_2$ topological theory \cite{Hellerman:2010fv}, producing the correct answer. The lattice definition we describe here is, however, more straightforward  and generalizes to other dimensions and values of $N$. It also holds lessons for the definition of the measure in Yang-Mills theory.} To describe it, we start with the Euclidean action  (\ref{z2action1}) discretized on a periodic two-dimensional square lattice:
\begin{eqnarray}\label{action2}
S = {2 \pi i \over 2} \sum_{p} b_p \prod_{\ell \in \partial p} a_{\ell} 
\end{eqnarray}
where the sum is over the plaquettes ($p$) of a two dimensional lattice and $\prod_{\ell \in \partial p} a_{\ell}$ is the usual lattice curvature of the link-based gauge field $a_{\ell}$. Both the plaquette-based Lagrange multiplier $b_p$ and the gauge field $a_\ell$ are $\Z_2$-valued, taking the values $1$ or $2$. The   partition function on the  discretized $\T^2$ is then defined as:
\begin{eqnarray}\label{z3}
Z_{\T^2} = {1 \over 2^{2 (\# p)}} \sum_{\{ b_p, a_\ell  =  \{1, 2\}\}} e^{S} 
\end{eqnarray}
where $\# p$ is the total number of plaquettes and the sum is over all possible configurations of the lattice gauge fields. 
To compute (\ref{z3}) one first  sums over $b_p$ $(= 1,2)$ to obtain  a factor of $2^{\# p}$, canceling one of the denominator factors. In addition, the sum gives rise to the constraint that $a_p$ be a flat $\Z_2$ gauge field, as indicated by the $\delta$-function (this is the result given after the first equality in (\ref{z4}) below). There are a total of $2 (\# p)$ $\Z_2$ lattice gauge fields $a_\ell$, but the flatness conditions eliminate  $\#p -1$ of them (the periodicity of the lattice guarantees that the total $\Z_2$ flux is zero, eliminating one constraint). Thus the sum over $a_\ell$ produces a factor of $2^{\# p + 1}$ in the numerator, giving the final answer $Z_{\T^2} = 2$. In summary, we found:
 \begin{eqnarray}\label{z4}
Z_{\T^2} = {1 \over 2^{ \# p}} \sum_{\{  a_\ell  =  \{1, 2\}\}}  \prod_{p} \delta(\prod_{\ell \in \partial p} a_{\ell}) = {2^{ \# p+1} \over 2^{ \# p}} = 2
\end{eqnarray}
The normalization factor of the measure appearing in (\ref{z3}) (as well as of the more general theories discussed in \cite{Choi:2021kmx}) can be understood as arising from the condition of locality and topological nature of the $\Z_2$ theory partition function, which, in particular, require that the partition function be independent on the number of lattice sites (here, $\# p$). 

The theory of interest to us, four-dimensional minimal super-Yang-Mills theory in the background of a 't Hooft flux, however, is not a topological field theory---although the Witten index is invariant under certain deformations, the theory can not be formulated on general manifolds while preserving supersymmetry. Our 
  goal now is to understand how one arrives at the definition of the measure from (\ref{z3}) starting from the Hilbert space description and to see if this allows us to draw lessons for the Yang-Mills case of interest.
  
To this end, we introduce the Hilbert space on a one dimensional periodic spatial lattice of $L$ sites. We label the sites  and the links to the right of each site  by $\ell$, $\ell =1,...L$, with $L+1 \equiv 1$. On each link we have a $\Z_2$ gauge field operator $\hat Z_\ell$, $\hat Z_\ell^2 = 1$ (e.g. represented by the Pauli matrix $\sigma_3$).  The Hilbert space is spanned by the vectors:
\begin{equation} \label{states} 
|s_1, s_2, ..., s_L \rangle, ~ \text{with} \; \hat Z_\ell |s_1, s_2, ..., s_L \rangle = |s_1, s_2, ..., s_L \rangle s_\ell, ~ s_\ell = \pm 1,
\end{equation}
labeled by the eigenvalue of the $\Z_2$ gauge field operator, $s_\ell$. At each site, we have $\langle s| s' \rangle = \delta_{s,s'}$.
The Hilbert space thus defined is $2^L$ dimensional.

The canonical momenta are $\hat X_\ell$, $\hat X_\ell^2 =1$ (with $\hat X_\ell$ represented by e.g. $\sigma_1$). The operator generating a $\Z_2$ gauge transformations on the site $\ell$ is $\hat g_\ell = \hat X_\ell \hat X_{\ell-1}$, with $\hat g_\ell^2 =1$.  Thus,  a general gauge transformation is labeled by a set of $\Z_2$ integers $(n_1, n_2, ..., n_L)$, one at each site, $n_\ell = 1,2$, such that 
\begin{equation}\label{gaugetransforms}
\hat G[n] = \prod\limits_{\ell =1}^L \hat g_\ell^{n_\ell} = \hat X_1^{n_1 + n_2} \hat X_2^{n_2 + n_3}... \hat X_L^{n_L + n_1}~.
\end{equation}
Clearly, the general gauge transformations (\ref{gaugetransforms}) act on the link fields as appropriate, 
$$ \hat G[n]\; \hat Z_\ell \;\hat G[n]^{-1} = (-1)^{n_\ell} \hat Z_\ell (-1)^{n_{\ell+1}}.$$ 
A projector on gauge invariant states can then be defined as 
\begin{eqnarray}\label{projector}
\hat P_{G} \equiv  {1 \over 2^{L}} \sum_{\{ n_\ell = 1,2 \}} \hat G[n]~,
\end{eqnarray}
where the normalization simply accounts for the fact that there are $2^L$ values of $n_\ell$ summed over. The partition function of this topological $\Z_2$ gauge theory, for a single time step in the periodic Euclidean time direction, i.e. for $\T^2$ of size $1 \times L$,  is then defined as a trace over the physical Hilbert space, 
\begin{eqnarray}\label{z5}
Z_{\T^2} &=& \tr {\bf{1}} =  \sum_{\{s_\ell = \pm 1\}} \langle s_1, s_2, ..., s_L|\hat P_{G} |s_1, s_2, ..., s_L \rangle  \\
&=&{1 \over 2^L} \sum_{\{s_\ell = \pm 1, n_\ell = 1,2\}}  \langle s_1, s_2, ..., s_L   | s_1 (-1)^{n_1 + n_2}, s_2 (-1)^{n_2 + n_3}, ..., s_L (-1)^{n_L + n_1} \rangle \nonumber~.
\end{eqnarray}
The sum above is over $\Z_2$ gauge fields living on the $L$ spatial links ($s_\ell$) and the $L$ timelike links $(n_\ell)$. As is familiar from Yang-Mills theory, and as we  review further below, the time-direction links arise from  the insertion of projection operators on gauge invariant states needed to define the partition function (gauge invariance requires that this  be done on at least one  time slice).
 
 Furthermore, for the gauge field configurations that give nonzero contribution to the r.h.s. of (\ref{z5}),  the $\Z_2$ flux  through every one of the $L$ plaquettes vanishes, because the sum is nonzero only if $n_\ell + n_{\ell+1} = 0$ for all $\ell$, which is precisely the flux through the $\ell$-th plaquette (since for each plaquette the spacelike links are the same, they do not contribute to the flux). Thus, eqn.~(\ref{z5}) is the same as eqn.~(\ref{z4}) with $\#p = L$. We can also see this explicitly, by finishing the computation of (\ref{z5}) (using $\langle s_1 | s_1 (-)^{n_1 + n_2}\rangle = \delta_{n_1 + n_2, 0}$):
 \begin{eqnarray}\label{z6}
Z_{\T^2} &=&  {1 \over 2^L} \sum_{\{s_\ell = \pm 1 \}} \sum_{\{ n_\ell = 1,2\}}\delta_{n_1+n_2, 0} \; \delta_{n_2+n_3, 0}\; ... \; \delta_{n_L+n_1, 0} = {2^L \over 2^L} \times  2  = 2 \end{eqnarray}
The last overall factor of $2$ arises because the product of delta functions requires that all $n_\ell$ be the same. We note that the two $n_1 = n_2 = ...= n_\ell$ configurations are precisely the saddle points of the continuum action on $\T^2$ eqn.~(\ref{z2action1}), with $\oint a^{(1)} = (0, \pi)$ in the Euclidean time direction.
In (\ref{z5}), these configurations are included in the path integral, but their contribution is divided out by the normalization of the projector $\hat P_{G}$.\footnote{\label{footnoteG}The clearest case is the one of $L=1$, where no gauge transformation acts nontrivially on the spin, yet the projector involves a sum over the two values of $n_1$, but divided by two.}

The  action of the projector on gauge invariant states $\hat P_{G}$ of eqn.~(\ref{projector}):
\begin{equation}\label{p1}
\hat P_{G} |s_1, s_2, ..., s_L \rangle = {1 \over 2^L}  \sum_{\{ n_\ell = 1,2 \}}  | s_1 (-1)^{n_1 + n_2}, s_2 (-1)^{n_2 + n_3}, ..., s_L (-1)^{n_L + n_1} \rangle,
\end{equation}
 shows that the two gauge transformations with $n_1=n_2 = ... = n_L$ act  trivially on all states.\footnote{Another peculiarity of this theory as that $\hat P_{G}$ projects on a two-dimensional space (there are $2^L$ vectors in Hilbert space but $2^{L-1}$ nontrivial gauge transforms, thus making the physical Hilbert space dimension $=2$).}  Another lesson from (\ref{p1}) is that two gauge transformations---one with $(n_1, n_2, ..., n_L)$  and the other with $(n_1+1, n_2+1, ..., n_L + 1)$ (i.e., the second  with $n_\ell \rightarrow n_\ell +1$, simultaneously for all $\ell$, of course all taken (mod $2$))---act identically on all states. Thus, we can define the same projector by summing over all $\{ n_\ell\}$ but identifying configurations where $n_\ell \equiv n_\ell + 1$, simultaneously for all $\ell$:
 \begin{equation}\label{p11}
\hat P_{G} |s_1, s_2, ..., s_L \rangle = {1 \over 2^{L-1}}  \sum_{\{ n_\ell = 1,2 \}}  | s_1 (-1)^{n_1 + n_2}, s_2 (-1)^{n_2 + n_3}, ..., s_L (-1)^{n_L + n_1} \rangle\bigg\vert_{n_1 +1, n_2 + 1,..., n_L+1 \equiv n_1, n_2, ... n_L},
\end{equation}
where we indicated that configurations where all $n_\ell$ differ by unity are not summed over. 
This identification reduces the number of configurations by two and is now accounted for in the normalization of the projector.
 
We shall now attempt to pursue the  analogy with  Yang-Mills theory. Owing to its nonabelian and non-topological nature, the analogy is not complete, but there is merit in studying the transition from the Hamiltonian to the Euclidean path integral formulation.

\subsection{From the Hilbert space to the path integral in the single-cube lattice Yang-Mills theory with 't Hooft flux}

\label{appx:g2}

\begin{figure}[t] %  figure placement: here, top, bottom, or page
   \centering
   \includegraphics[width=3in]{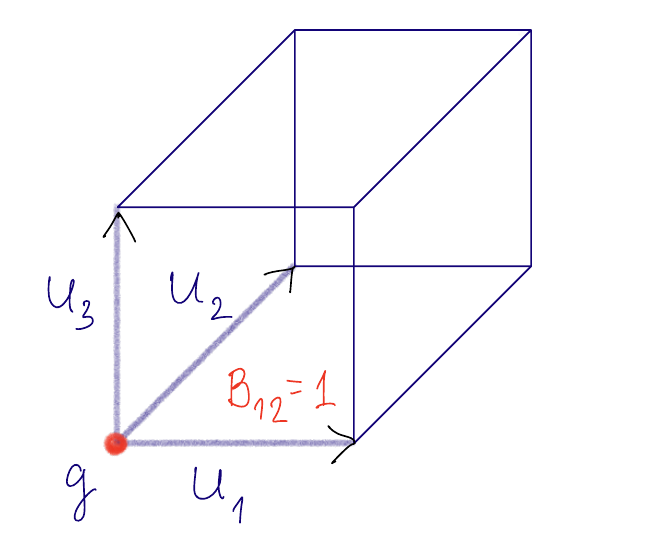} 
   \caption{The single cube lattice  Yang-Mills theory with a unit 't Hooft flux (represented by a two-form $\Z_2$  background field, $B_{12} = 1$, and reflected in the change of sign in the $1-2$ plane plaquette potential energy).  There is only one gauge transformation parameter $g$ in this single-cube theory.  }
   \label{single cube}
\end{figure}

Consider $SU(2)$ Yang-Mills theory with a single unit of 't Hooft flux $n_{12}=1$ defined on a single-cube spatial lattice. There are three link variables and a single gauge transformation parameter $g$, see Figure \ref{single cube}.  The   lattice Kogut-Susskind Hamiltonian is as follows:
\begin{eqnarray}\label{kogutsusskind1}
\hat H= {g^2 \over 2 a} \hat J_i^a \hat J_i^a - {2 \over g^2 a} \left(- \tr \hat U_1 \hat U_2 \hat U_1^{-1} \hat U_2^{-1} + \tr \hat U_2 \hat U_3 \hat U_2^{-1} \hat U_3^{-1} + \tr \hat U_1 \hat U_3 \hat U_1^{-1} \hat U_3^{-1} \right)
\end{eqnarray}
where $\hat J^a_i$ are the electric field operators (classically, $\sim (\dot U_i U_i^{-1})^a$) and $a$ is the lattice spacing (we recalled that traces of group elements are real in $SU(2)$ gauge theories). We work in the $g^2 \rightarrow 0$ limit, thus focusing on minimizing the classical potential energy. The minus sign in front of the $1-2$ plane plaquette represents the insertion of a unit 't Hooft flux. 

Clearly, the classical potential is minimized when the $2-3$ and $1-3$ plaquette traces are $+2$ while the $1-2$ plaquette is $-2$, thus requiring: 
\begin{eqnarray}\label{minima1}
U_1 U_2 &=& - U_2 U_1, \nonumber \\
U_2 U_3 &=& U_3 U_2, \\
U_1 U_3 &=& U_3 U_1. \nonumber 
\end{eqnarray}
Clearly, up to gauge transformations, we have that 
\begin{eqnarray}\label{minima2} 
U_1 = i \sigma_1, U_2 = i \sigma_2, U_3 = \pm 1~.
\end{eqnarray}
The gauge invariant characterization of these states is that $W_1 ={1\over 2}  \tr U_1 = W_2 = {1\over 2} \tr U_2 = 0$ and $W_3 = {1\over 2} \tr U_3 = \pm 1$. These are exactly the two vacua with classically broken center symmetry along $x_3$ that contribute to the Witten index shown in (\ref{wilson witten})---which can be recovered in this simple single-cube world with a 't Hooft twist. 
Thus, the classical states can be denoted by
\begin{eqnarray}
|U_1, U_2, U_3\rangle_{class.} = |i \sigma_1, i \sigma_2, \pm 1\rangle,
\end{eqnarray}
 where we use position space eigenvectors, $\hat U_1 |U_1 \rangle = |U_1\rangle U_1$, etc., similar to the $\Z_2$ theory of the previous section.
 
Now, we proceed to define a single time-step, $\epsilon$, partition function. To this end, we  need a projector on gauge invariant states, which acts as 
\begin{eqnarray}
\hat P_{G} |U_1, U_2, U_3\rangle = \int\limits_{SU(2)} d g \; | g U_1 g^{-1}, g U_2 g^{-1}, g U_3 g^{-1} \rangle, ~\rm{with} ~ \int\limits_{SU(2)} dg =1,
\end{eqnarray}
where the integral is defined with the $SU(2)$ Haar measure. 
Notice, however, that for all $U_i$, the gauge transformations $g$ and $-g$ act identically.\footnote{The action of this single gauge transformation corresponds to   the action of $n_1 = n_2 = n_3 =... = n_L = 1$ and $n_1 = n_2 = ... = n_L = -1$ in the $\Z_2$ gauge theory of the previous section (the analogy is more pronounced if one takes $L=1$ there, recall footnote \ref{footnoteG}). The difference is, however, that due to the abelian nature there, none of these gauge transformations act on any states, while here they do.} Thus, we could also define the projector by restricting $g$ to be in $PSU(2)=SO(3)$, since the  equation below is an identity:
\begin{eqnarray}\label{projector9}
\hat P_{G} |U_1, U_2, U_3\rangle &=& \int\limits_{SU(2)} d g \; | g U_1 g^{-1}, g U_2 g^{-1}, g U_3 g^{-1} \rangle = \int\limits_{PSU(2)} d g' \; | g' U_1 g^{' \;-1}, g' U_2 g^{'\; -1}, g' U_3 g^{' \;-1} \rangle, \nonumber \\
  &&~\rm{provided \;we \; normalize} ~ \int\limits_{PSU(2)} dg' =1 \; ~\rm{and} \; \int\limits_{SU(2)} dg =1.
\end{eqnarray}
The identity follows from the fact that the Haar measures differ by restricting one of the Euler angles and the fact that the integrand is the same in the second cover of $PSU(2)$. We note that eqn.~(\ref{projector9}) is the counterpart of the two definitions of the projector in the $\Z_N$ theory (recall eqns.~(\ref{p1}, \ref{p11})).

The  transfer matrix, written without specifying whether we use $g$ or $g'$ from (\ref{projector9}), is the matrix element of $e^{- \epsilon \hat H}$ between  general Hilbert space vectors on two neighboring time slices, with a projector on gauge invariant states inserted:
\begin{eqnarray}
\langle U_1', U_2', U_3'| e^{- \epsilon \hat H} \hat P_G |U_1, U_2, U_3\rangle = \int d g \; \langle U_1', U_2', U_3'| e^{- \epsilon \hat H}  |g U_1 g^{-1}, g U_2 g^{-1}, g U_3 g^{-1} \rangle.
\end{eqnarray}
Then we use the well known expression for the matrix element of the transfer matrix \cite{Creutz:1976ch,Luscher:1976ms} 
$$\langle U'_1 U'_2 U'_3  | e^{- \epsilon \hat H} | U_1 U_2 U_3\rangle =  e^{ {2 a \over g^2 \epsilon} \sum_{i=1}^3 \tr U'_i U_i^{-1} + {2 \epsilon \over a g^2}    \left(- \tr U_1 U_2 U_1^{-1} U_2^{-1} + \tr U_2 U_3 U_2^{-1} U_3^{-1} + \tr U_1 U_3 U_1^{-1} U_3^{-1} \right)}$$
where we skip the overall constant (which can be written out but is not informative; in addition, in super-Yang-Mills it should cancel with the fermion contribution, in an ideal supersymmetric-lattice world).
Thus, we
 obtain for the single time step partition function, defined as a trace over the physical Hilbert space:
\begin{eqnarray} \label{z11}
Z &=& \tr e^{- \epsilon \hat H} = \int dU_1 dU_2 dU_3\;  \langle U_1 U_2 U_3 | e^{- \epsilon \hat H} \hat P_G | U_1 U_2 U_3\rangle  \\
&=&  \int dU_1 dU_2 dU_3 d g \; \langle U_1 U_2 U_3 | e^{- \epsilon \hat H}    |g U_1 g^{-1}, g U_2 g^{-1}, g U_3 g^{-1} \rangle  \nonumber \\
&=& \int dU_1 dU_2 dU_3 d g \; e^{ {2 a \over g^2  \epsilon}  \sum_{i=1}^3 \tr U_i g U_i^{-1} g^{-1} + {2 \epsilon \over a g^2}    \left(- \tr U_1 U_2 U_1^{-1} U_2^{-1} + \tr U_2 U_3 U_2^{-1} U_3^{-1} + \tr U_1 U_3 U_1^{-1} U_3^{-1} \right)} \nonumber
\end{eqnarray}
Semiclassically, at small coupling $g^2$, the spatial plaquettes in the exponent are maximized when the $U_i$'s obey (\ref{minima1}, \ref{minima2})---this is clear, since potential energy is same as in (\ref{kogutsusskind1}). On the other hand,  the kinetic term is maximal only when $g$ commutes with all $U_i$ saddles (\ref{minima2}). However, since they are proportional to $\sigma_{1,2,3}$ in the three directions, it must be that $g$ be $\pm 1$, if we allow $g$ in $SU(2)$ or $g=+1$ if we use $PSU(2)$. But in view of (\ref{projector9}) there shouldn't be a difference, since the two cases should be identical.
 
 Recalling that ${1 \over 2} \tr g$ is our fourth direction $W_4$, 
we notice that the minimum action saddles that we found above are precisely the ones of (\ref{zeroaction many}) for $N=2$, with $W_1=W_2=0$ and $W_{3}= \pm 1$, $W_4 = \pm 1$. The definition of the path integral measure suggests that the contributions of the two $W_4 = \pm 1$ saddles is counted, but divided out in the partition function.

Finally, we note that in this single-hypercube world, one can go further and study the ``fractional instantons,"
by including a second 't Hooft twist in the $3-4$ plane.  To this end,  the twisted partition function is, relabeling $g \rightarrow U_4$:  \begin{eqnarray}\label{z53}
 \tr e^{- \epsilon \hat H} \hat T_3 &=& \\
 &=& \int dU_1 dU_2 dU_3 d g \; e^{ {2 a \over g^2  \epsilon}   \left(  \tr U_1 U_4 U_1^{-1} U_4^{-1} + \tr U_2 U_4 U_2^{-1} U_4^{-1} - \tr U_3 U_4 U_3^{-1} U_4^{-1} \right)} \nonumber \\
 && \qquad \qquad \qquad \times \; e^{ {2 \epsilon \over a g^2}     \left(- \tr U_1 U_2 U_1^{-1} U_2^{-1} + \tr U_2 U_3 U_2^{-1} U_3^{-1} + \tr U_1 U_3 U_1^{-1} U_3^{-1} \right)}~. \nonumber
 \end{eqnarray}
The action is now frustrated by the $3$-$4$ plaquette twist and---as far as we know---the best one can do analytically is to prove that the action is strictly larger than the minimum action in  (\ref{z11}) (i.e. show that a configuration where all terms in the exponent  in (\ref{z53}), taken with the appropriate signs,  $\pm \Pi_{ij} \equiv   \pm \tr U_i U_j U_i^{-1} U_j^{-1}$ achieve their maximum value $=2$, for all $i,j=1,...4$, is impossible).

 Taking $a=\epsilon$, one can now
 ask what are the single-hypercube analogues of the fractional instantons. The minimization of the classical action can be performed numerically, as in \cite{Wandler:2024hsq}---where it was done on larger lattices, with results close to the continuum limit and agreeing with it within errors. The result\footnote{We are grateful to Andrew Cox for performing this numerical simulation.} is that there are $8$ minimum action ``fractional instanton" configurations in the single-hypercube twisted torus. In all of them the values of the plaquettes $\Pi_{ij} \equiv \tr U_i U_j U_i^{-1} U_j^{-1}$ are $\Pi_{13} = \Pi_{14} = \Pi_{23} = \Pi_{24} = 2$ (i.e. they take their maximum value, thus maximizing their contribution to the exponent in (\ref{z53})). In the  $1-2$ and $3-4$ plaquettes, however, two distinct types of configurations occur
 \begin{eqnarray} \label{g33}
 \text{type} \; 1:&& \Pi_{12} = -2, \Pi_{34} =  2, \tr U_3 = \pm 1, \tr U_4 = \pm 1, \tr U_1 = \tr U_2 = 0, \nonumber \\
 \text{type} \; 2:&&\Pi_{12} =  2,  \Pi_{34} =  -2, \tr U_1 = \pm 1, \tr U_2 = \pm 1, \tr U_3 = \tr U_4 = 0.
 \end{eqnarray}
 Clearly, the action is the same in all these configurations. 
  As opposed to the continuum constant flux fractional instanton of 
' t Hooft, in each ``instanton" shown in (\ref{g33}) there is unbroken center symmetry in the two directions with nonminimal contribution to the action (``flux"), so there are only four center symmetry copies of each instanton.\footnote{These two types of configuration  occur equally often, starting the minimization algorithm from a random initial value, after generating thousands of minimum action configurations. In addition, further traces $\tr U_i U_j$, etc.,  were studied to corroborate the statement that in each configuration there is unbroken center symmetry.} Here again, the definition of the measure (\ref{projector9}) suggests that the two configurations with $\tr U_4 = \pm 1$ are summed over, with their contribution divided out in the path integral (or identified), as in the $\Z_N$ gauge theory. We take this to suggest that images of the fractional instantons under center symmetry in the $x_4$ direction should not be counted also in the continuum super-Yang-Mills theory.

  \bibliography{ReferencesALLrevsd.bib}
  
  \bibliographystyle{JHEP}
  \end{document}